\documentclass[11pt,a4paper]{article}
\pdfoutput=1

\usepackage[utf8]{inputenc} 
\usepackage{bbold}
\usepackage{slashed}
\usepackage{a4wide}
\usepackage{amsmath}
\usepackage{mathrsfs}
\usepackage{cite}       
\usepackage[table]{xcolor}
\usepackage{amssymb}
\usepackage{amsfonts}
\usepackage{booktabs}
\usepackage{makecell}
\usepackage{multirow}
\usepackage{pdflscape}
\usepackage{textcomp}
\usepackage{gensymb}
\usepackage{bigstrut}
\usepackage{array}
\usepackage{enumerate}
\usepackage{enumitem}
\usepackage[small,bf]{caption}
\usepackage{subcaption}
\usepackage{diagbox}
\usepackage{graphicx}
\usepackage{mathbbol}
\usepackage{appendix}
\usepackage{verbatim}
\usepackage[top=2.5cm,bottom=2.5cm,left=3cm,right=3cm]{geometry}
\usepackage{makecell}
\usepackage{listings}
\usepackage{mathtools}
\usepackage{dsfont}
\usepackage{braket}
\usepackage{longtable}
\usepackage{threeparttable}
\usepackage{floatpag}
\usepackage[colorlinks=true,urlcolor=myred,anchorcolor=black,citecolor=myred,filecolor=black,linkcolor=myred,menucolor=blue,linktocpage=true]{hyperref}
\usepackage[normalem]{ulem}
\usepackage{multicol}
\newcommand{\be}{\begin{equation}}
\newcommand{\ee}{\end{equation}}
\usepackage{afterpage}
\usepackage{colortbl}
\usepackage{nicematrix}

\usepackage[multiple]{footmisc}

\usepackage[capitalise]{cleveref} 
\crefname{table}{Table}{tables}

\definecolor{myred}{rgb}{0.8, 0, 0.4}
\definecolor{myblue}{cmyk}{0.8, 0.4, 0, 0.2}
\definecolor{mygreen}{rgb}{0.27, 0.64, 0.28}
\definecolor{mygray}{gray}{.95}
\definecolor{verdes}{cmyk}{0.92,0,0.59,0.4}

\DeclareMathOperator{\diag}{diag}
\DeclareMathOperator{\tr}{Tr}
\DeclareMathOperator{\im}{Im}

\newcolumntype{x}[1]{>{\centering\arraybackslash\hspace{0pt}}p{#1}}



\allowdisplaybreaks
\numberwithin{equation}{section}

\begin{document}

\thispagestyle{empty}

\vspace*{-15mm}
\begin{flushright}
\end{flushright}
\vspace*{5mm}

\vspace{2cm}

\renewcommand*{\thefootnote}{\fnsymbol{footnote}}

\begin{center}
{\bf 
{\LARGE Quark hierarchies and CP violation\\[2mm] from the Siegel modular group}
}\\[8mm]
M.~Carducci$^{\,a,b~}$\footnote{E-mail: \href{mailto:marco.carducci@uniroma3.it}{\texttt{marco.carducci@uniroma3.it}}},
D.~Meloni$^{\,a,b~}$\footnote{E-mail: \href{mailto:davide.meloni@uniroma3.it}{\texttt{davide.meloni@uniroma3.it}}},
M.~Parriciatu$^{\,a,b~}$\footnote{E-mail: \href{mailto:matteo.parriciatu@uniroma3.it}{\texttt{matteo.parriciatu@uniroma3.it}}},
J.~T.~Penedo$^{\,b~}$\footnote{E-mail: \href{mailto:jpenedo@roma3.infn.it}{\texttt{jpenedo@roma3.infn.it}}} \\[2mm]
$^{a}$\,{\it \small  Dipartimento di Matematica e Fisica, Università di Roma Tre,\\
Via della Vasca Navale 84, 00146, Roma, Italy} \\
$^{b}$\,{\it \small INFN Sezione di Roma Tre, Via della Vasca Navale 84, 00146, Roma, Italy} \\

\end{center}

\vskip 7mm      
\begin{center}
\textbf{Abstract}
\end{center}
\vspace{-.5em} 
We investigate theories of flavour based on genus $g=2$ modular invariance and analyze how fermion mass hierarchies can be generated in this context, in the vicinity of invariant points or regions in moduli space where a residual symmetry is preserved.
We apply this mechanism of modular proximity-induced hierarchies to the quark sector, with the vacuum expectation values of the moduli being the only sources of spontaneous breaking of the flavour and CP symmetries.
We present a benchmark model where quark mass hierarchies and CP violation are explained, with mass ratios vanishing in the symmetric limit, and quark mixing is reproduced. In this model, the values of the moduli turn out to be close to special values such as $\tau_1 \simeq \omega$ and $\tau_2 \simeq \omega, i$. 

\renewcommand*{\thefootnote}{\arabic{footnote}}
\setcounter{footnote}{0}

%
\clearpage
\vfill
\tableofcontents
\vskip 1cm
\hrule
\vskip 1cm
\section{Introduction}
\label{sec:intro}
%
Understanding the peculiar fermion mixing patterns, mass hierarchies, and the origin of the CP violation (CPV) encoded in the Yukawa sector of the Standard Model (SM) remains one of the most profound open questions in particle physics.
The situation is especially challenging for the quark sector, where hierarchies are ubiquitous, and mixing angles and the CPV phase have been measured with extraordinary precision, at or below the few percent level~\cite{ParticleDataGroup:2024cfk}.

The flavour puzzle has been studied from the point of view of several approaches, see e.g.~\cite{Xing:2020ijf,Feruglio:2025ztj} for recent reviews.
The symmetric approach is arguably one of the most promising, in virtue of its potential unifying and explanatory power~\cite{Altarelli:2010gt,Ishimori:2010au,King:2014nza,Petcov:2017ggy,Feruglio:2019ybq,Ding:2024ozt}.
A putative family (or horizontal) symmetry should select particular directions in the space of family replicas, producing the distinctive structures one observes, i.e.~fermion masses spanning nearly six orders of magnitude and quark (lepton) mixing angles that are highly suppressed (rather sizeable) as low-energy relics.
Nevertheless, the lack of experimental signatures from Beyond-Standard-Model (BSM) physics suggests that the agents of such a flavour mechanism must be decoupled from SM dynamics by many orders of magnitude in energy.
This fact potentially leaves open too large a space of possibilities, reducing the appeal of the approach. 
Moreover, when confronted with the traditional literature on the subject, some common sentiments emerge:
\begin{enumerate}
  \item It seems that too large a number of models is able to accommodate the data. Usually, 
  the success of a model hinges on finding the correct tuning of ingredients, with the choice of symmetry group, symmetry breaking sector, field charges and irreducible representations being apparently limited only by one's imagination.
  \item \label{it:2} On the other hand, it seems that fermion data is not compatible with too much symmetry, i.e.~the most minimal models with too few free parameters are not able to accommodate the data. For this reason, it can be argued that in some of the traditional approaches, symmetry acts merely as a tool to slightly reduce the number of free parameters when compared to the SM. 
  \item \label{it:3} In some cases, the complexity contained in the set of free parameters (either in its size or tuned nature) is traded by apparently ad-hoc structures, such as additional heavy fields and/or direct products of symmetry groups $G_1\times G_2\times\ldots$, where each $G_i$ addresses a different aspect of the puzzle. A common example of the latter strategy is to have one group being responsible for the mixing, while another one induces the mass hierarchies. It can be argued that this \emph{modus operandi} is, in many cases, in no way ``more natural" than the artificial tuning of the parameters in a simpler model.
\end{enumerate}
It is a formidable task to avoid these unpleasant features in ``bottom-up'' flavour model building altogether.
However, one could start by insisting on the intended unifying power of the approach, and follow a direction orthogonal to~\cref{it:3}. Namely, one may ask for the explanation of multiple aspects of the flavour puzzle using fewer common ingredients. Facing this direction, one is led to consider the following thought: traditional group-theoretical considerations may not be sufficient. The flavour group is usually chosen first, and the symmetry-breaking sector is constructed later in order to obtain the desired relic structures in the mass matrices. Reversing the logic, one can look for physically and mathematically motivated symmetry-breaking sectors and then derive the compatible symmetries~\cite{Feruglio:2022cgv}. A choice of the topology of the symmetry-breaking space is one way to do it, in what one may call ``topology-driven'' flavour model building.
Topological concepts have assumed an increasingly important role across theoretical physics: from the underlying differential-geometric structure of the spacetime manifold in general relativity, to string theory, where the compactification geometry determines the Yukawa couplings, and to the ADM formalism which constitutes an essential ingredient in loop quantum gravity (see e.g.~\cite{Rovelli:2014ssa}).

\vskip 2mm
An example of topology-driven flavour model building is provided by the modular invariance program~\cite{Feruglio:2017spp}, as proposed in 2017 (see~\cite{Kobayashi:2023zzc,Ding:2023htn} for reviews).
The symmetry breaking is determined by a choice of the shape of the torus, which is the assumed underlying geometry of flavour space. In particular, the broken symmetry is $SL(2,\mathbb{Z})$, and the torus shape is described by the vacuum expectation value (VEV) of the modulus $\tau$, which is assumed to be a complex spurion and a SM gauge singlet, living in the upper half plane $\text{Im}\,\tau>0$.
From the top-down, this is motivated since fermion masses usually depend on multiple moduli in theories of superstring compactifications, see e.g.~\cite{Blumenhagen:2013fgp,Greene:1986qva,Bailin:1999nk}.
From the bottom-up perspective, the Yukawa couplings of the Higgs sector are assumed to be modular forms of a certain weight, which can be organized in multiplets of the finite modular group $\Gamma_N$, isomorphic for each level $N$ to traditional, non-Abelian discrete groups like $S_3$, $A_4$, $S_4$, and $A_5$.
However, unlike the more traditional approaches involving flavons, here the internal structure of a given vector of modular forms of a certain weight is entirely determined by the numerical VEV of the modulus. Indeed, the assumed symmetry is non-linearly realized: the modulus controls the modular forms via their Fourier expansion coefficients as $\exp(2\pi i\,\tau)$. For this reason, in minimal setups, a modular theory of flavour can be written at the renormalizable level.

Striving for minimality, one can impose a generalized CP symmetry (gCP) on the theory, which will be spontaneously broken by the modulus VEV. This choice is motivated by both top-down and bottom-up perspectives. For the former, in string theory it has been conjectured that only the complex VEV of fields can break CP, which is a gauge symmetry of the four-dimensional theory~\cite{Dine:1992ya,Choi:1992xp,Leigh:1993ae}.%
\footnote{See also~\cite{Acharya:1995ag,Dent:2001cc,Giedt:2002ns,Baur:2019kwi} for the action of CP on the modulus in string-inspired models.}
For the latter, it was shown that a generalized CP symmetry is compatible with the modular flavour framework, with a notable consequence being that all superpotential parameters have to be real in an appropriate basis~\cite{Novichkov:2019sqv}, further constraining the models.

Modular flavour symmetry has also been employed in understanding fermion mass hierarchies, using the approach of residual symmetries~\cite{Feruglio:2021dte,Novichkov:2021evw}. By stabilizing the VEV of the modulus at a small distance $\epsilon\ll 1$ from special (fixed) points inside the moduli space, it was shown that specific assignments of matter irreducible representations and weights are able to generate mass spectra with a hierarchical structure $\sim(1,\epsilon^a,\epsilon^b)$, where $a$ and $b$ are integers which depend also on the chosen point of residual symmetry. Importantly, the coefficients within this structure are dictated by symmetry constraints, the modulus VEV and a (potentially) small number of parameters, which is in stark contrast with the Froggatt-Nielsen mechanism~\cite{Froggatt:1978nt}, where mass matrix entries are uncorrelated in general. We refer to this mechanism as \textit{Modular Proximity-Induced Hierarchies} (MPIH) in what follows.

Given all its features, it does not come as a surprise that, in recent years, modular invariance has also been used to tackle other issues adjacent to the flavour puzzle, e.g.~the strong CP problem~\cite{Feruglio:2023uof,Penedo:2024gtb,Feruglio:2024ytl,Feruglio:2025ajb}, baryogenesis~\cite{Duch:2025abl} and inflation~\cite{Abe:2023ylh,Ding:2024neh,Ding:2024euc,Aoki:2025wld}.
For instance, it has been shown that the same $\epsilon$ which governs charged-lepton mass hierarchies via the MPIH mechanism can be responsible for the breaking of lepton number $L$ in a symmetry-protected scenario~\cite{Granelli:2025lds}. In this context, the small, common parameter can thus source light Majorana neutrino masses, flavour patterns and the splitting of a pseudo-Dirac heavy neutrino pair, providing a bridge to the phenomenology of heavy neutral leptons.

\vskip 2mm
To fully exploit the explanatory and unifying potential of the modular approach, one must confront a crucial issue regarding MPIH.
The special points of $SL(2,\mathbb{Z})$, i.e.~$\tau_{\text{sym}}=\{\omega, i , i\infty\}$, are CP-conserving. In a theory where CP is spontaneously violated by $\tau$, as previously discussed, this is an important hurdle for realistic model building in the quark sector. It seems that the required departure $\epsilon\sim \mathcal{O}(10^{-2})$ from the special points is not enough to generate sufficiently large CPV in MPIH quark models~\cite{Petcov:2022fjf,Petcov:2023vws,deMedeirosVarzielas:2023crv,Petcov:2026mdx},%
\footnote{This issue has also been discussed in Ref.~\cite{Kikuchi:2023jap}.} where the ratios of quark masses approach zero in the symmetric limit.\footnote{During the development of this work, Ref.~\cite{deMedeirosVarzielas:2026lfw} appeared, in which a fit close to $\tau\simeq i$ was found in a gCP scenario. However, the mass hierarchies are explained by the proximity to this point only partially, since the residual symmetry is a $\mathbb{Z}_2$.}
Insisting on a top-down motivated gCP symmetry, one is led to the following consideration: using a single modulus may be an oversimplification. Indeed, and as previously stated, in top-down approaches fermion masses often depend on multiple moduli. 
While the single modulus case corresponds to a space of genus $g=1$, the natural extended framework would rely on genus $g=2$ modular invariance.

At $g=2$, the modular group is the symplectic group $Sp(4,\mathbb{Z})$, the moduli space is the Siegel upper-half space, and special points are generalized to (potentially larger-dimensional) fixed regions where CP may be broken. 
This symplectic or Siegel approach to the flavour problem was recently discussed in~\cite{Ding:2020zxw}, further developed from the point of view of gCP in~\cite{Ding:2021iqp}, employed for generating lepton masses and mixing near fixed points in~\cite{Ding:2024xhz} and most recently considered within an inflation model~\cite{Jiang:2025qbi} (see also~\cite{Kikuchi:2023dow} for flavour model building at $g=3$). 
While Ref.~\cite{Ding:2024xhz} has examined the lepton sector in the vicinity of special points, with the symmetry realized through triplets representations, here we investigate the quark sector, which may favour a $\mathbf{2}\oplus \mathbf{1}$ family assignment. We have systematically analyzed the MPIH mechanism at genus two, showing that it goes beyond the naïve intuition of a direct product of two tori. For the first time in the modular literature, we have obtained a fit to quark data implementing both MPIH and gCP, with all quark mass ratios vanishing in the symmetric limit.

This paper is organized as follows: in~\cref{sec:topology} we briefly review the Siegel space, symplectic transformations, Siegel modular forms, gCP and the model building framework; in~\cref{sec:mpih} we detail the MPIH mechanism at genus two; in~\cref{sec:model} we provide three benchmarks within a viable quark model based on MPIH and gCP; finally, in~\cref{sec:summary} we summarize our findings.

\section{A glimpse into the Siegel framework}
\label{sec:topology}
In the genus-one scenario, the relevant coset space $G/K=SL(2,\mathbb{R})/SO(2,\mathbb{R})$, by virtue of the Iwasawa decomposition~\cite{Fleig:2015vky}, can be identified with the upper-half plane ${\cal H}$:
\begin{equation}
  {\cal H} \;=\; \left\{ \tau \in \mathbb{C} \;\middle|\; \mathrm{Im}\,\tau > 0 \right\}\,,
\end{equation}
and the left action of an element $\gamma\in G$ on the modulus $\tau\in {\cal H}$ yields:
\begin{equation}
  \tau\;\longmapsto\;\frac{a\tau+b}{c\tau+d}\, , \qquad\begin{pmatrix}
    a&b\\c&d
  \end{pmatrix}\in SL(2,\mathbb{R})\,.
\end{equation}
In this framework, the complex structure emerges since $SO(2)\cong U(1)$ and modular forms are associated with the discrete subgroup $SL(2,\mathbb{Z})\subset SL(2,\mathbb{R})$.

The higher-genus generalization ($g>1$) is provided by the Siegel upper-half space of genus $g$, defined as
\begin{equation}
  {\cal H}_g   \;=\;
  \left\{  \tau_{(g)} \in GL(g,\mathbb{C})
  \;\middle|\;  \tau_{(g)} = \tau_{(g)}^T,
  \,\,\mathrm{Im}\,\tau_{(g)} > 0  \right\}\,,
\end{equation}
where $\tau_{(g)}$, the generalization of the modulus $\tau$~\cite{Hamidi:1986vh}, is called the period matrix, and its imaginary part is required to be positive definite. This space can be identified with the coset $Sp(2g,\mathbb{R})/U(g)$, whose complex structure is due to the maximal compact subgroup $U(g)$~\cite{Alvarez-Gaume:1986bwm}.
The symplectic group $Sp(2g,\mathbb{R})$ acts on ${\cal H}_g$ via%
\footnote{With the appropriate conditions on the submatrices $A$, $B$, $C$ and $D$ in order to leave invariant the symplectic form:
$$
J=\begin{pmatrix}
0_g&\mathbb{1}_g\\
-\mathbb{1}_g & 0_g
\end{pmatrix}
\,.
$$
}
\begin{equation} \label{eq:transf_g}
  \tau_{(g)} \;\longmapsto\;
  (A\tau_{(g)} + B)(C\tau_{(g)} + D)^{-1}\,, \qquad
  \begin{pmatrix}
    A & B \\
    C & D
  \end{pmatrix} \in Sp(2g,\mathbb{R})\,,
\end{equation}
and, in analogy with the genus-one case, modular forms are now generalized to Siegel modular forms, which are associated with the discrete subgroup $Sp(2g,\mathbb{Z})\equiv \Gamma_g$.
A Riemannian metric for the space $\mathcal{H}_g$, left invariant by $\Gamma_g$, is given by~\cite{Siegel:1943}: 
\begin{equation}
    \label{eq:Riemannmetric}
    ds^2=\tr\left(Y^{-1}d\tau_{(g)}\,Y^{-1}d\bar{\tau}_{(g)}\right)\,,
\end{equation}
where $Y\equiv \im\,\tau_{(g)}$.
The length of the geodesic connecting two points $\tau_{(g)}$ and $\tau'_{(g)}$, induced by the metric~\eqref{eq:Riemannmetric}, is invariant under the Siegel transformations of~\cref{eq:transf_g}. It is given by~\cite{Siegel:1943}:
\begin{equation} \label{eq:geolength}
  d(\tau_{(g)},\tau_{(g)}')=\left(\sum_{k=1}^g\ln^2\frac{1+\sqrt{\lambda_k}}{1-\sqrt{\lambda_k}}\right)^{1/2}\,,
\end{equation}
where $\lambda_k$ are the eigenvalues of the Hermitian cross-ratio $\mathcal{D}_g$, defined as
\begin{equation} \label{eq:crossratio}
  \mathcal{D}_g(\tau_{(g)},\tau_{(g)}')=(\tau_{(g)}-\tau_{(g)}')(\tau_{(g)}-\bar\tau_{(g)}')^{-1}(\bar{\tau}_{(g)}-\bar{\tau}_{(g)}')(\bar{\tau}_{(g)}-\tau_{(g)}')^{-1}\,.
\end{equation}
These quantities thus provide a notion of distance in the Siegel upper-half space. In particular, there exists a transformation $\gamma\in \Gamma_g$ connecting two pairs of points in $\mathcal{H}_g$ if and only if these pairs share the same $\mathcal{D}_g$ eigenvalues.

\vskip 2mm
Using the symplectic group for the characterization of the moduli space on which Yukawa couplings depend does not necessarily imply a specific string theory realization, as pointed out in~\cite{Ding:2020zxw}. On the other hand, the presence of $Sp(2g,\mathbb{R})$ has been considered in BSM theories, e.g.~in supergravity scenarios~\cite{Gaillard:1981rj}. 
In the context of string theory, the description of spinors in higher-genus Riemann surfaces depends on the assignment of a set of boundary conditions, that introduces new parameters in the characterization of the surface~\cite{Castellani:1991et,Castellani:1991eu,Castellani:1991ev}.%
\footnote{A rigorous and formal introduction to these topics requires a substantial amount of machinery, such as Calabi-Yau manifolds~\cite{Yau:1977ms}, orbifolds and line bundles~\cite{Marcus:1982fr,Candelas:1985en,Narain:1986am,Bianchi:1991eu,Giveon:1994fu,Sen:1995ff,Witten:1997bs,Hollowood:2003cv,Blumenhagen:2006ci,Dijkgraaf:2007sw,Antoniadis:2009bg,Cordova:2012xk,Hebecker:2017lxm}. The importance of the toroidal orbifolding is outlined, e.g., in~\cite{Reffert:2006du}. Indeed, the toroidal orbifolds incorporate many features of the SM, such as non-Abelian gauge groups, chiral fermions and family repetition~\cite{Dixon:1985jw,Dixon:1986jc}. Phenomenologically-viable Yukawa couplings, in the presence of fluxes and magnetized branes, can be found, e.g., in~\cite{Cremades:2004wa,Kikuchi:2020frp,Kikuchi:2020nxn}.}
Moreover, it turns out that the set of boundary conditions one can assign is in a one-to-one correspondence with what is known as the characteristics of the Riemann theta functions~\cite{Obers:1999es,DHoker:1988pdl}.%
\footnote{The technical challenges involved in constructing a consistent chiral measure for the superstring vacuum-to-vacuum amplitude, together with their resolution in the genus-two case, can be found in~\cite{DHoker:2001kkt,DHoker:2001qqx,DHoker:2001foj,DHoker:2001jaf}.}
A theta function on the Siegel space can be defined as~\cite{Igusaa,DallaPiazza:2008qoi}:
\begin{equation}
\label{eq:thfunction}
    \theta\left[\begin{smallmatrix} a \\ b \end{smallmatrix}\right](\tau_{(g)},z)\equiv\sum_{m\in \mathbb{Z}^g}e^{\pi i\left((m+a/2)\tau_{(g)}(m+a/2)^T+(m+a/2)b^T (z+b/2)\right)}\, ,
\end{equation}
where $a$, $b$, and $m$ are $g$-dimensional row vectors, with $a_i,b_i\in \{0,1\}$ and $m_i \in \mathbb{Z}$.
The matrix $\Delta\equiv\left[\begin{smallmatrix} a \\ b \end{smallmatrix}\right]$ is called a characteristic and the function $e(\Delta)\equiv (-1)^{a\cdot b}$ is called the parity of the characteristic. Moreover, a characteristic is called even if $e(\Delta)=1$ and odd if $e(\Delta)=-1$. 
It can be shown that on a genus $g$ surface there are a total of $2^{2g}$ characteristics, $2^{g-1}(2^g+1)$ are even and $2^{g-1}(2^g-1)$ are odd. The Siegel modular forms at genus $g=2$ will be constructed from the so-called second-order theta constants $\Theta$, which are obtained from the theta functions of~\cref{eq:thfunction} for $z=0$ (see the following subsection).

\subsection{Flavour at genus \texorpdfstring{$g=2$}{g=2}}
\label{sec:flavourgenus2}
In the context of the present work, we are interested in the features of the genus $g=2$ scenario and, thus, in the symplectic group $Sp(4,\mathbb{Z})\equiv \Gamma_2$. Its generators are:
\begin{equation} \label{eq:generators}
    S=
\begin{pmatrix}
0&\mathbb{1}_2\\
-\mathbb{1}_2&0
\end{pmatrix}
\,,\qquad    T_i=
\begin{pmatrix}
\mathbb{1}_2&B_i\\
0&\mathbb{1}_2
\end{pmatrix}
\,,
\end{equation}
where 
\begin{equation}
B_1=\begin{pmatrix}
1&0\\
0&0
\end{pmatrix}
\,,\quad
B_2=\begin{pmatrix}
0&0\\
0&1
\end{pmatrix}\,, 
\quad
B_3=\begin{pmatrix}
0&1\\
1&0
\end{pmatrix}
\,.    
\end{equation}
At this genus, moduli are described by a $2\times 2$ complex symmetric matrix $\tau$:
\begin{equation} \label{eq:modulusg2}
\tau=\begin{pmatrix}
\tau_1&\tau_3\\
\tau_3&\tau_2    
\end{pmatrix}\,,
\qquad \det \im \tau >0\,,\quad\tr \im \tau>0\,,
\end{equation}
and live in the Siegel upper-half space ${\cal H}_2$. 

When looking for a theory of flavour with unitary irreducible representations (irreps) of a finite, compact group, in analogy to the genus $g=1$ case, one could consider the finite version $\Gamma_{2,n}$ of $Sp(4,\mathbb{Z})$, known as the finite Siegel modular group and defined as the quotient $\Gamma_{2,n}\equiv Sp(4,\mathbb{Z})/\Gamma(n)$. Here, $\Gamma(n)$
is the principal congruence subgroup $\Gamma(n)$ of level $n$:
\begin{equation} \label{eq:pcs}
\Gamma(n)=\Big\{\gamma \in Sp(4,\mathbb{Z}) \,\Big|\,  \gamma \equiv \mathbb{1} \,\,(\text{mod}\, n)\Big\}\,,
\end{equation}
with $n\in\mathbb{N}^+$. 
Thus, a Siegel modular form will be a holomorphic function which transforms, under $\gamma\in \Gamma(n)$, as:
\begin{equation} \label{eq:siegmodular}
    f(\tau) \;\longmapsto\; f(\gamma\,\tau)=[\det(C\tau+D)]^kf(\tau)\,,
\end{equation}
where $k$ is an integer, and the determinant allows for the generalization of the $g=1$ automorphy factor $(c\tau+d)^k$.
However, the naïve $g=1$ analogy on the choice of the finite version of our modular group comes with a price. The rich structure of the Siegel space manifests itself in how rapidly the order of the finite group $\Gamma_{2,n}$ grows with the level $n$:
\begin{equation}
\label{eq:finiteord}
|\Gamma_{2,n}|=n^{10}\prod_{p|n}\left(1-\frac{1}{p^2}\right)\left(1-\frac{1}{p^4}\right)\,.
\end{equation}
For instance, we have $|\Gamma_{2,2}|=|\Gamma_{2,4}|=720$, $|\Gamma_{2,3}|=51840$, and $|\Gamma_{2,5}|= 9.36\times 10^6$. 

Choosing the simplest option by modding out for $\Gamma(2)$, one finds the isomorphism $\Gamma_{2,2}\cong S_6$. The emergence of the finite group $S_6$ significantly simplifies the analysis of the structure of a theory, for instance in the context of two-loop superstring amplitudes~\cite{Cacciatori:2007vk}, as studying its finite set of representations is considerably easier than working directly with the full symplectic group. The classical Siegel modular forms of even weight on $\Gamma(2)$, can be written in terms of the second-order theta constants $\Theta[\sigma](\tau_{(g)})\equiv \theta\!\left[\begin{smallmatrix} \sigma \\ 0 \end{smallmatrix}\right](2\tau_{(g)},0)$. At genus two, there are 10 theta functions with even characteristic. By employing the several relations between $\Theta$ and $\theta$, it can be shown that, at level two, the fourth powers of the theta functions form a five-dimensional vector space $\mathcal{M}_k$ of weight $k=2$, i.e.~$\mathcal{M}_2(\Gamma(2))$ spanned by the polynomials:
\begin{equation} \label{eq:w2formsgen}
\begin{aligned}
p_0 &= \Theta^4[00] + \Theta^4[01] + \Theta^4[10] + \Theta^4[11], \\[4pt]
p_1 &= 2\bigl( \Theta^2[00]\Theta^2[01] + \Theta^2[10]\Theta^2[11] \bigr), \\[4pt]
p_2 &= 2\bigl( \Theta^2[00]\Theta^2[10] + \Theta^2[01]\Theta^2[11] \bigr), \\[4pt]
p_3 &= 2\bigl( \Theta^2[00]\Theta^2[11] + \Theta^2[01]\Theta^2[10] \bigr), \\[4pt]
p_4 &= 4\,\Theta[00]\Theta[01]\Theta[10]\Theta[11]\,.
\end{aligned}
\end{equation}
However, as noted in~\cite{Ding:2020zxw}, the smallest non-1D irrep of $S_6$ is the quintet $\mathbf{5}$. At level $n=3$, the corresponding irrep of $\Gamma_{2,3}$ is a quartet $\mathbf{4}$. Such irreps are clearly unsuited for our flavour program with three chiral generations.
Thus, focusing on discrete groups of manageable size, one can argue that the multi-dimensional complex space described by $\tau$ calls for additional constraints, which at the moment do not have a clear top-down justification (see however~\cite{Nilles:2021glx}).
One the other hand, consistent constraints can also be applied from a bottom-up perspective. As proposed in~\cite{Ding:2020zxw}, one may restrict $\tau$ to a region $\Sigma$ of $\mathcal{H}_2$ which enjoys special properties under a subgroup of $Sp(4,\mathbb{Z})$.
One starts by defining the stabilizer group $H\subset Sp(4,\mathbb{Z})$ as the set of transformations that leave certain values of $\tau$ invariant,
\begin{equation} \label{eq:stab}
    H\tau=\tau\,,\qquad \tau\in \Sigma\,.
\end{equation}
It follows that the whole ``fixed'' or ``invariant'' region $\Sigma$ will be mapped to itself by transformations $\gamma\in Sp(4,\mathbb{Z})$ satisfying:
\begin{equation} \label{eq:norm}
\gamma^{-1}H\gamma=H\,.
\end{equation}
Such transformations define the normalizer $N(H)$, and of course $H\subseteq N(H)$. Note that $H$ and $N(H)$ coincide when $\Sigma$ has dimension 0, i.e.~corresponds to a single point, as is the case for genus $g=1$ and its symmetric or fixed points $\tau=\{i,\,\omega,\,i\infty\}$.

In what follows, we consider restrictions of the kind $\mathcal{H}_2 \to \Sigma$, trading also the full set of Siegel transformations for a restricted one, $\Gamma_2 \to N(H)$.
By applying the same reasoning as before, one can further obtain a finite version of our restricted modular group, $\Gamma_{2,n} \to N_n(H)$, namely by modding out a principal congruence subgroup of $N(H)$, which, for level $n$, is defined as
\begin{equation} \label{eq:pcsN}
N(H,n)=\Big\{\gamma \in N(H) \,\Big|\,  \gamma \equiv \mathbb{1} \,\,(\text{mod}\, n)\Big\}\,.
\end{equation}
We thus obtain a group $N_n(H) \equiv N(H)/N(H,n)$, which is smaller than $\Gamma_{2,n}$ and may in principle be associated with the desired lower-dimensional irreps, i.e.~doublets and triplets. 
We will further concentrate on level $n=2$, having $|N_2(H)| < 100$ for all possible $\Sigma$ choices.

\subsection{The modular framework at genus \texorpdfstring{$g=2$}{g=2}}

We assume ${\cal N}=1$ rigid SUSY invariance and focus on the Yukawa sector. Introducing the matter supermultiplets, $\Phi=(\tau,\varphi^{(I)})$ (here the fields inside $\tau$ are assumed to be dimensionless gauge singlets), the modular flavour group $N(H)$, acts on $\Phi$ as~\cite{Ding:2020zxw}:
\begin{equation}\label{eq:transf}
\begin{cases}
\tau\mapsto \gamma \tau=(A \tau+B)(C\tau +D)^{-1}\,,
\\[0.2 cm]
\varphi^{(I)}\mapsto [\det(C\tau+D)]^{-k_I} \rho_I(\gamma) \varphi^{(I)}~\,,    
\end{cases}
\qquad\gamma=\begin{pmatrix}
A & B \\
C & D
\end{pmatrix}\in N(H)\,.
\end{equation}
The weight $k_I$ is taken to be an integer, for simplicity, and $\rho_I(\gamma)$ is a unitary representation of the finite group $N_n(H)$.
By analogy with the $g=1$ case, the Yukawa couplings of the theory $Y_{(I_1\ldots I_p)}(\tau)$ should transform as Siegel modular forms with weight $k_Y$ in the representation $\rho_{Y}$ of $N_n(H)$:
\begin{equation}
Y(\gamma\tau) \;\longmapsto\;
Y(\gamma\tau)=[\det(C\tau+D)]^{k_Y}\rho_{Y}(\gamma)\, Y(\tau)\,,
\end{equation}
for $\gamma\in N(H)$.
Thus, one may obtain a modular-invariant superpotential, provided the total weight of each operator is zero and there exists a singlet contraction among all the tensor products of the irreps $\rho_{I_1},\,\ldots,\,\rho_{I_p}$ and $\rho_{Y}$.

Concerning the Kähler potential, we assume its minimal form for both the moduli fields and matter fields:
\begin{align}
\label{eq:kahlmod}
    K_\tau &= -h \Lambda^2 \log \det (-i\tau+i\tau^\dagger)\,,
    \\[2mm]
    \label{eq:kahlfields}
    K_\varphi &= \sum_I[\det(-i\tau+i\tau^\dagger)]^{k_I}|\varphi^{(I)}|^2\,,
\end{align}
where $h$ is a positive constant and $\Lambda$ represents a mass scale for the moduli. It was shown in~\cite{Ding:2020zxw} that these choices are invariant under $\Gamma_2$ up to a Kähler shift of the type $K(\phi,\bar\phi)\to K(\phi,\bar\phi)+f(\phi)+\bar{f}(\bar\phi)$.

\subsection{Generalized CP symmetry}
\label{sec:gCP}
It was shown in~\cite{Ding:2021iqp} that a consistent implementation of CP symmetry in a SUSY theory involving multiple moduli is described by the transformations:
\begin{equation} \label{eq:cp_transf}
    \tau \,\xrightarrow{\mathcal{CP}}\,-\tau^*\,,\qquad \varphi^{(I)}(x)\,\xrightarrow{\mathcal{CP}}\,
    X_\mathbf{r}\,\overline{\varphi^{(I)}}(x_\mathcal{P})\,,\qquad Y_a(\tau)\,\xrightarrow{\mathcal{CP}} \, Y_a(-\tau^*)=\lambda_a^bX_\mathbf{r}Y_b^*(\tau)\,,
\end{equation}
where $X_\mathbf{r}$ is unitary and $\lambda$ is a matrix in the space of modular form multiplets sharing a common weight and irrep. 
In particular,
for a standard choice of the CP outer automorphism acting as an involution in field space,
it is always possible to move to a basis where the $X_\mathbf{r}=\mathbb{1}$ and where the matrices representing the generators ($\rho(S)$, $\rho(T_i)$) are symmetric. 
Moreover, in this case, one is free to select a basis for modular form multiplets where $\lambda^b_a=\delta^b_a$.
Provided Clebsch-Gordan coefficients are also real in the chosen basis, the CP symmetry condition greatly simplifies to the transparent requirement of having real superpotential free parameters.

From the above considerations, it follows that in a minimal modular- and gCP-invariant setup, CP may only be spontaneously broken by the VEVs of the moduli, which encode the only complex phases in the theory. If the period matrix VEV satisfies:
\begin{equation}\label{eq:Cp_sym}
-\tau^*=\tau\,,
\end{equation}
then the low-energy theory is clearly CP-conserving. More generally, a point will be CP-conserving if there exists a modular transformation $\gamma$ such that:
\begin{equation}\label{eq:cp_sym_mod}
    -\tau^*=\gamma\,\tau\,.
\end{equation}
Indeed, in the case where $\tau$ belongs to a fixed region $\Sigma$, it may well be that $-\tau^* \notin \Sigma$. To obtain a suitable CP transformation on a point $\tau\in\Sigma$ it can be requested that~\cite{Ding:2021iqp}:
\begin{equation}\label{eq:cp_sym_fixed}
\tau\,\xrightarrow{\gamma\,\circ\,\mathcal{CP}}\,
(-A\tau^*+B)(-C\tau^*+D)^{-1}=\tau'\in\Sigma\,,
\end{equation}
i.e.~the CP transformation maps $\tau$ into another point within the same fixed region. If one further finds that $\tau'=\tau$, then the corresponding point is CP-conserving. The genus-2 inequivalent fixed regions of varying complex dimension have been classified by Gottschling~\cite{gottschling1961fixpunkte} 
and are summarized in the leftmost column of~\cref{tab:FixedPoints}. 
Representative choices for the CP transformations operating within each fixed region can be found in~\cite{Ding:2021iqp}.

\newenvironment{tightpmatrix}
  {\begingroup\renewcommand{\arraystretch}{1}\begin{pmatrix}}
  {\end{pmatrix}\endgroup}

\begin{table}[t]
\centering
\renewcommand{\arraystretch}{1.4}
\begin{tabular}{cccc} \toprule
& Fixed region $\Sigma$
&  
{ }\qquad Stabilizer $\bar{H} \equiv H/\{\pm\mathbb{1}\}$\qquad{ } &
{ }\qquad $N_2(H)$ \qquad { }\\
 \midrule
$\mathcal{T}_1$:& $\begin{tightpmatrix} \tau_1 & 0 \\ 0 & \tau_2 \end{tightpmatrix}$ 
& $\mathbb{Z}_2$ & $(S_3\times S_3)\rtimes \mathbb{Z}_2$ \\
$\mathcal{T}_2$:& $\begin{tightpmatrix} \tau_1 & \tau_3 \\ \tau_3 & \tau_1 \end{tightpmatrix}$ 
& $\mathbb{Z}_2$ & $S_4 \times \mathbb{Z}_2$  \\
 \midrule
$\mathcal{O}_1$:& $\begin{tightpmatrix} i & 0 \\ 0 & \tau_2 \end{tightpmatrix}$
& $\mathbb{Z}_4$ & $D_6$  \\  
$\mathcal{O}_2$:& $\begin{tightpmatrix} \omega & 0 \\ 0 & \tau_2 \end{tightpmatrix}$
& $\mathbb{Z}_6$ & $S_3 \times \mathbb{Z}_3$  \\ 
$\mathcal{O}_3$:& $\begin{tightpmatrix} \tau_1 & 0 \\ 0 & \tau_1 \end{tightpmatrix}$
& $\mathbb{Z}_2\times \mathbb{Z}_2$ & $D_6$  \\ 
$\mathcal{O}_4$:& $\begin{tightpmatrix} \tau_1 & 1/2 \\ 1/2 & \tau_1 \end{tightpmatrix}$
& $\mathbb{Z}_2\times \mathbb{Z}_2$ & $D_4\times \mathbb{Z}_2$ \\ 
$\mathcal{O}_5$:& $\begin{tightpmatrix} \tau_1 & \tau_1 /2 \\ \tau_1 /2 & \tau_1 \end{tightpmatrix}$
& $S_3$ & $S_3 \times S_3$  \\ 
 \midrule
$\mathcal{Z}_1$:& $\begin{tightpmatrix} \zeta & \zeta+\zeta^{-2} \\ \zeta+\zeta^{-2} & -\zeta^{-1} \end{tightpmatrix} $
& $\mathbb{Z}_5$ & $\mathbb{Z}_{5}$    \\
$\mathcal{Z}_2$:& $\begin{tightpmatrix} \xi & \frac{1}{2}(\xi -1) \\ \frac{1}{2}(\xi -1) & \xi \end{tightpmatrix} $
& $S_4$ & $S_4$  \\ 
$\mathcal{Z}_3$:& $\begin{tightpmatrix} i & 0 \\ 0 & i \end{tightpmatrix}$
& $(\mathbb{Z}_4\times \mathbb{Z}_2)\rtimes \mathbb{Z}_2$ & $D_4$ \\ 
$\mathcal{Z}_4$:& $\begin{tightpmatrix} \omega & 0 \\ 0 & \omega \end{tightpmatrix}$
& $S_3\times \mathbb{Z}_6$ & $\mathbb{Z}_3\times S_3$ \\
$\mathcal{Z}_5$:& $\dfrac{i\sqrt{3}}{3}\begin{tightpmatrix} 2 & 1 \\ 1 & 2 \end{tightpmatrix}$
& $D_{6}$ & $D_{6}$  \\
$\mathcal{Z}_6$:& $\begin{tightpmatrix} \omega & 0 \\ 0 & i \end{tightpmatrix}$
& $\mathbb{Z}_{12}$ & $\mathbb{Z}_6$ \\ 
 \bottomrule
\end{tabular}
\caption{All inequivalent fixed regions and points of $Sp(4,\mathbb{Z})$ in the Siegel upper-half space $\mathcal{H}_2$. We use the labels $\mathcal{T}_i$ to identify the regions of complex dimension two, $\mathcal{O}_i$ for the regions of complex dimension one, and $\mathcal{Z}_i$ for the zero-dimensional points. Here $\zeta= e^{2\pi i /5},~\xi=\frac{1}{3}(1+i2\sqrt{2}),~\omega= e^{2\pi i/3}$. The corresponding finite version of the modular group at level $n=2$, $N_2(H)$, is given in the last column. 
We use $D_n \cong \mathbb{Z}_n \rtimes \mathbb{Z}_2$ to denote the dihedral group of order $2n$, with $D_3  \cong S_3$ and $D_6 \cong  S_3 \times \mathbb{Z}_2$.
For a complete reference on the stabilizers $H$ and normalizers $N(H)$ associated with these regions, see~\cite{Ding:2020zxw}.}
\label{tab:FixedPoints}
\end{table}

\section{Fermion mass hierarchies at genus \texorpdfstring{$g=2$}{g=2}}
\label{sec:mpih}
As stated in the introduction, we aim at a topology-driven model building strategy where fermion mass hierarchies, mixing and CP violation are influenced mainly by the choice of the vacuum (the VEV of the moduli contained in $\tau$), which completely breaks the modular symmetry. Motivated by the aforementioned model building constraints (see~\cref{sec:flavourgenus2}), we have restricted the full Siegel group $\Gamma_2$ to a normalizer $N(H)$ associated to a given fixed or invariant subregion $\Sigma$ of the Siegel upper-half space $\mathcal{H}_2$ ($\Sigma \neq \mathcal{H}_2$). The possible regions $\Sigma$ are reported in~\cref{tab:FixedPoints}.
As can be seen, the invariant regions have a maximum of two complex dimensions (labelled $\mathcal{T}_1$ and $\mathcal{T}_2$ in~\cref{tab:FixedPoints}), and a minimum of zero (which we call fixed points $\mathcal{Z}_i$, $i=1,\ldots,6$). The one-dimensional regions are labelled $\mathcal{O}_i$ ($i=1,\ldots,5$). It is important to emphasize that all the zero-dimensional fixed points are CP-conserving~\cite{Ding:2021iqp}.

\vskip 2mm

We now discuss how to realize the MPIH mechanism in the genus 2 framework. The generation of fermion mass hierarchies relies on an \emph{approximately unbroken residual symmetry group}, which is restored as one approaches a point (or region) $\Sigma^* \subset \Sigma$.
We are thus interested in ``trajectories'' $\Sigma \rightarrowtail \Sigma^*$. We denote by $H\,(\subseteq N(H))$ and $H^*$ the stabilizers associated with $\Sigma$ and $\Sigma^*$, respectively. A bar is used to denote the quotients of these stabilizers by $\{\pm\mathbb{1}\}$.
Two considerations are in order:
\begin{itemize}
    \item It may turn out that the stabilizer $H^*$ of the smaller region that one approaches, $\Sigma^*$, contains transformations that do not leave invariant the larger region $\Sigma$ defining the flavour group, i.e.~$H^* \not\subset N(H)$. Hence, one is interested in identifying $H^*_0 \subseteq H^*$, the largest subgroup of the stabilizer which is fully contained in the normalizer of $\Sigma$, i.e.~$H^*_0 \subset N(H)$.
    \item One must disregard those transformations in $H^*_0$ which also belong to $H$, as these are always unbroken and do not affect the generation of mass hierarchies.%
    \footnote{Consider e.g.~the generator $R=S^2$ in the $g=1$ case~\cite{Novichkov:2021evw}.}
    Hence the group that plays a central role is the genus-2 MPIH mechanism is the quotient of $H^*_0$ by (the normal closure of its intersection with) $H$, denoted $H^*_0/\langle H\rangle \cong \bar{H}^*_0/\langle\bar{H}\rangle$.
\end{itemize}
We summarize in~\cref{tab:trajectories} the possible trajectories $\Sigma \rightarrowtail \Sigma^*$ in the Siegel half-space, the corresponding stabilizers $H$ and $H^*$, the active finite modular group $N_2(H)$, and the groups $H_0^*$ and $H^*_0/\langle H\rangle$ relevant for the generation of mass hierarchies. In the last column, we also indicate $\Delta\dim = \dim(\Sigma)-\dim(\Sigma^*)$, which corresponds to the number of small independent (complex) parameters describing the deviation.

\definecolor{lightgraycol}{gray}{0.95}

\afterpage{
\begin{landscape}    
\begin{table}[p]
\thisfloatpagestyle{empty}
    \centering
    \begin{NiceTabular}{ccccccccccc}%
[code-before=
    \columncolor{lightgraycol}{3,6,7,10,11} 
]
    \toprule
       \multicolumn{2}{c}{$\Sigma \,\,\rightarrowtail\,\,\Sigma^*$} &
       $N_2(H)$ &  $H$ &  $\bar{H}$  & $H^*$ &$\bar{H}^*$ &  $H^*_0 \,(\subset N(H))$ & $\bar{H}^*_0$ & $H^*_0/\langle H \rangle$  & $\Delta \dim$ \\
       \midrule
       $\mathcal{T}_2$  & \multirow{1}{*}{$\mathcal{Z}_2$} 
            &  $S_4 \times \mathbb{Z}_2$ 
            & $\mathbb{Z}_2\times \mathbb{Z}_2$
            & $\mathbb{Z}_2$
            & \multirow{1}{*}{$GL(2,3)$} 
            & \multirow{1}{*}{$S_4$} 
                & $D_4$
                & $\mathbb{Z}_2\times\mathbb{Z}_2$
                & $\mathbb{Z}_2$  & 2 \\
       \midrule
       $\mathcal{T}_1$  & \multirow{4}{*}{$\mathcal{Z}_3$} 
            & $(S_3 \times S_3) \rtimes \mathbb{Z}_2$ 
            & $\mathbb{Z}_2\times \mathbb{Z}_2$           
            & $\mathbb{Z}_2$
            & \multirow{4}{*}{$(\mathbb{Z}_4 \times \mathbb{Z}_4) \rtimes \mathbb{Z}_2$} 
            & \multirow{4}{*}{$(\mathbb{Z}_4 \times \mathbb{Z}_2) \rtimes \mathbb{Z}_2$}
                & $(\mathbb{Z}_4 \times \mathbb{Z}_4) \rtimes \mathbb{Z}_2$ 
                & $(\mathbb{Z}_4 \times \mathbb{Z}_2) \rtimes \mathbb{Z}_2$
                & $D_4$ & 2  \\
       $\mathcal{T}_2$  & & $S_4 \times \mathbb{Z}_2$
            & $\mathbb{Z}_2\times \mathbb{Z}_2$
            & $\mathbb{Z}_2$ 
                &&& $(\mathbb{Z}_4 \times \mathbb{Z}_2) \rtimes \mathbb{Z}_2$
                & $\mathbb{Z}_2 \times \mathbb{Z}_2 \times \mathbb{Z}_2$
                & $\mathbb{Z}_2 \times \mathbb{Z}_2$ & 2\\
       $\mathcal{O}_1$  & & $S_3 \times \mathbb{Z}_2$ 
            & $\mathbb{Z}_4\times \mathbb{Z}_2$
            & $\mathbb{Z}_4$
                &&& $\mathbb{Z}_4\times \mathbb{Z}_4$
                & $\mathbb{Z}_4\times \mathbb{Z}_2$
                & $\mathbb{Z}_2$ & 1\\ 
       $\mathcal{O}_3$  & & $S_3 \times \mathbb{Z}_2$
            & $D_4$
            & $\mathbb{Z}_2 \times \mathbb{Z}_2$
                &&& $(\mathbb{Z}_4 \times \mathbb{Z}_2) \rtimes \mathbb{Z}_2$
                & $\mathbb{Z}_2 \times \mathbb{Z}_2 \times \mathbb{Z}_2$
                & $\mathbb{Z}_2$ & 1\\
       \midrule
       $\mathcal{T}_1$  & \multirow{4}{*}{$\mathcal{Z}_4$} 
            & $(S_3 \times S_3) \rtimes \mathbb{Z}_2$ 
            & $\mathbb{Z}_2\times \mathbb{Z}_2$
            & $\mathbb{Z}_2$
            & \multirow{4}{*}{$(\mathbb{Z}_6\times\mathbb{Z}_6)\rtimes \mathbb{Z}_2$} 
            & \multirow{4}{*}{$S_3 \times \mathbb{Z}_6$} 
                & $(\mathbb{Z}_6\times\mathbb{Z}_6)\rtimes \mathbb{Z}_2$
                & $S_3 \times \mathbb{Z}_6$
                & $S_3 \times \mathbb{Z}_3$ & 2\\
       $\mathcal{T}_2$  & & $S_4 \times \mathbb{Z}_2$ 
            & $\mathbb{Z}_2\times \mathbb{Z}_2$
            & $\mathbb{Z}_2$
                &&& $D_4 \times \mathbb{Z}_3$
                & $\mathbb{Z}_6 \times \mathbb{Z}_2$
                & $\mathbb{Z}_6$ & 2\\
       $\mathcal{O}_2$  & & $S_3 \times \mathbb{Z}_3$ 
            & $\mathbb{Z}_6\times \mathbb{Z}_2$
            & $\mathbb{Z}_6$
                &&& $\mathbb{Z}_6\times \mathbb{Z}_6$
                & $\mathbb{Z}_6\times \mathbb{Z}_3$
                & $\mathbb{Z}_3$ & 1\\ 
       $\mathcal{O}_3$  & & $S_3 \times \mathbb{Z}_2$ 
            & $D_4$
            & $\mathbb{Z}_2 \times \mathbb{Z}_2$
                &&& $D_4 \times \mathbb{Z}_3$
                & $\mathbb{Z}_6 \times \mathbb{Z}_2$
                & $\mathbb{Z}_3$ & 1\\ 
       \midrule
       $\mathcal{T}_2$  & \multirow{2}{*}{$\mathcal{Z}_5$}
            & $S_4 \times \mathbb{Z}_2$ 
            & $\mathbb{Z}_2\times \mathbb{Z}_2$
            & $\mathbb{Z}_2$
            & \multirow{2}{*}{$(\mathbb{Z}_6\times\mathbb{Z}_2)\rtimes \mathbb{Z}_2$}  
            & \multirow{2}{*}{$S_3 \times \mathbb{Z}_2$}
                & $D_4$
                & $\mathbb{Z}_2\times \mathbb{Z}_2$
                & $\mathbb{Z}_2$ & 2\\
       $\mathcal{O}_5$  & & $S_3 \times S_3$ 
            & $S_3 \times \mathbb{Z}_2$
            & $S_3$
                &&& $(\mathbb{Z}_6\times\mathbb{Z}_2)\rtimes \mathbb{Z}_2$
                & $S_3 \times \mathbb{Z}_2$
                & $\mathbb{Z}_2$ & 1\\ 
       \midrule
       $\mathcal{T}_1$  & \multirow{3}{*}{$\mathcal{Z}_6$}
            & $(S_3 \times S_3) \rtimes \mathbb{Z}_2$ 
            & $\mathbb{Z}_2\times \mathbb{Z}_2$
            & $\mathbb{Z}_2$
            & \multirow{3}{*}{$\mathbb{Z}_{12}\times \mathbb{Z}_2$}  
            & \multirow{3}{*}{$\mathbb{Z}_{12}$} 
                & $\mathbb{Z}_{12}\times \mathbb{Z}_2$
                & $\mathbb{Z}_{12}$
                & $\mathbb{Z}_{6}$ & 2\\
       $\mathcal{O}_1$  & & $S_3 \times \mathbb{Z}_2$ 
            & $\mathbb{Z}_4\times \mathbb{Z}_2$
            & $\mathbb{Z}_4$
                &&& $\mathbb{Z}_{12}\times \mathbb{Z}_2$
                & $\mathbb{Z}_{12}$
                & $\mathbb{Z}_3$ & 1\\ 
       $\mathcal{O}_2$  & & $S_3 \times \mathbb{Z}_3$ 
            & $\mathbb{Z}_6\times \mathbb{Z}_2$
            & $\mathbb{Z}_6$
                &&& $\mathbb{Z}_{12}\times \mathbb{Z}_2$
                & $\mathbb{Z}_{12}$
                & $\mathbb{Z}_2$ & 1\\ 
       \midrule
       $\mathcal{T}_1$  & \multirow{1}{*}{$\mathcal{O}_1$} 
            & $(S_3 \times S_3) \rtimes \mathbb{Z}_2$ 
            & $\mathbb{Z}_2\times \mathbb{Z}_2$
            & $\mathbb{Z}_2$
            & \multirow{1}{*}{$\mathbb{Z}_4\times \mathbb{Z}_2$} 
            & \multirow{1}{*}{$\mathbb{Z}_4$}
                & $\mathbb{Z}_4\times \mathbb{Z}_2$
                & $\mathbb{Z}_4$
                & $\mathbb{Z}_2$ & 1\\
       \midrule
       $\mathcal{T}_1$  & \multirow{1}{*}{$\mathcal{O}_2$}
            & $(S_3 \times S_3) \rtimes \mathbb{Z}_2$ 
            & $\mathbb{Z}_2\times \mathbb{Z}_2$
            & $\mathbb{Z}_2$
            & \multirow{1}{*}{$\mathbb{Z}_6\times \mathbb{Z}_2$} 
            & \multirow{1}{*}{$\mathbb{Z}_6$}
                & $\mathbb{Z}_{6} \times \mathbb{Z}_{2}$
                & $\mathbb{Z}_{6}$
                & $\mathbb{Z}_{3}$ & 1\\
       \midrule
       $\mathcal{T}_1$  & \multirow{2}{*}{$\mathcal{O}_3$}
            & $(S_3 \times S_3) \rtimes \mathbb{Z}_2$ 
            & $\mathbb{Z}_2\times \mathbb{Z}_2$
            & $\mathbb{Z}_2$
            & \multirow{2}{*}{$D_4$} 
            & \multirow{2}{*}{$\mathbb{Z}_2 \times \mathbb{Z}_2$}
                & $D_4$
                & $\mathbb{Z}_2 \times \mathbb{Z}_2$
                & $\mathbb{Z}_2 $ & 1\\
       $\mathcal{T}_2$  & 
            & $S_4 \times \mathbb{Z}_2$ 
            & $\mathbb{Z}_2\times \mathbb{Z}_2$
            & $\mathbb{Z}_2$
                &&& $D_4$
                & $\mathbb{Z}_2 \times \mathbb{Z}_2$
                & $\mathbb{Z}_2 $ & 1\\
       \midrule
       $\mathcal{T}_2$  & \multirow{1}{*}{$\mathcal{O}_4$}
            & $S_4 \times \mathbb{Z}_2$ 
            & $\mathbb{Z}_2\times \mathbb{Z}_2$
            & $\mathbb{Z}_2$
            & \multirow{1}{*}{$D_4$} 
            & \multirow{1}{*}{$\mathbb{Z}_2 \times \mathbb{Z}_2$}
                & $D_4$
                & $\mathbb{Z}_2 \times \mathbb{Z}_2$
                & $\mathbb{Z}_2$ & 1\\
       \midrule
       $\mathcal{T}_2$  & \multirow{1}{*}{$\mathcal{O}_5$}
            & $S_4 \times \mathbb{Z}_2$ 
            & $\mathbb{Z}_2\times \mathbb{Z}_2$
            & $\mathbb{Z}_2$
            & \multirow{1}{*}{$S_3 \times \mathbb{Z}_2$} 
            & \multirow{1}{*}{$S_3$} 
                & $\mathbb{Z}_2 \times \mathbb{Z}_2$
                & $\mathbb{Z}_2$
                & 1 & 1
        \\
         \bottomrule
\end{NiceTabular}
    \caption{Summary of the possible trajectories $\Sigma \rightarrowtail \Sigma^*$, where a point or region $\Sigma^*$ is approached within a higher-dimensional region $\Sigma$; and the associated groups. $N_2(H)$ is the modular flavour group, $H^{(*)}$ is the stabilizer of $\Sigma^{(*)}$, $H^*_0$ is the part of $H^*$ available within $\Sigma$, and $H^*_0/\langle H\rangle$ is the group relevant for the mechanism of modular proximity-induced hierarchies. The last column shows the difference of complex dimensions $\Delta\dim = \dim(\Sigma)-\dim(\Sigma^*)$. }
    \label{tab:trajectories}
\end{table}
\end{landscape}
}

\vskip 2mm
We are interested in generating a hierarchical fermion mass spectrum. Consider first the cases with $\Delta \dim = 1$, where one approaches a region with one fewer dimension. The relevant part of the residual symmetry group, namely $H_0^*/\langle H\rangle$, must then include at least one element of order $\geq 3$ to obtain a spectrum of the type $\sim (1,\epsilon^a,\epsilon^b)$ where the positive integers $a$ and $b$ may satisfy $a\neq b$, mirroring the rationale of the $g=1$ MPIH mechanism. Here, there is a common $\epsilon$ parameterizing the small departure from the fixed point or region $\Sigma^*$, in terms of the distance described in~\cref{eq:geolength}. Only four $\Delta \dim = 1$ trajectories are thus selected: $\mathcal{T}_{1}\rightarrowtail \mathcal{O}_2$, $\mathcal{O}_{1}\rightarrowtail \mathcal{Z}_6$, and $\mathcal{O}_{2, 3}\rightarrowtail \mathcal{Z}_4$.

For the cases with $\Delta\dim = 2$, corresponding to trajectories of the type $\mathcal{T}_i \rightarrowtail \mathcal{Z}_j$, there are two small independent parameters $\epsilon_1$ and $\epsilon_2$ at play. Hence, even $H_0^*/\langle H\rangle = \mathbb{Z}_2$ may be enough to produce a hierarchical spectrum, since a vanishing mass matrix element may be lifted at order $\epsilon_1 \,\epsilon_2 \sim \epsilon^2$, allowing, at least in principle, to accommodate a hierarchical spectrum $\sim(1,\epsilon,\epsilon^2)$. The seven possible trajectories are, explicitly: $\mathcal{T}_{1}\rightarrowtail \mathcal{Z}_{3,4,6}$ and $\mathcal{T}_{2}\rightarrowtail \mathcal{Z}_{2,3,4,5}$.

\vskip 2mm
Trajectories of the type $\mathcal{O}_i \rightarrowtail \mathcal{Z}_j$ are expected to lead to scenarios analogous to those found at genus $g=1$~\cite{Novichkov:2021evw}. Each of the four trajectories starting from $\Sigma = \mathcal{T}_2$ has instead been considered in~\cite{Ding:2024xhz}, in the context of models of the lepton sector involving triplets. These models rely on the flavour group $N_2(H) \cong S_4 \times \mathbb{Z}_2$,
associated with the $\tau_1 = \tau_2$ subspace $\mathcal{T}_2$.%
\footnote{This finite Siegel modular flavour group has also been used to construct non-MPIH models of the lepton sector~\cite{Ding:2020zxw,Ding:2021iqp,RickyDevi:2024ijc} and of the quark sector with explicit CP violation~\cite{Ding:2020zxw}.}
Note that the hierarchical structure of charged-lepton matrices found therein is only partially explained by the genus-2 MPIH mechanism. 
In what follows, we focus on trajectories starting from $\Sigma = \mathcal{T}_1$, the $\tau_3 = 0$ invariant subspace, associated with the flavour group $N_2(H) \cong (S_3 \times S_3) \rtimes \mathbb{Z}_2$, seeking to realize the Siegel MPIH flavour program in the quark sector. We start by describing the $\mathcal{T}_1$ subspace in~\cref{sec:T1subspace}, and consider each of the four trajectories
\begin{equation*}
    \begin{pmatrix} \tau_1 & 0\\ 0 &\tau_2 \end{pmatrix}
    \quad
    \rightarrowtail
    \quad
    \begin{pmatrix} \omega & 0 \\ 0 & \tau_2 \end{pmatrix}\,,
    \begin{pmatrix} \omega & 0 \\ 0 & i \end{pmatrix}\,,
    \begin{pmatrix} \omega & 0 \\ 0 & \omega \end{pmatrix}\,,
    \begin{pmatrix} i & 0 \\ 0 & i \end{pmatrix}\,,
\end{equation*}
in turn, in~\cref{sec:analytics}. A viable model of the quark sector based on the genus-2 MPIH mechanism with $\Sigma = \mathcal{T}_1$ is then presented in~\cref{sec:model}.

\subsection{The \texorpdfstring{$\tau_3=0$}{tau3=0} subspace}
\label{sec:T1subspace}
Consider the $\mathcal{T}_1$ fixed region where $\tau_3=0$:
\begin{equation}
    \tau=\begin{pmatrix}
        \tau_1& 0\\ 0 & \tau_2
    \end{pmatrix}\,,
\end{equation}
which is a moduli subspace where at level $n=2$ we have $N_2(H)\cong(S_3\times S_3)\rtimes \mathbb{Z}_2 $. The stabilizer is $H=\{\pm \mathbb{1}, \pm h\}$ with:
\begin{equation}
    h=\begin{pmatrix}
        1& 0 &0 &0 \\ 0& -1 &0 &0 \\ 0& 0 &1 &0 \\ 0& 0 &0 &-1 \\
    \end{pmatrix}\,.
\end{equation}
As discussed in~\cite{Ding:2020zxw}, the most general normalizer transformations $\hat{\gamma}\in N(H)$ have two forms: $\hat{\gamma}_+$ and $\hat{\gamma}_-$. They act on the subspace as:
\begin{eqnarray}
\nonumber&&
\hat{\gamma}_{+}\begin{pmatrix} \tau_1 & 0 \\ 0 & \tau_2 \end{pmatrix} = \begin{pmatrix}
\dfrac{a_{1}\tau_1 +b_{1}}{c_{1}\tau_1 +d_{1}} & 0 \\
0 & \dfrac{a_{4}\tau_2 + b_{4}}{c_{4}\tau_2 + d_{4}}
\end{pmatrix}\,,\\
&&
\hat{\gamma}_{-}\begin{pmatrix} \tau_1 & 0 \\ 0 & \tau_2 \end{pmatrix} = \begin{pmatrix}
\dfrac{a_{4}\tau_2 + b_{4}}{c_{4}\tau_2 + d_{4}} & 0 \\
0 & \dfrac{a_{1}\tau_1 +b_{1}}{c_{1}\tau_1 +d_{1}}
\end{pmatrix}\,,
\end{eqnarray}
thus, the action of $\hat{\gamma}_+$ is equivalent to two independent $SL(2,\mathbb{Z})$ transformations on $\tau_1$ and $\tau_2$, whereas $\hat{\gamma}_-$ swaps them. 
The normalizer $N(H)$ is generated by:
\begin{equation}
\begin{aligned}
G_1&=\begin{pmatrix} 1&0&0&0 \\ 0&0&0&1 \\ 0&0&1&0 \\ 0&-1&0&0\end{pmatrix}\,,\quad
G'_1=\begin{pmatrix} 0&0&1&0\\ 0&1&0&0 \\ -1&0&0&0 \\ 0&0&0&1 \end{pmatrix},\\[2mm]
G_2&=\begin{pmatrix} 1&0&0&0\\ 0&1&0&1 \\ 0&0&1&0 \\ 0&0&0&1 \end{pmatrix}\,,\quad
G'_2=\begin{pmatrix} 1&0&1&0\\ 0&1&0&0 \\ 0&0&1&0 \\ 0&0&0&1 \end{pmatrix}\,,\quad
G_3=\begin{pmatrix} 0&1&0&0\\ 1&0&0&0 \\ 0&0&0&1 \\ 0&0&1&0 \end{pmatrix}\,.
\end{aligned}
\end{equation}
Note that the $\hat{\gamma}_+$ and $\hat{\gamma}_-$ transformations are related by $\hat{\gamma}_+=G_3\,\hat{\gamma}_-$.
On the other hand, the generators of the finite group $N_2(H)$ fulfil the relations:
\begin{equation} \label{eq:S3xS3:Z2}
\begin{aligned}
&G^2_1=(G_1G_2)^3=G^2_2=1\,, \quad G'^2_1=(G'_1G'_2)^3=G'^2_2=1\,,\quad G^2_3=1\,,\\
&G_1G'_1=G'_1G_1\,,\quad G_1G'_2=G'_2G_1\,,\quad G_2G'_1=G'_1G_2\,,\quad G_2G'_2=G'_2G_2\,,\\
&G_3G_1G^{-1}_3=G'_1\,,\quad G_3G_2G^{-1}_3=G'_2\,.
\end{aligned}
\end{equation}
The group $N_2(H) \cong (S_3\times S_3)\rtimes \mathbb{Z}_2$ is characterized by four singlets, $\mathbf{1},\mathbf{1'},\mathbf{1''},\mathbf{1'''}$, one doublet $\mathbf{2}$ and four quartets $\mathbf{4},\mathbf{4'},\mathbf{4''},\mathbf{4'''}$. The tensor rules, Clebsch-Gordan coefficients, and corresponding modular forms are listed in~\cref{app:groupth,app:modforms}.
Crucially, $G_3$ generates the $\mathbb{Z}_2$ factor and its action swaps $\tau_1\leftrightarrow \tau_2$. If one were to restrict the available $N(H)$ transformations to $\hat{\gamma}\in\{\hat{\gamma}_+\}$, one would effectively exclude $G_3$ at the level of $N_2(H)$, restricting the finite flavour group to $S_3\times S_3$, with the moduli subspace describing a product of two tori, $\mathbf{T}^2\times \mathbf{T}^2$. The presence of $G_3$ already distinguishes the genus-two scenario from a mere duplication of the genus-one construction.

\subsection{MPIH within the \texorpdfstring{$\tau_3=0$}{tau3=0} subspace}
\label{sec:analytics}

Within the $\mathcal{T}_1$ subspace, defined by $\tau_3 = 0$, the cross-ratio of~\cref{eq:crossratio} simplifies to
\begin{equation} \label{eq:crsimple}
    \mathcal{D}_2(\tau,\tau')\Big|_{\tau_3=0}=\begin{pmatrix}
        \left|\frac{\tau_1-\tau_{1}'}{\tau_1-\bar{\tau}_{1}'}\right|^2&0
        \\
        0 & \left|\frac{\tau_2-\tau_{2}'}{\tau_2-\bar{\tau}_{2}'}\right|^2
    \end{pmatrix}\,,
\end{equation}
while the metric of~\cref{eq:Riemannmetric} simplifies to
\begin{equation} \label{eq:rmsimple}
ds^2=\frac{|d\tau_1|^2}{(\text{Im}\,\tau_1)^2}+\frac{|d\tau_2|^2}{(\text{Im}\,\tau_2)^2}\,,
\end{equation}
i.e.~it decomposes as the sum of two $g=1$ (Poincaré) metrics.
From~\cref{eq:crsimple}, it is clear that the eigenvalues $\lambda_{k}$ ($k=1,2$) required to define a Siegel-invariant distance between $\tau$ and $\tau'$ correspond to the absolute values of the complex variables
\begin{equation} \label{eq:proxvars}
    u_k(\tau_k,\tau_k')=\frac{\tau_k-\tau_k'}{\tau_k-\overline{\tau}_k'}\,.
\end{equation}
For a fixed $\tau'$, these $u_k$ generalize the Möbius transformation 
\begin{equation}
    u(z,z_0)=\frac{z-z_0}{z-\overline{z}_0 }\,,
\end{equation}
that, at genus $g=1$, provides a useful map to describe a modular theory in the vicinity of a fixed point~\cite{Novichkov:2021evw}.

Modular invariance severely constrains the functional dependence of the elements $M_{ij}$ of a mass matrix on the $u_k$ variables. 
In particular, consider a modular-invariant bilinear $\psi^c_i \,M_{ij}(\tau) \,\psi_j$, factoring out Higgs fields which are taken as weightless flavour singlets without loss of generality.%
\footnote{In determining the transformation properties of elements of the mass matrix, any non-trivial charge of a Higgs superfield can be absorbed in the weight and irrep of one of the matter fields within the bilinear.}
Given the modular transformation properties of the reducible%
\footnote{Since we are interested in the case of three fermion generations, the matter fields $\psi^{(c)}$ transform as a direct sum of three singlets $\mathbf{1}_a \oplus \mathbf{1}_b \oplus \mathbf{1}_c$, or of a singlet and the doublet $\mathbf{1}_a \oplus \mathbf{2}$ of the finite modular group $(S_3\times S_3)\rtimes \mathbb{Z}_2$.}
matter superfields $\psi^{(c)}$, namely
\begin{equation}
\begin{aligned}
\psi^c_i&\,\,\mapsto\,\, [\det(C\tau+D)]^{-k^c_i}\, \rho^c_{ij}(\gamma)\, \psi^c_j\,,\\
\psi_i&\,\,\mapsto\,\, [\det(C\tau+D)]^{-k_i} \,\rho_{ij}(\gamma) \,\psi_j\,,
\end{aligned}
\end{equation}
modular invariance imposes
\begin{equation}
    M_{ij}(\tau) \,\,\mapsto\,\, M_{ij}(\gamma\tau)
    \,=\, [\det(C\tau+D)]^{k_{ij}}\, \rho^{c*}_{ik}(\gamma)\,\rho^{*}_{jl}(\gamma)\, M_{kl}(\tau)
    \,,
\end{equation}
with $k_{ij} \equiv k_i^c + k_j$,
for any $\gamma \in N(H)$.
Note that, for a given $\gamma$, in a symmetry basis where the representation matrices $\rho^{(c)}(\gamma)$ are diagonal, this further simplifies to
\begin{equation}
M_{ij}(\gamma\tau)
    \,=\, [\det(C\tau+D)]^{k_{ij}}\, (\rho^c_i\rho_j)^*\, M_{ij}(\tau)
    \,,
\end{equation}
where $\rho^{(c)}_i \equiv \rho^{(c)}_{ii} (\gamma)$.
The transformation properties of the $M_{ij}(u_k)$ under elements from the (relevant part of the) residual symmetry group, $\gamma \in H_0^*/\langle H\rangle$, are especially informative. These constrain the order at which zero elements $M_{ij}(0) = 0$ can be lifted, when departing from symmetric values with $|u_k| \ll 1$.
One is thus able to identify the possible hierarchical structures that may arise from the genus-2 MPIH mechanism.
Below, we illustrate this explicitly for the four aforementioned choices of the VEV of $\tau$. 
Our results are summarized in~\cref{tab:epsilon}.

\definecolor{darkgraycol}{gray}{0.10}

\begin{table}[t]
\centering
\renewcommand{\arraystretch}{1.4}
\begin{NiceTabular}{llllcccccc}%
[code-before=
    \rowcolor{lightgraycol}{1-2} 
    \columncolor{lightgraycol}{1-4} 
]
\toprule
& \multicolumn{3}{r}{$k_{ij} \,\,\text{mod}\,3$}                                                                                              & \multicolumn{2}{c}{0}                               & \multicolumn{2}{c}{1}                             & \multicolumn{2}{c}{2}                             \\[-1mm]
& \multicolumn{3}{r}{$k_{ij}/2 \,\,\text{mod}\,2$}                                                                                            & 0                        & 1                        & 0                       & 1                       & 0                       & 1                       \\
\midrule
$\tau \simeq \begin{tightpmatrix} \omega & \\ & \tau_2 \end{tightpmatrix}$   & \multicolumn{2}{c}{\parbox{2.5cm}{\footnotesize \color{darkgraycol} [any basis]}} &
& \multicolumn{2}{c}{1} & \multicolumn{2}{c}{$\epsilon$} & \multicolumn{2}{c}{$\epsilon^2$}                  
\\
\midrule
\multirow{2}{*}{$\tau \simeq\begin{tightpmatrix} \omega & \\ & i \end{tightpmatrix}$}      & \multicolumn{2}{c}{\multirow{2}{*}{\parbox{2.5cm}{\footnotesize \color{darkgraycol} [basis of \cref{tab:basis}]}}}
& \multicolumn{1}{l}{$\bar{b}_{ij} = +1$}  & 1                        & $\epsilon$               & $\epsilon$              & $\epsilon^2$            & $\epsilon^2$            & $\epsilon^3$            \\
       &&& \multicolumn{1}{l}{$\bar{b}_{ij} = -1$}  & $\epsilon$               & 1                        & $\epsilon^2$            & $\epsilon$              & $\epsilon^3$            & $\epsilon^2$        
\\
\midrule
\multirow{2}{*}{$\tau \simeq\begin{tightpmatrix}\omega & \\ & \omega \end{tightpmatrix}$} &  \multicolumn{2}{c}{\multirow{2}{*}{\parbox{2.5cm}{\footnotesize \color{darkgraycol} [basis of\\ \cref{foot:basis}]}}} & \multicolumn{1}{l}{$\bar{a}_{ij} = +1$} & \multicolumn{2}{c}{1}                               & \multicolumn{2}{c}{$\epsilon^2$}                  & \multicolumn{2}{c}{$\epsilon^4$}                  
\\
       & &&\multicolumn{1}{l}{$\bar{a}_{ij} = -1$} & \multicolumn{2}{c}{$\lesssim\epsilon^3$}            & \multicolumn{2}{c}{$\lesssim\epsilon^5$}          & \multicolumn{2}{c}{$\lesssim\epsilon^7$}
\\
\midrule
\multirow{6}[25]{*}{\parbox{2cm}{$\tau \simeq\begin{tightpmatrix} i & \\ & i \end{tightpmatrix}$\\[5mm]
\footnotesize \color{darkgraycol} [basis of\\ \cref{foot:basis}]}
}           & \multirow{4}{*}{$\mathbf{1}_a\otimes\mathbf{1}_b$} & \multirow{2}{*}{$\bar{a}_{ij} = +1$} & 
    $\bar{b}_{ij} = +1$ &                         &            & 1                       & $\epsilon^2$            &                       &          \\
       &            &      &  
    $\bar{b}_{ij} = -1$ &&                     & $\epsilon^2$            & 1                       & &                    \\
       \cmidrule(l){3-10} 
       & & \multirow{2}{*}{$\bar{a}_{ij} = -1$} &   
    $\bar{b}_{ij} = +1$ & \hphantom{$\,\,\,\epsilon^X$} &  \hphantom{$\,\,\,\epsilon^X$}   & $\lesssim\epsilon^2$    & $\lesssim\epsilon^4$    & \hphantom{$\,\,\,\epsilon^X$}  & \hphantom{$\,\,\,\epsilon^X$}   \\
       &               &    &  
    $\bar{b}_{ij} = -1$ &&  & $\lesssim\epsilon^4$    & $\lesssim\epsilon^2$    & &   \\
       \cmidrule(l){2-10} 
       & \multicolumn{3}{c}{$\mathbf{1}_a\otimes\mathbf{2}$} & \multicolumn{6}{c}{$\begin{tightpmatrix} \lesssim\epsilon & \lesssim\epsilon \end{tightpmatrix}$}                           \\
       \cmidrule(l){2-10} 
       & \multicolumn{3}{c}{$\mathbf{2}\otimes\mathbf{2}$} & \multicolumn{6}{c}{$\begin{tightpmatrix} 1 & \lesssim\epsilon^2\\ \lesssim\epsilon^2 & 1 \end{tightpmatrix}$}
\\
\bottomrule
\end{NiceTabular}
\caption{Expected leading-order behaviour of a mass matrix element $M_{ij}$ under the genus-2 MPIH mechanism, in an expansion in the deviation parameters $|u_k| \sim \epsilon$ of a common magnitude. 
Here,  $\bar{a}_{ij} = (\rho^c_i\rho_j)^{(*)}_{G_3}$ and $\bar{b}_{ij} = (\rho^c_i\rho_j)^{(*)}_{G_1}$ in the indicated basis. The notation $\lesssim$ signals that the element can be further suppressed (see text). Results for the last fixed point  are independent of $k_{ij}\,\,\text{mod}\,3$, but depend instead on $k_{ij}/2\,\,\text{mod}\,2$ and on the irreps within $\psi^c \otimes \psi$ contributing to the mass matrix element or sub-block.}
\label{tab:epsilon}
\end{table}

\subsubsection{\texorpdfstring{$\tau \simeq \begin{pmatrix}
    \omega & 0\\ 0 & \tau_2
\end{pmatrix}$}{T1 >-> O2}}

Keeping $\tau_2 = \tau_2'$ generic, the proximity of the VEV of $\tau$ to the one-dimensional region $\mathcal{O}_2$ can be assessed via the absolute value of $u$, defined as
\begin{equation}
    u\,=\,\frac{\tau_1-\omega}{\tau_1-\omega^2}
    \qquad \Leftrightarrow \qquad
    \tau_1\,=\,\omega\,\frac{1-\omega \,u}{1-u}\,,
\end{equation}
following~\cref{eq:proxvars}.
In this case, the relevant group is $H_0^*/\langle H\rangle \cong \mathbb{Z}_3$, as one may expect from an analogy with the $g=1$ case. It can be generated by the transformation
\begin{equation}
    h_{\mathcal{O}_2}=\begin{pmatrix}
        0 & 0 & 1 & 0\\
        0 & 1 & 0 & 0\\
        -1 & 0 & -1 & 0\\
        0 & 0 & 0 & 1
    \end{pmatrix}
    = G_1'G_2' \,,
\end{equation}
which acts on the moduli and on $u$ as 
\begin{equation}
    h_{\mathcal{O}_2}\tau
    \,=\,
    \begin{pmatrix}
        -\frac{1}{1+\tau_1} & 0\\
        0 & \tau_2
    \end{pmatrix}
    \quad \Rightarrow \quad
    u\,\xrightarrow{h_{\mathcal{O}_2}}\, \omega^2 u\,.
\end{equation}
At the level of the finite modular group $(S_3\times S_3)\rtimes \mathbb{Z}_2$, this transformation is represented by $\rho(G_1'G_2') = \mathbb{1}$ in the cases of interest, cf.~\cref{tab:basis}.
Hence, a mass matrix element obeys
\begin{equation} \label{eq:mass_transf}
    M_{ij}(h_{\mathcal{O}_2}\tau)=[-(\tau_1+1)]^{k_{ij}} M_{ij}(\tau)\,,
\end{equation}
independently of the symmetry basis.
Treating the $M_{ij}$ as analytic functions of $u$ and $\tau_2$,~\cref{eq:mass_transf} becomes
\begin{equation}
    M_{ij}(\omega^2 u,\tau_2)=\omega^{2k_{ij}}\left(\frac{1-\omega^2u}{1-u}\right)^{k_{ij}} M_{ij}(u,\tau_2)\,,
\end{equation}
which further simplifies to
\begin{equation}
    \tilde{M}_{ij}(\omega^2u,\tau_2)=\omega^{2k_{ij}}\tilde{M}_{ij}(u,\tau_2)\,,
\end{equation}
where we have defined $\tilde{M}_{ij}(u,\tau_2)\equiv (1-u)^{-k_{ij}}M_{ij}(u,\tau_2)$, as in the $g=1$ case~\cite{Novichkov:2021evw}.
By comparing the Taylor expansion of each side near $u\to0$, one finds, at each order,
\begin{equation}
\left(\omega^{2n}-\omega^{2k_{ij}}\right)\,
\tilde{M}^{(n)}_{ij}(0,\tau_2)=0\,,
\end{equation}
where $\tilde{M}^{(n)}_{ij}$ denotes the $n$-th derivative of $\tilde{M}_{ij}(u,\tau_2)$ with respect to $u$. Hence, unless $n \equiv k_{ij}\,(\text{mod}\,3)$, one has $\tilde{M}^{(n)}_{ij}(0,\tau_2) = 0$ necessarily. In other words, one expects the leading term in the $u$-expansions of $\tilde{M}_{ij}$ and $M_{ij} \simeq \tilde{M}_{ij}$ to be of order $\mathcal{O}(|u|^\ell)$, where $\ell = (k_{ij}\,\text{mod}\,3) = 0,1,2$.
This allows us to potentially realize the mass pattern $(1,\epsilon,\epsilon^2)$, where
$\epsilon = |u|$, with an appropriate weight assignment $k_{ij}$.

\subsubsection{\texorpdfstring{$\tau \simeq \begin{pmatrix}
    \omega & 0\\ 0 & i
\end{pmatrix}$}{T1 >-> Z6}}

In this case, 
the cross-ratio of~\cref{eq:crossratio} reads 
\begin{equation}
    \mathcal{D}_2=\begin{pmatrix}
        \left|\frac{\tau_1-\omega}{\tau_1-\omega^2}\right|^2&0
        \\
        0 & \left|\frac{\tau_2-i}{\tau_2+i}\right|^2
    \end{pmatrix}=\begin{pmatrix}
        \left|u\right|^2&0
        \\
        0 &  \left|s\right|^2
    \end{pmatrix}\,,
\end{equation}
and the proximity of the VEV of $\tau$ to the point $\mathcal{Z}_6$ can be assessed via the absolute values of $u$ and $s$, defined here as
\begin{equation}
    u=\frac{\tau_1-\omega}{\tau_1-\omega^2}
    \,\, \Leftrightarrow \,\,
    \tau_1=\omega\,\frac{1-\omega \,u}{1-u}\,,
    \qquad\quad
    s=\frac{\tau_2-i}{\tau_2+i}
    \,\, \Leftrightarrow \,\,
    \tau_2=i \frac{1+s}{1-s}\,.
\end{equation}
The relevant group here is $H_0^*/\langle H\rangle \cong \mathbb{Z}_6 \cong \mathbb{Z}_2 \times \mathbb{Z}_3$, generated by
\begin{equation}
    h_{\mathcal{Z}_6}=\begin{pmatrix}
        0 & 0 & 1 & 0\\
        0 & 0 & 0 & 1\\
        -1 & 0 & -1 & 0\\
        0 & -1 & 0 & 0
    \end{pmatrix}
    = G_1'G_2'G_1 \,,
\end{equation}
which acts on the moduli, on $u$ and on $s$ as 
\begin{equation}
    h_{\mathcal{Z}_6}\tau
    \,=\,
    \begin{pmatrix}
        -\frac{1}{1+\tau_1} & 0\\
        0 & -\frac{1}{\tau_2}
    \end{pmatrix}
    \quad \Rightarrow \quad
    \begin{cases}
    u\,\xrightarrow{h_{\mathcal{Z}_6}}\, \omega^2 u \\
    s\,\xrightarrow{h_{\mathcal{Z}_6}}\, -s
    \end{cases}\,.
\end{equation}
Note that $h_{\mathcal{Z}_6}^4=h_{\mathcal{O}_2}$ acts only on $\tau_1 \mapsto -1/(1+\tau_1) $ and generates the $\mathbb{Z}_3$, while $h_{\mathcal{Z}_6}^3$ acts only on $\tau_2 \mapsto -1/\tau_2$ and generates the $\mathbb{Z}_2$.
At the level of the finite modular group, $h_{\mathcal{Z}_6}$ is represented by $\rho(G_1'G_2'G_1) = \rho(G_1)$ in the cases of interest, which is a diagonal matrix in the basis of~\cref{tab:basis}.
In this basis, the transformation of the mass matrix element produces
\begin{equation}
    M_{ij}(h_{\mathcal{Z}_6}\tau)=[(\tau_1+1)\tau_2]^{k_{ij}}(\rho^c_i\rho_j)^*M_{ij}(\tau)\,,
\end{equation}
where $(\rho^c_i\rho_j)^* = \pm 1$.
Treating the $M_{ij}$ as analytic functions of $u$ and $s$ and using the fact that $k_{ij}$ is even, this constraint becomes
\begin{equation}
    M_{ij}(\omega^2u,-s)=\omega^{2k_{ij}} i^{k_{ij}}
    \left(\frac{1-\omega^2u}{1-u}\right)^{k_{ij}}
    \left(\frac{1+s}{1-s}\right)^{k_{ij}}
    (\rho^c_i\rho_j)^*M_{ij}(u,s)\,,
\end{equation}
which further simplifies to
\begin{equation}
    \tilde{M}_{ij}(\omega^2u,-s)=\omega^{2k_{ij}}i^{k_{ij}}(\rho^c_i\rho_j)^* \tilde{M}_{ij}(u,s)\,,
\end{equation}
where we have defined $\tilde{M}_{ij}(u,s) \equiv (1-u)^{-k_{ij}}(1-s)^{-k_{ij}}M_{ij}(u,s)$.
Denoting by $\tilde{M}^{(n,m)}_{ij}$ the mixed partial derivative of $\tilde{M}_{ij}(u,s)$ of order $n$ with respect to $u$ and $m$ with respect to $s$, we obtain
\begin{equation} \label{eq:doublesym}
    \left(\omega^{2n}i^{2m} -\omega^{2k_{ij}}i^{k_{ij}}(\rho^c_i\rho_j)^*\right)\,
    \tilde{M}^{(n,m)}_{ij}(0,0)=0\,.
\end{equation}
Consider first the case $(\rho^c_i\rho_j)^*= +1$.
To allow for $\tilde{M}^{(n,m)}_{ij}(0,0) \neq 0$, one must then have
\begin{equation}
    \begin{cases}
        k_{ij} \equiv n\,\,\,(\text{mod}\,3) \\
        k_{ij} \equiv 2m\,\,\,(\text{mod}\,4)
    \end{cases}
     \Rightarrow\quad
     k_{ij} \equiv 4n+6m\,\,\,(\text{mod}\,12)\,,
\end{equation}
which follows from the Chinese remainder theorem.
If instead $(\rho^c_i\rho_j)^*= -1$, having $\tilde{M}^{(n,m)}_{ij}(0,0) \neq 0$ implies 
\begin{equation}
    \begin{cases}
        k_{ij} \equiv n\,\,\,(\text{mod}\,3) \\
        k_{ij} \equiv 2(m+1)\,\,\,(\text{mod}\,4)
    \end{cases}
     \Rightarrow\quad
     k_{ij} \equiv 6+4n+6m\,\,\,(\text{mod}\,12)\,.
\end{equation}
Given that $k_{ij}$ is even, there are six possibilities for the leading term in the $(u,s)$-expansion of $M_{ij} \simeq \tilde{M}_{ij}$.
In particular, this entry can be of order $\mathcal{O}(1)$ in two cases: if $(\rho^c_i\rho_j)^*=+1$, then the weights need to satisfy $k_{ij}\equiv 0\,(\text{mod}\,12)$; while if $(\rho^c_i\rho_j)^*=-1$, the condition becomes $k_{ij}\equiv 6\,(\text{mod}\,12)$.
For generic weights, one finds
\begin{equation} \label{eq:leadingz6}
    M_{ij} \sim \mathcal{O}\left(|u|^\ell\, |s|^p\right)\,,
\end{equation} 
where $\ell = (k_{ij}\,\,\text{mod}\,3)= 0,1,2$ and $p = 0,1$. Here, the value of $p$ is determined by $p = (k_{ij}/2\,\,\text{mod}\,2)$ if $(\rho^c_i\rho_j)^*=+1$, or by $p = (k_{ij}/2-1\,\,\text{mod}\,2)$ if $(\rho^c_i\rho_j)^*=-1$ instead.
From~\cref{eq:leadingz6} one sees that, if $|u|$ and $|s|$ have the same order of magnitude $\epsilon$, one can have the following options: $M_{ij} \sim\mathcal{O}(1)$, $M_{ij} \sim \epsilon \in\{|u|,\, |s|\}$, $M_{ij} \sim \epsilon^2 \in\{|u|^2,\,|u|\,|s|\}$, or $M_{ij} \sim \epsilon^3 \in \{ |u|^2|s|\}$.
Thus, a $(1,\epsilon,\epsilon^2)$ spectrum induced by $\tau_1\simeq \omega$ can be enhanced by the simultaneous proximity of $\tau_2\simeq i$, resulting e.g.~in a spectrum of the type $(1,\epsilon^2,\epsilon^2)$ or $(1,\epsilon,\epsilon^3)$.

\subsubsection{\texorpdfstring{$\tau \simeq \begin{pmatrix}
    \omega & 0\\ 0 & \omega
\end{pmatrix}$}{T1 >-> Z4}}

In this case, we have
\begin{equation}
    \mathcal{D}_2=\begin{pmatrix}
        \left|\frac{\tau_1-\omega}{\tau_1-\omega^2}\right|^2&0
        \\
        0 & \left|\frac{\tau_2-\omega}{\tau_2-\omega^2}\right|^2
    \end{pmatrix}= \begin{pmatrix}
        \left|u_1\right|^2&0
        \\
        0 &  \left|u_2\right|^2
    \end{pmatrix}\,,
\end{equation}
where
\begin{equation}
    u_i=\frac{\tau_i-\omega}{\tau_i-\omega^2}
    \quad \Leftrightarrow \quad
    \tau_i=\omega\,\frac{1-\omega \,u_i}{1-u_i}\,.
\end{equation}
The relevant group here is $H_0^*/\langle H\rangle \cong S_3 \times \mathbb{Z}_3$, generated by
\begin{equation}
\begin{aligned}
h_{\mathcal{Z}_4,1} &=
\begin{pmatrix}
    0 & 0 & 0 & -1\\
    1 & 0 &1 &0\\
    0 & 1 & 0 & 1\\
    -1&0 & 0 &0
\end{pmatrix}
= (G_1')^{-1}G_2^{-1}G_1G_3G_2
\,, \quad 
h_{\mathcal{Z}_4,2}=
\begin{pmatrix}
    0 & 1 & 0 & 0\\
    1 & 0 &0 &0\\
    0 & 0 & 0 & 1\\
    0&0 & 1 &0
\end{pmatrix}
= G_3
\,, \\
h_{\mathcal{Z}_4,3}&=
\begin{pmatrix}
    0 & 0 & 1 & 0\\
    0 & 0 &0 &-1\\
    -1 & 0 & -1 & 0\\
    0&1 & 0 &1
\end{pmatrix}
= G_1^{-1}G_2G_1'G_2'
\,,
\end{aligned}
\end{equation}
which act on the moduli and on the $u_i$ as 
\begin{equation}
\begin{array}{ll}    
    h_{\mathcal{Z}_4,1}\,\tau
    \,=\,
    \begin{pmatrix}
        -\frac{1}{1+\tau_2} & 0\\
        0 & -\frac{1+\tau_1}{\tau_1}
    \end{pmatrix}
    \quad &\Rightarrow \quad
    \begin{cases}
    u_1\,\xrightarrow{h_{\mathcal{Z}_4,1}}\, \omega^2 u_2 \\
    u_2\,\xrightarrow{h_{\mathcal{Z}_4,1}}\, \omega\, u_1
    \end{cases}\,,
    \\[6mm]
    h_{\mathcal{Z}_4,2}\,\tau
    \,=\,
    \begin{pmatrix}
        \tau_2 & 0\\
        0 & \tau_1
    \end{pmatrix}
    \quad &\Rightarrow \quad
    \begin{cases}
    u_1\,\xrightarrow{h_{\mathcal{Z}_4,2}}\, u_2 \\
    u_2\,\xrightarrow{h_{\mathcal{Z}_4,2}}\, u_1
    \end{cases}\,,
    \\[6mm]
    h_{\mathcal{Z}_4,3}\,\tau
    \,=\,
    \begin{pmatrix}
        -\frac{1}{1+\tau_1} & 0\\
        0 & -\frac{1}{1+\tau_2}
    \end{pmatrix}
    \quad &\Rightarrow \quad
    \begin{cases}
    u_1\,\xrightarrow{h_{\mathcal{Z}_4,3}}\, \omega^2 u_1 \\
    u_2\,\xrightarrow{h_{\mathcal{Z}_4,3}}\, \omega^2 u_2
    \end{cases}\,.
\end{array}
\end{equation}
Note that $h_{\mathcal{Z}_4,1}$ and $h_{\mathcal{Z}_4,2}$ generate the $S_3$ factor, with $\tau$ being invariant under $h_{\mathcal{Z}_4,1}^2$, $h_{\mathcal{Z}_4,2}^2$, and $(h_{\mathcal{Z}_4,1}\,h_{\mathcal{Z}_4,2})^3$, while $h_{\mathcal{Z}_4,3}$ generates the $\mathbb{Z}_3$ factor, with $\tau$ being invariant under $h_{\mathcal{Z}_4,3}^3$.
At the level of the finite group $(S_3\times S_3)\rtimes \mathbb{Z}_2$, $h_{\mathcal{Z}_4,1}$ and $h_{\mathcal{Z}_4,2}$ are represented by $\rho(G_3)$, while
$h_{\mathcal{Z}_4,3}$ is represented by $\rho(G_1G_2G_1'G_2') = \mathbb{1}$,
in the cases of interest.
There exists a $G_3$-diagonal basis where both matrices are diagonal, which we consider in what follows.%
\footnote{\label{foot:basis}%
This basis is related to that of~\cref{tab:basis} via the change-of-basis matrix $U = \frac{1}{\sqrt{2}} \left(\begin{smallmatrix}
    1 & -1 \\ 1 & 1
\end{smallmatrix}\right)$.}
Then, the transformation properties of a mass matrix element imply
\begin{equation}
\begin{cases}
    M_{ij}(h_{\mathcal{Z}_4,1}\tau)\,=\,
    [\tau_1(\tau_2+1)]^{k_{ij}}\,\bar{a}_{ij}\,M_{ij}(\tau)\,,
    \\[1mm]
    M_{ij}(h_{\mathcal{Z}_4,2}\tau)\,=\,
    \bar{a}_{ij}\,M_{ij}(\tau)\,,
    \\[1mm]
    M_{ij}(h_{\mathcal{Z}_4,3}\tau)\,=\,
    [(\tau_1+1)(\tau_2+1)]^{k_{ij}} \,M_{ij}(\tau)\,,
\end{cases}
\end{equation}
where $\bar{a}_{ij} \equiv (\rho^c_i\rho_j)^*_{\gamma = G_3} = \pm 1$, and  we have used the fact that $k_{ij}$ is even.
Treating the $M_{ij}$ as analytic functions of the $u_i$, these constraints become
\begin{equation}
\begin{cases}
    M_{ij}(u_2,u_1)\,=\,
    \bar{a}_{ij}\,M_{ij}(u_1,u_2)\,,
    \\[1mm]
    M_{ij}(\omega^2u_2,\omega\, u_1)
    \,=\,
    \left(\frac{1-\omega\, u_1}{1-u_1}\right)^{k_{ij}}
    \left(\frac{1-\omega^2 u_2}{1-u_2}\right)^{k_{ij}}
    \,\bar{a}_{ij}\,M_{ij}(u_1,u_2)\,,
    \\[1mm]
    M_{ij}(\omega^2u_1,\omega^2u_2)\,=\,
    \omega^{k_{ij}} 
    \left(\frac{1-\omega^2\, u_1}{1-u_1}\right)^{k_{ij}}
    \left(\frac{1-\omega^2 u_2}{1-u_2}\right)^{k_{ij}}
    \,M_{ij}(u_1,u_2)\,,
\end{cases}
\end{equation}
which further simplify to
\begin{equation} \label{eq:z4conds0}
\begin{cases}
    \tilde{M}_{ij}(u_2,u_1)\,=\,
    \bar{a}_{ij}\,\tilde{M}_{ij}(u_1,u_2)\,,
    \\[1mm]
    \tilde{M}_{ij}(\omega\,u_1,\omega^2 u_2)
    \,=\,
    \tilde{M}_{ij}(u_1,u_2)\,,
    \\[1mm]
    \tilde{M}_{ij}(\omega^2u_1,\omega^2u_2)\,=\,
    \omega^{k_{ij}} 
    \,\tilde{M}_{ij}(u_1,u_2)\,,
\end{cases}
\end{equation}
where we have defined $\tilde{M}_{ij}(u_1,u_2) \equiv (1-u_1)^{-k_{ij}}(1-u_2)^{-k_{ij}}M_{ij}(u_1,u_2)$.
Denoting by $\tilde{M}^{(n,m)}_{ij}$ the mixed partial derivative of $\tilde{M}_{ij}(u_1,u_2)$ of order $n$ with respect to $u_1$ and $m$ with respect to $u_2$, we obtain
\begin{equation} \label{eq:z4swapcond}
        \tilde{M}^{(m,n)}_{ij}(0,0)\,=\,\bar{a}_{ij}\, \tilde{M}^{(n,m)}_{ij}(0,0)\,,
\end{equation}
and
\begin{equation} \label{eq:z4conds}
\begin{cases}
    \left(\omega^{n+2m} -1\right)\,
    \tilde{M}^{(n,m)}_{ij}(0,0)=0\,,
    \\[2mm]
    \left(\omega^{2n+2m} -\omega^{k_{ij}} \right)\,
    \tilde{M}^{(n,m)}_{ij}(0,0)=0\,.
\end{cases}
\end{equation}
To allow for $\tilde{M}^{(n,m)}_{ij}(0,0) \neq 0$,~\cref{eq:z4conds} implies that one must have
\begin{equation}
        n \equiv m\,\,\,(\text{mod}\,3) 
        \quad\wedge\quad
        2(n+m) \equiv k_{ij} \,\,\,(\text{mod}\,3)\,.
\end{equation}
If $\bar{a}_{ij}= -1$,
the condition in~\cref{eq:z4swapcond} additionally imposes $\tilde{M}^{(n,n)}_{ij}(0,0) = 0$ for all $n$.
There are thus six possibilities for the leading term in the $(u_1,u_2)$-expansion of $M_{ij} \simeq \tilde{M}_{ij}$.
In particular, this mass matrix element can be $\mathcal{O}(1)$ only if $k_{ij}\equiv 0\,(\text{mod}\, 3)$ and $\bar{a}_{ij}= +1$.
For generic weights and for $\bar{a}_{ij}= +1$, we expect
\begin{equation}
    M_{ij} \sim \mathcal{O}\left(|u_1|^\ell\, |u_2|^\ell\right)\,,
\end{equation} 
where $\ell = (k_{ij}\,\,\text{mod}\,3)= 0,1,2$.
If $|u_1|$ and $|u_2|$ share the same order of magnitude $\epsilon$, one can then have $M_{ij} \sim \epsilon^2$ or $M_{ij} \sim \epsilon^4$.
This picture changes if, instead, $\bar{a}_{ij}= -1$. In this case,~\cref{eq:z4swapcond} forbids terms of the type $u_1^n u_2^n$, as it imposes that the expansion coefficients of $u_1^n u_2^m$ and $u_1^m u_2^n$ differ by a sign. We then find, at leading order 
\begin{equation} \label{eq:leadingz4minus}
    M_{ij} \sim \mathcal{O}\left(|u_1|^\ell\, |u_2|^\ell\, |u_1^3-u_2^3|\right)\,,
\end{equation} 
where once again $\ell = (k_{ij}\,\,\text{mod}\,3)= 0,1,2$.
If $|u_1|$ and $|u_2|$ are of the same order $\epsilon$, one can have at most $M_{ij} \sim \epsilon^3$, $M_{ij} \sim \epsilon^5$, or $M_{ij} \sim \epsilon^7$. Thus, hierarchical mass patterns such as $(1,\epsilon^2,\epsilon^4)$, $(1,\epsilon^3,\epsilon^5)$, and $(1,\epsilon^4,\epsilon^7)$ can potentially be realized in the vicinity of $\mathcal{Z}_4$.

Before proceeding, we note that the factor $u_1^3-u_2^3 = (u_1-u_2)(u_1-\omega u_2)(u_1-\omega^2 u_2)$ in~\cref{eq:leadingz4minus} indicates that a further suppression of $M_{ij}$ is possible for $\bar{a}_{ij}= -1$, provided $|u_1-u_2| \ll \epsilon$, $|u_1- \omega\, u_2| \ll \epsilon$, or $|u_1- \omega^2 u_2| \ll \epsilon$, i.e.~if one can naturally arrange for $u_1 \simeq u_2$, $u_1 \simeq \omega\, u_2$, or $u_1 \simeq \omega^2 u_2$. If one of these holds strictly instead of approximately, then $M_{ij}$ vanishes identically, as can be seen from the first line of~\cref{eq:z4conds0}.

\subsubsection{\texorpdfstring{$\tau \simeq \begin{pmatrix}
    i & 0\\ 0 & i
\end{pmatrix}$}{T1 >-> Z3}}

In this case, we have
\begin{equation}
    \mathcal{D}_2=\begin{pmatrix}
        \left|\frac{\tau_1-i}{\tau_1+i}\right|^2&0
        \\
        0 & \left|\frac{\tau_2-i}{\tau_2+i}\right|^2
    \end{pmatrix}= \begin{pmatrix}
        \left|s_1\right|^2&0
        \\
        0 &  \left|s_2\right|^2
    \end{pmatrix}\,,
\end{equation}
where
\begin{equation}
    s_i=\frac{\tau_i-i}{\tau_i+i}
    \quad \Leftrightarrow \quad
    \tau_i=i\,\frac{1+ s_i}{1-s_i}\,.
\end{equation}
The relevant group here is $H_0^*/\langle H\rangle \cong D_4 \cong \mathbb{Z}_4 \rtimes \mathbb{Z}_2$, generated by
\begin{equation}
h_{\mathcal{Z}_3,1} =
\begin{pmatrix}
    0 & 1 & 0 & 0\\
    0 & 0 & -1 & 0\\
    0 & 0 & 0 & 1\\
    1 & 0 & 0 & 0
\end{pmatrix}
= G_1^3 G_3
\,, \quad 
h_{\mathcal{Z}_3,2}=
\begin{pmatrix}
    0 & 1 & 0 & 0\\
    1 & 0 &0 &0\\
    0 & 0 & 0 & 1\\
    0&0 & 1 &0
\end{pmatrix}
= G_3
\,, 
\end{equation}
which act on the moduli and on the $s_i$ as 
\begin{equation}
\begin{array}{ll}    
    h_{\mathcal{Z}_3,1}\,\tau
    \,=\,
    \begin{pmatrix}
        \tau_2 & 0\\
        0 & -\frac{1}{\tau_1}
    \end{pmatrix}
    \quad &\Rightarrow \quad
    \begin{cases}
    s_1\,\xrightarrow{h_{\mathcal{Z}_3,1}}\, s_2 \\
    s_2\,\xrightarrow{h_{\mathcal{Z}_3,1}}\, -s_1
    \end{cases}\,,
    \\[6mm]
    h_{\mathcal{Z}_3,2}\,\tau
    \,=\,
    \begin{pmatrix}
        \tau_2 & 0\\
        0 & \tau_1
    \end{pmatrix}
    \quad &\Rightarrow \quad
    \begin{cases}
    s_1\,\xrightarrow{h_{\mathcal{Z}_3,2}}\, s_2 \\
    s_2\,\xrightarrow{h_{\mathcal{Z}_3,2}}\, s_1
    \end{cases}\,.
\end{array}
\end{equation}
Note that $h_{\mathcal{Z}_3,1}$ generates the $\mathbb{Z}_4$ factor, with $\tau$ being invariant under $h_{\mathcal{Z}_3,1}^4 = -\mathbb{1}$, while $h_{\mathcal{Z}_3,2}$ coincides with  $h_{\mathcal{Z}_4,2}$ from the previous case and generates the $\mathbb{Z}_2$ factor, with $\tau$ being invariant under $h_{\mathcal{Z}_3,2}^2 = \mathbb{1}$.
At the level of the finite modular group, the transformation $h_{\mathcal{Z}_3,1}$ is represented by $\rho(G_1 G_3)$,
while $h_{\mathcal{Z}_3,2}$ is represented by $\rho(G_3)$. 
In the cases of interest, there is no doublet basis in which these matrices are both diagonal. In what follows, we work in the $G_3$-diagonal basis of the previous section, with $\rho_\mathbf{2}(G_3) = \diag(1,-1)$.

To render the discussion more straightforward, we define the transformations $b \equiv h_{\mathcal{Z}_3,1}\,h_{\mathcal{Z}_3,2}$ and  $b' \equiv h_{\mathcal{Z}_3,2}\,h_{\mathcal{Z}_3,1}$, represented by $\rho(G_1)$ and $\rho(G_3G_1G_3) = \rho(G_1')$, respectively. In the considered basis,
\begin{equation}
    \rho_\mathbf{2}(b) =\rho_\mathbf{2}(G_1) = \begin{pmatrix}
        0 & -1 \\ -1 & 0
    \end{pmatrix}\,,\qquad
    \rho_\mathbf{2}(b') =\rho_\mathbf{2}(G_1') = \begin{pmatrix}
        0 & 1 \\ 1 & 0
    \end{pmatrix} \,.
\end{equation}
While $\rho_\mathbf{2}(b) = -\rho_\mathbf{2}(b')$, one has
$\rho_{\mathbf{1}_a}(b) = \rho_{\mathbf{1}_a}(b')$ for all 1D irreps.
Note that $b$ acts only on $\tau_2 \mapsto -1/\tau_2$ ($s_2 \mapsto -s_2$), while $b'$ acts only on $\tau_1 \mapsto -1/\tau_1$ ($s_1 \mapsto -s_1$).
Together with the swap transformation $a \equiv h_{\mathcal{Z}_3,2
}=G_3$, they generate the $D_4$ group. 
Then, in the $G_3$-diagonal basis, the transformation properties of a mass matrix element imply
\begin{equation}
\begin{cases}
    M_{ij}(a\,\tau)\,=\,
    \bar{a}_{ij}\,M_{ij}(\tau)\,,
    \\[1mm]
    M_{ij}(b\,\tau)\,=\,
    \tau_2^{k_{ij}}\,
    (\rho^c_{ik}\rho_{jl})^*_{G_1}\, M_{kl}(\tau)\,,
    \\[1mm]
    M_{ij}(b'\tau)\,=\,
    \tau_1^{k_{ij}}\,
    (\rho^c_{ik}\rho_{jl})^*_{G_1'}\, M_{kl}(\tau)\,,
\end{cases}
\end{equation}
where $\bar{a}_{ij} \equiv (\rho^c_i\rho_j)^*_{G_3} = \pm 1$.
Treating the $M_{ij}$ as analytic functions of the $s_i$, these constraints become
\begin{equation}
\begin{cases}
    \tilde{M}_{ij}(s_2,s_1)\,=\,
    \bar{a}_{ij}\,\tilde{M}_{ij}(s_1,s_2)\,,
    \\[1mm]
    \tilde{M}_{ij}(s_1,-s_2)
    \,=\,
    i^{k_{ij}}
    \,(\rho^c_{ik}\rho_{jl})^*_{G_1}\, \tilde{M}_{kl}(s_1,s_2)\,,
    \\[1mm]
    M_{ij}(-s_1,s_2)\,=\,
    i^{k_{ij}} 
    \,(\rho^c_{ik}\rho_{jl})^*_{G_1'}\, \tilde{M}_{kl}(s_1,s_2)\,,
\end{cases}
\end{equation}
where we have defined $\tilde{M}_{ij}(s_1,s_2) \equiv (1-s_1)^{-k_{ij}}(1-s_2)^{-k_{ij}}M_{ij}(s_1,s_2)$.
Denoting by $\tilde{M}^{(n,m)}_{ij}$ the mixed partial derivative of $\tilde{M}_{ij}(s_1,s_2)$ of order $n$ with respect to $s_1$ and $m$ with respect to $s_2$, we first obtain
\begin{equation} \label{eq:z3swapcond}
        \tilde{M}^{(m,n)}_{ij}(0,0)\,=\,\bar{a}_{ij}\, \tilde{M}^{(n,m)}_{ij}(0,0)\,.
\end{equation}
Depending on the irreps assigned to the (parts of) $\psi$ and $\psi^c$ that contribute to the mass matrix element(s) under scrutiny, we also find:
\begin{equation} \label{eq:z3cond11}
    (-1)^n \tilde{M}^{(n,m)}_{ij}(0,0)=
    (-1)^m \tilde{M}^{(n,m)}_{ij}(0,0)=
    i^{k_{ij}}\,\bar{b}_{ij}\, \tilde{M}^{(n,m)}_{ij}(0,0)\,,
\end{equation}
where $\bar{b}_{ij} \equiv (\rho^c_i\rho_j)^*_{G_1} = (\rho^c_i\rho_j)^*_{G_1'} = \pm 1$,
if $\psi$ and $\psi^c$ are singlets;
\begin{equation} \label{eq:z3cond22}
    (-1)^n \tilde{M}^{(n,m)}_{ij}(0,0)=
    (-1)^m \tilde{M}^{(n,m)}_{ij}(0,0)=
    i^{k_{ij}}\,(1- \delta_{ik})(1- \delta_{jl})\, \tilde{M}^{(n,m)}_{kl}(0,0)\,,
\end{equation}
if both $\psi$ and $\psi^c$ are doublets; and
\begin{equation} \label{eq:z3cond12}
    (-1)^n \tilde{M}^{(n,m)}_{ij}(0,0)=
    -(-1)^m \tilde{M}^{(n,m)}_{ij}(0,0)=
    i^{k_{ij}}\,\bar{b}_i(1- \delta_{jk})(1- \delta_{jl})\, \tilde{M}^{(n,m)}_{ik}(0,0)\,,
\end{equation}
in the mixed case, where
$\bar{b}_i \equiv \rho^{c*}_i(G_1) = \rho^{c*}_i(G_1') = \pm 1$
and we have taken $\psi^c \sim \mathbf{1}_a$ and $\psi \sim \mathbf{2}$ without loss of generality.

In the case where both $\psi$ and $\psi^c$ are singlets,~\cref{eq:z3cond11} and $\tilde{M}^{(n,m)}_{ij}(0,0) \neq 0$ imply that $n$ and $m$ have the same parity.
Consider first the case $\bar{a}_{ij}= +1$.
If $\bar{b}_{ij}= +1$, one further has $n\equiv m \equiv k_{ij}/2\,\,\,(\text{mod}\,2)$, so that one expects 
\begin{equation} \label{eq:leadingz3plus}
    M_{ij} \sim \mathcal{O}\left(|s_1|^p\, |s_2|^p\right)\,,
\end{equation} 
where $p = (k_{ij}/2\,\,\text{mod}\,2)= 0,1$.
If instead $\bar{b}_{ij}= -1$,~\cref{eq:leadingz3plus} still holds, but with $p = (k_{ij}/2-1\,\,\text{mod}\,2)= 0,1$.
If $|s_1|$ and $|s_2|$ are of the same order $\epsilon$, one can have $M_{ij} \sim \mathcal{O}(1)$ or $M_{ij} \sim \epsilon^2$.
For the case $\bar{a}_{ij}= -1$,
the condition in~\cref{eq:z3swapcond} additionally imposes $\tilde{M}^{(n,n)}_{ij}(0,0) = 0$ for all $n$, so that 
\begin{equation} \label{eq:leadingz3minus}
    M_{ij} \sim \mathcal{O}\left(|s_1|^p\, |s_2|^p\, |s_1^2-s_2^2|\right)\,,
\end{equation} 
is expected at leading order, with $p$ as described above.
If $|s_1|$ and $|s_2|$ are of the same order $\epsilon$, one can have $M_{ij} \sim \epsilon^2$ or $M_{ij} \sim \epsilon^4$. In the case of $\bar{a}_{ij}= -1$, the factor $s_1^2-s_2^2 = (s_1-s_2)(s_1+ s_2)$ in~\cref{eq:leadingz3minus} indicates that a further suppression of $M_{ij}$ is possible, provided $s_1 \simeq s_2$ or $s_1 \simeq -s_2$. As in the previous section, if one of these holds strictly instead of approximately, then $M_{ij}$ vanishes identically.

In the case where both $\psi$ and $\psi^c$ are doublets,~\cref{eq:z3cond22} and $\tilde{M}^{(n,m)}_{ij}(0,0) \neq 0$ imply once again that $n$ and $m$ have the same parity.
Within the relevant $2\times 2$ block of the mass matrix, one finds $\tilde{M}_{11}^{(n,m)} = (-1)^{n+k/2} \tilde{M}_{22}^{(n,m)}$ and $\tilde{M}_{12}^{(n,m)} = (-1)^{n+k/2} \tilde{M}_{21}^{(n,m)}$, where $k$ is the common weight $k=k_{ij}$ ($i,j=1,2$). Since $\bar{a}_{ii} = +1$, we expect $M_{ii} \sim \mathcal{O}(1)$. Instead, for $i\neq j$ one has $\bar{a}_{ij} = -1$ and~\cref{eq:z3swapcond} forbids the zeroth term, once again relating the expansion coefficients of $s_1^ns_2^m$ and $s_1^ms_2^n$, so that one expects
\begin{equation}
    M \sim
    \begin{pmatrix}
        \mathcal{O}(1) & \mathcal{O}\left(|s_1^2-s_2^2|\right)\\[2mm]
        \mathcal{O}\left(|s_1^2-s_2^2|\right) & \mathcal{O}(1)
    \end{pmatrix}\,,
\end{equation} 
for the relevant block, at leading order. 
If $|s_1|$ and $|s_2|$ are of the same order $\epsilon$, the off-diagonal elements are generically of order $\epsilon^2$. 
Further suppression of these elements is once again possible if $s_1 \simeq s_2$ or $s_1 \simeq -s_2$.

Finally, in the mixed case with $\psi^c \sim \mathbf{1}_a$ and $\psi \sim \mathbf{2}$,~\cref{eq:z3cond12} implies opposite parities for $n$ and $m$ whenever $\tilde{M}^{(n,m)}_{ij}(0,0) \neq 0$. The $(s_1,s_2)$-expansions of the two relevant mass matrix elements are related by $\tilde{M}_{11}^{(n,m)} = \bar{b}_1(-1)^{n+k/2} \tilde{M}_{12}^{(n,m)}$.
Together with~\cref{eq:z3swapcond} and the fact that $\bar{a}_{11} = -\bar{a}_{12}$, one arrives at the expectation
\begin{equation}
    M \sim
    \begin{pmatrix}
        \mathcal{O}\left(|s_1+s_2|\right) & \mathcal{O}\left(|s_1-s_2|\right)
    \end{pmatrix}\,,
\end{equation} 
for the relevant $1\times 2$ block, if $\bar{a}_{11} =+1$. If $\bar{a}_{11} =-1$ instead, the expectations for these entries are swapped.
Then, for $|s_1|$ and $|s_2|$ of the same order $\epsilon$, these entries are generically of order $\epsilon$. 
Further suppression of one of these elements is possible if $s_1 \simeq s_2$ or $s_1 \simeq -s_2$.

\section{A viable model for quarks}
\label{sec:model}

Denoting with $Q$ the quark superfields associated with the electroweak doublets, and using $u^c$ and $d^c$ for the singlets of the up and down sectors respectively, the superpotential of the Yukawa sector reads:
\begin{equation} \label{eq:W}
    \mathcal{W}=(Y_u)_{\alpha \beta}Q_\alpha u^c_\beta H_u+(Y_d)_{\alpha \gamma}Q_\alpha d^c_\gamma H_d\,.
\end{equation}
In order to realize the MPIH mechanism within the region $\Sigma = \mathcal{T}_1$ that we singled out in~\cref{sec:mpih}, it must be noted that the lowest-weighted doublet of modular forms for the group $N_2(H) \cong (S_3\times S_3)\rtimes \mathbb{Z}_2$ corresponds to $k=6$. From the discussion on the weights required for the mass hierarchy patterns near the mentioned regions or points $\Sigma^*$ and looking at the available modular forms listed in~\cref{app:groupth}, it follows that the usage of modular forms of weight $k=\{8,10,12\}$ is one of the few options for realizing our program with the following requirements: 1) a non-identically-zero determinant, 2) a suitable hierarchy pattern, and 3) a not-too-rigid structure to avoid the cancellation of the CP-violating phases among the modular forms. Although there are other possibilities (such as assigning both $u^c$ and $d^c$ to the same irreps), we have explored, as a benchmark, the modular group assignments summarized in~\cref{tab:charges}, which allow to fit the quark data in the vicinities of $\mathcal{O}_2$, $\mathcal{Z}_4$, and $\mathcal{Z}_6$ within the same model.
The corresponding Yukawa matrices, written in the left-right convention of~\cref{eq:W}, are:
\begin{equation}
Y_u=  \begin{pmatrix}
    \alpha_1(Y_{\mathbf{2}}^{(10)})_2&\alpha_2(Y_{\mathbf{2},1}^{(12)})_1+\alpha_3(Y_{\mathbf{2},2}^{(12)})_1  &
    \alpha_4(Y_{\mathbf{2}}^{(10)})_1 
    \\
    -\alpha_1(Y_{\mathbf{2}}^{(10)})_1 &\alpha_2(Y_{\mathbf{2},1}^{(12)})_2+\alpha_3(Y_{\mathbf{2},2}^{(12)})_2  &
   \alpha_4(Y_{\mathbf{2}}^{(10)})_2
    \\
     0 & \alpha_5Y_{\mathbf{1}}^{(10)}& \alpha_6 Y_{\mathbf{1}}^{(8)}
\end{pmatrix}\,,
\end{equation}
\begin{equation}
Y_d=\begin{pmatrix}
    \beta_1(Y_{\mathbf{2}}^{(10)})_1&\beta_2(Y_{\mathbf{2},1}^{(12)})_2+\beta_3(Y_{\mathbf{2},2}^{(12)})_2  &
    \beta_4(Y_{\mathbf{2}}^{(10)})_1 
    \\
    -\beta_1(Y_{\mathbf{2}}^{(10)})_2 &\beta_2(Y_{\mathbf{2},1}^{(12)})_1+\beta_3(Y_{\mathbf{2},2}^{(12)})_1  &
    \beta_4(Y_{\mathbf{2}}^{(10)})_2
    \\
     0 & \beta_5 Y_{\mathbf{1''}}^{(10)}&   \beta_6 Y_{\mathbf{1}}^{(8)}
\end{pmatrix}\,,
\end{equation}
where $\alpha_i,\beta_i$ are free parameters in the superpotential, assumed to already include the effects of the canonical rescaling of fields due to the choice of Kähler potential.
Depending on the considered region, near $\mathcal{O}_2$  or $\mathcal{Z}_{4,6}$, these matrices are described by a total of 12 or 10 dimensionless parameters, respectively. The counting is made with the assumption that there exists a moduli stabilization mechanism which fixes $\epsilon_j=|\epsilon_j|e^{i\phi_j}$ for both $\tau_1$ and $\tau_2$,%
\footnote{These complex variables coincide with the $u_k$ defined in~\cref{eq:proxvars}. When referring to quark mass spectra, we use the compact notation $\epsilon\equiv |\epsilon|$.}
whenever one or both approach a symmetric value.
An example of how such a mechanism may arise at genus $g=1$, for the proximity to $\tau_{\text{sym}}=\omega$, has been shown in~\cite{Novichkov:2022wvg}.
In our numerical analysis, we allowed $|\epsilon_2|$ and $\phi_2$ to vary freely as opposed to $|\epsilon_1|$ and $\phi_1$, i.e.~we explored the parameter space always close to $\tau_1\simeq \omega$.
While the model has an overall limited predictive power, it fulfils our aim as a proof of concept of how MPIH constrains viable Yukawa structures in the quark sector.

\begin{table}[t]
\centering
\renewcommand{\arraystretch}{1.4}
\begin{tabular}{cccc}
\toprule
 &$Q$&$ u^c $ & $d^c$ \\ 
\midrule
$(S_3\times S_3)\rtimes \mathbb{Z}_2$ & $\mathbf{2}\oplus \mathbf{1}$  & $ \mathbf{1'}\oplus  \mathbf{1} \oplus  \mathbf{1}$   & $\mathbf{1'''}\oplus  \mathbf{1''} \oplus  \mathbf{1}$    \\ 
$k_I$ & $(6,4)$
& $(4,6,4)$ & $(4,6,4)$\\ 
\bottomrule
\end{tabular}
\caption{Irrep and weight assignments for the quark superfields in our viable model. It is assumed that the Higgs fields $H_u$ and $H_d$ are weightless flavour singlets.}
\label{tab:charges}
\end{table}

Since the fitted regions near $\mathcal{O}_2$, $\mathcal{Z}_4$, and $\mathcal{Z}_6$ share the feature that $\tau_1\simeq \omega$, we first focus on the Yukawa matrix structure for generic $\tau_2$:
\begin{equation}\label{eq:matstruct}
Y_q\, \lesssim
\begin{pmatrix}
    \epsilon & 1 &  \epsilon \\
        \epsilon & 1 &  \epsilon \\
           0 & \epsilon & \epsilon^2
\end{pmatrix}\,.
\end{equation}
Here the notation $\lesssim$ accounts for modifications to the power-counting when one additionally has $\tau_2\simeq \tau_{\text{sym}} \in\{\omega,i\}$, which potentially changes the leading order of the determinant as $\det{Y_q}\sim\epsilon^{3+r}$ with $r>0$ if $|\epsilon_2|\sim |\epsilon_1| \equiv \epsilon$. For example, in the case of $\mathcal{O}_2$ with generic $\tau_2$, it can be shown that if $\tau_2$ is far from $\{\omega,i\}$, then the leading order of the determinant is $\det{Y_q}\sim \epsilon^3$. Thus, as can be inferred from~\cref{eq:matstruct}, the starting point is, generically, a $(1,\epsilon,\epsilon^2)$ spectrum. Then, the latter can be suppressed due to the proximity of $\tau_2$ to $\tau_{\text{sym}}$. Choosing $\tau_2\simeq i$ at a comparable distance to that between $\tau_1$ and $\omega$, and assuming $\mathcal{O}(1)$ parameters, we get the following distinct modifications in the two sectors:
\begin{equation}
\label{eq:mass_omega_i0}
Y_u\sim\begin{pmatrix}
    \epsilon & 1 &  \epsilon^2 \\
        \epsilon^2 & \epsilon &  \epsilon \\
           0 & \epsilon^2 & \epsilon^2
\end{pmatrix}\,,
\quad \quad Y_d\sim\begin{pmatrix}
    \epsilon^2 & \epsilon &  \epsilon^2 \\
        \epsilon & 1 &  \epsilon \\
           0 & \epsilon & \epsilon^2
\end{pmatrix}\,,
\end{equation}
and we now have $\det{Y_u}\sim \det{Y_d}\sim \epsilon^4$. The resulting mass pattern for the up quarks is $(1,\epsilon,\epsilon^3)$, while for the down quarks it is $(1,\epsilon^2,\epsilon^2)$. This represents a particular MPIH realization in the vicinity of $\mathcal{Z}_6$.

If we instead consider a regime with $\tau_2\simeq \omega$ and $|\epsilon_2|\sim |\epsilon_1|$, we find:
\begin{equation} \label{eq:mass_omega_i1}
Y_u\sim\begin{pmatrix}
    \epsilon^2 & 1 &  \epsilon^2 \\
        \epsilon^2 & 1 &  \epsilon^2 \\
           0 & \epsilon^2 & \epsilon^4
\end{pmatrix}\,,
\quad \quad Y_d\sim\begin{pmatrix}
    \epsilon^2 & 1 &  \epsilon^2 \\
        \epsilon^2 & 1 &  \epsilon^2 \\
           0 & \epsilon^2 & \epsilon^4
\end{pmatrix}\,.
\end{equation}
We now have $\det{Y_u}\sim \det{Y_d}\sim \epsilon^6$ and the resulting mass pattern for both sectors is $(1,\epsilon^2,\epsilon^4)$. This represents a particular MPIH realization near $\mathcal{Z}_4$.

It must be stressed that the three fits were obtained without imposing a bias on the initial value of $\tau_2$. For this reason, it is noteworthy that the numerical analysis spontaneously clustered near the points $\mathcal{Z}_4$ and $\mathcal{Z}_6$, with $\tau_2\simeq \omega$ and $\tau_2\simeq i$ respectively, hinting at what seems to be a trend in the modular literature~\cite{Feruglio:2022koo,Feruglio:2023mii}. Furthermore, for both $\mathcal{Z}_4$ and $\mathcal{Z}_6$ the fit selected $|\epsilon_2|\simeq |\epsilon_1|$.

\begin{table}[t]
\centering
\renewcommand{\arraystretch}{1.3}
\begin{tabular}{cc}
\toprule
\textbf{Observable} & \textbf{Value $\pm$ $1\,\sigma$}  \\
\midrule
$y_u/y_t$ & $(5.468\pm 3.407)\cdot10^{-6}$\\
$y_c/y_t$ & $(2.678\pm 0.254)\cdot10^{-3}$\\
$y_d/y_b$ & $(6.921\pm 1.535)\cdot10^{-4}$\\
$y_s/y_b$ & $(1.370\pm 0.152)\cdot10^{-2}$\\
$\theta_{12}^q$ & $(2.274\pm 0.010)\cdot10^{-1}$\\
$\theta_{13}^q$ & $(3.145\pm 0.490)\cdot10^{-3}$\\
$\theta_{23}^q$ & $(3.585\pm0.670 )\cdot10^{-2}$\\
$\delta_\text{CP}^q$ & $1.208\pm 0.108$\\
\bottomrule
\end{tabular}
\caption{Quark observables at the GUT scale of $2\times 10^{16}$ GeV with $\tan\beta=5$, taken from~\cite{Okada:2020rjb}.}
\label{tab:observables} 
\end{table}

We performed the numerical analysis by fitting the 8 dimensionless observables listed in~\cref{tab:observables}.
As a figure of merit for the exploration of the parameter space, we adopted the Gaussian approximated
\begin{equation}
    \label{eq:gaussian}
    \chi^2=\sum_{i=1}^8\left(\frac{q_i-q_i^{\text{b-f}}}{\sigma_i}\right)^2\,,
    \end{equation}
where $q_i^{\text{b-f}}$ are the best-fit values of the observables from~\cref{tab:observables}. The input free parameters are initially selected from flat distributions, and the parameter space is explored using the algorithm outlined in~\cite{Novichkov:2018ovf,Novichkov:2021cgl}. The central values obtained when fitting near $\mathcal{O}_2,\mathcal{Z}_2$ and $\mathcal{Z}_4$ are listed in~\cref{tab:results}.

\begin{table}[t]
\small
\centering
  \renewcommand{\arraystretch}{1.2}
  \begin{tabular}{lccc}
  \toprule
Region &  $\mathcal{O}_2$ & $\mathcal{Z}_4$ & $\mathcal{Z}_6$ \\
\midrule
$\tau_1$                            
& $\omega +0.047$
& $\omega +0.072$
& $\omega+0.019$
\\
$\tau_2$                            
& $-0.196+2.220\,i$
& $-0.537+0.927\,i$
& $0.029+0.996\,i$

\\

$\alpha_1 /\alpha_5$
& $-0.751$
& $-0.388$
& $-0.286$ 

\\
$\alpha_2 /\alpha_5$
& $56.00$ 
& $1.205$
& $ -12.08$ 

\\
$\alpha_3 /\alpha_5$             
& $ 1.303$ 
& $-7.124$
& $-3.685$
\\
$\alpha_4 /\alpha_5$
& $1.594$ 
& $-1.178$
& $-0.464$ 

\\

$\alpha_6 /\alpha_5$
& $0.035$ 
& $0.188$
& $0.067$ 

\\

$\beta_1 /\beta_6$
& $-1.209$
& $0.076$
& $-0.076$ 

\\
$\beta_2 /\beta_6$
& $0.034$ 
& $-55.42$
& $ 0.206$ 

\\
$\beta_3 /\beta_6$             
& $ 0.046$ 
& $0.977$ 
& $14.05$
\\
$\beta_4 /\beta_6$
& $30.22$ 
& $7.492$
& $0.291$ 

\\

$\beta_5 /\beta_6$
& $2.444$ 
& $0.223$
& $-0.084$ 

\\

$v_u\,\alpha_5$, GeV                  
&  $15.30$  
&  $16.04$  
&  $23.23$ 

\\
$v_d\,\beta_6 $, GeV     
&  $0.370$  
&  $0.250$  
&  $0.462$  

\\
\midrule
$y_u / y_t$            
& $5.383\cdot 10^{-6}$ 
&  $5.396\cdot 10^{-6}$
& $2.304\cdot 10^{-6}$ 

\\
$y_c / y_t$     
&  $2.674\cdot 10^{-3}$ 
&  $2.650\cdot 10^{-3}$ 
& $2.687\cdot 10^{-3}$ 
\\

$y_d / y_b$            
&  $6.711\cdot 10^{-4}$
&  $7.185\cdot 10^{-4}$
& $7.370\cdot 10^{-4}$

\\
$y_s / y_b$     
&  $1.367\cdot 10^{-2}$ 
&  $1.406\cdot 10^{-2}$  
& $1.359\cdot 10^{-2}$ 
\\

$\theta_{12}$             
& $2.274\cdot 10^{-1}$
&  $2.275\cdot 10^{-1}$
& $2.273\cdot 10^{-1}$

\\
$ \theta_{13}$          
&  $3.167\cdot 10^{-3}$ 
&  $2.770\cdot 10^{-3}$  
& $3.201\cdot 10^{-3}$ 

\\
$\theta_{23}$              
&  $3.784\cdot 10^{-2}$  
&  $4.150\cdot 10^{-2}$
& $3.562\cdot 10^{-2}$
 
\\

$\delta_\text{CP}$
&  $1.211$ 
&  $1.204$
& $1.217$

\\

\midrule
$|\epsilon_1|$
& 0.027
& 0.042
& $0.011$

\\

$|\epsilon_2|$
& --
& 0.040
& $0.015$

\\

\midrule
$\chi^2_{\text{min}}$
& $0.11$
& $1.41$
& $0.97$
\\
\bottomrule
  \end{tabular}
  \caption{Benchmark values of parameters and observables for the fitted regions near $\mathcal{O}_2$, $\mathcal{Z}_4$ and $\mathcal{Z}_6$.}
  \label{tab:results}
\end{table}

Since we rely on the MPIH mechanism, in the absence of a standard normalization procedure at genus $g=2$ we chose to normalize the weight $2$ modular forms of $\Gamma_1(2)$, i.e.~the building blocks of our modular finite group reported in~\cref{app:modforms}, with the phenomenological choice for the arbitrary constant $c\simeq 0.9$. In this way, their expansion in $\epsilon_i$ near the fixed points can be considered reliable, i.e.~the numerical coefficients $b_a$ of an expansion $\sum_a b_a\epsilon_i^a$ do not considerably spoil the order of magnitude $\mathcal{O}(\epsilon_i^a)$. While this choice can in principle be reabsorbed in the definition of the superpotential free parameters, it was made in order to assist the numerical search in the parameter space, since we expect the mass hierarchies to be mainly driven by $\epsilon$ and not by parameters of order $\mathcal{O}(10^{\pm 3})$. As an \emph{a posteriori} quality measure in what concerns parameter spread, one can use the logarithmic amplitude from~\cite{Penedo:2024gtb}:
\begin{equation}
    \label{eq:PPmeasure}
    \mathcal{A}_q\equiv \text{max}\log_{10}|\alpha_{ij}^q|-\text{min}\log_{10}|\alpha_{ij}^q|\,,
\end{equation}
where $\alpha_{ij}^q$ are ratios of superpotential free parameters for each sector. A value of $\mathcal{A}_q\sim 1$ signals that the constants differ at most by $\mathcal{O}(10)$. By considering $\mathcal{A}\equiv (\mathcal{A}_u,\mathcal{A}_d)$, in our case we have:
\begin{equation}
    \label{eq:models_amplitudes}
    \mathcal{A}^{\mathcal{O}_2}\simeq(3.2,2.9)\,,\quad \mathcal{A}^{\mathcal{Z}_4}\simeq(1.6,2.9)\,,\quad \mathcal{A}^{\mathcal{Z}_6}\simeq(2.3,2.3)\,.
\end{equation}
Note that the realization near $\mathcal{Z}_4$ scores slightly better in one sector: it is characterized by the most suppressed spectrum $(1,\epsilon^2,\epsilon^4)$. Indeed, with $\epsilon\sim \mathcal{O}(10^{-2})$ being the characteristic scale required by the mass hierarchies within the fit, in the up sector we correctly have $m_u/m_c\sim m_c/m_t\sim \epsilon^2$. However, when considering both sectors, the $\mathcal{Z}_4$ realization can hardly be considered the most natural: a greater amount of fine-tuning in the free parameters enters in the down sector, since we have $m_d/m_s\sim \epsilon^2$. Thus, the Gatto-Sartori-Tonin~\cite{Gatto:1968ss} empirical relation $\lambda = \sin\theta_{12}\simeq \sqrt{m_d/m_s}$, where $\theta_{12}$ is the Cabibbo angle and $\lambda\simeq 0.22$ is the Wolfenstein expansion parameter, cannot be naturally satisfied. 

These results illustrate how quark mixing may drive the fit towards a tuned regime, even when mass hierarchies are dictated by residual symmetry. Consider, for instance, the fit around $\mathcal{Z}_6$. In the symmetric limit $\epsilon=0$, the CKM matrix $V$ is the identity matrix. Departing from this limit, one has the expansion $V=\mathbb{1}+V_1\,\epsilon+V_2\,\epsilon^2+\mathcal{O}(\epsilon^3)$. To fit the data, the matrices $V_i$ are inevitably characterized by big numerical coefficients, in order to counterbalance the smallness of $\epsilon$. In particular, one can derive the analytical expression of the CKM matrix by solving the degenerate perturbation theory problem using the Kato-Bloch method~\cite{kato:1949ts,bloch:1958cl}. From this retrospective analysis on $\mathcal{Z}_6$, we obtain:
\begin{equation} \label{eq:vckm}
    |V|\sim \begin{pmatrix}
       1-\mathcal{O}(10^2)|\epsilon|^2 & \mathcal{O}(10)|\epsilon| & \mathcal{O}(10)|\epsilon|^2\\
       \mathcal{O}(10)|\epsilon| & 1-\mathcal{O}(10^2)|\epsilon|^2 & |\epsilon| \\
       \mathcal{O}(10^2)|\epsilon|^2 & |\epsilon| &   1-\mathcal{O}(10)|\epsilon|^2
    \end{pmatrix}+\mathcal{O}(|\epsilon|^3)\,,
\end{equation}
where the numerical coefficients, for which we specified the order of magnitude, derive from some combination of the input free parameters listed in~\cref{tab:results}, and from the numerical coefficients of the Fourier expansions of the modular forms. For the fit around $\mathcal{Z}_6$ we also derived the analytical expression of the Jarlskog invariant, which at leading order turns out to be given by:
\begin{equation}
    \label{eq:jarlskog}
    J_\text{CP}=|\text{Im}(V_{11}^*V_{12}V_{21}V_{22}^*)|\sim \mathcal{O}(10^3)\,|\epsilon|^4\,.
\end{equation}
This correctly reproduces the value $J_\text{CP}\sim 10^{-5}$ for $|\epsilon|\sim 10^{-2}$, illustrating however that there is not a clear connection between the magnitudes of mixing angles and the powers of $\epsilon$. 
From this discussion, it follows that if, e.g., there exists a particular residual symmetry which is able to provide a mass hierarchy of the type $(1,\epsilon^4, \epsilon^8)$ for the up sector, and $(1,\epsilon^2, \epsilon^4)$ for the down sector, then it would be possible to have $\epsilon \sim \mathcal{O}(\lambda)$, and the correct CKM matrix would follow with natural sizes of the matrix elements, entirely driven by adequate powers of $\epsilon$. We leave this insight to future work.

\section{Summary and conclusions}
\label{sec:summary}
Modular invariance is a powerful tool to investigate the flavour puzzle from a topology-driven point of view.  In this context, the proximity to criticality (fixed regions or fixed points of a residual symmetry) may explain the mass spectra of fundamental fermions. We refer to this mechanism as Modular Proximity-Induced Hierarchies (MPIH).

We have analyzed a generalization of the MPIH mechanism by considering the most natural extension of the modular invariance framework: the Siegel modular group $Sp(4,\mathbb{Z})$. This choice is motivated by the requirement that both the mass hierarchies and the CP violation of the theory arise from the same VEVs which break the flavour symmetry.
We discuss how to quantify the proximity to a fixed point or region in the Siegel half-space, which correspond to: six fixed points (dimension zero, denoted~$\mathcal{Z}_i$), five regions of complex dimension one (denoted $\mathcal{O}_i$), and two regions of complex dimension two (denoted $\mathcal{T}_i$), summarized in~\cref{tab:FixedPoints}.

To realize the flavour program, one starts by choosing a phenomenologically favoured Siegel subspace $\Sigma$, associated with a finite flavour group admitting irreducible representations of adequate dimensions.
Postulating that, within a region $\Sigma$ of the moduli space, a particular mechanism stabilizes the moduli in the vicinity of a fixed point or region $\Sigma^*$, the small breaking of the residual symmetry group at $\Sigma^*$ determines the hierarchical pattern of the mass matrix entries, in some cases with a richer structure than the genus-one case. Indeed, the Siegel extension of the MPIH cannot be thought as simply the direct product of two tori. For instance, for the $\Sigma = \mathcal{T}_1$ subspace, a novel transformation exchanging the moduli is present and may provide additional constraints on the mass matrix elements.
The possible ``trajectories'' $\Sigma \rightarrowtail \Sigma^*$, where one approaches $\Sigma^*$ within $\Sigma$, are summarized in~\cref{tab:trajectories}. In its last two columns, we present the relevant residual symmetry group and the number of available (complex) degrees of freedom in moduli space, $\Delta \dim$. We find that, to obtain the desired fermion mass hierarchies, when $\Delta \dim = 1$ one must focus on fixed points where at least a $\mathbb{Z}_3$ residual symmetry is preserved. This is in line with what is already found at genus $g=1$. On the other hand, when considering $\Delta\dim=2$, a residual $\mathbb{Z}_2$ is sometimes enough to obtain the required mass hierarchies, which are functions of the two deviation parameters, extending the space of possibilities with respect to the $g=1$ case. 
In particular, we have exhausted the viable MPIH scenarios for the case of $\Sigma = \mathcal{T}_1$, summarizing the possible hierarchical patterns in~\cref{tab:epsilon}.

Finally, as a proof of concept, we have provided a benchmark gCP-invariant model for the quark sector which fits the data in the proximity of $\Sigma^* \in \{\mathcal{O}_2, \mathcal{Z}_4, \mathcal{Z}_6\}$, with $\tau_1 \simeq \omega$. This was done by choosing as a flavour symmetry the modular finite group of $\mathcal{T}_1$, namely $(S_3\times S_3)\rtimes \mathbb{Z}_2$. It is remarkable that the fits around $\mathcal{Z}_4$ and $\mathcal{Z}_6$ were obtained automatically, i.e.~the unconstrained numerical search on $\tau_2$ led to $\tau_2\simeq \{\omega , i\}$. Moreover, due to the more complex structure of the Siegel space, for the first time in the modular literature, the CP violation from a vacuum which controls all quark mass ratios (that approach zero near the symmetric limit) is enough to fit also the CKM phase. On the other hand, a predictive model free of significant fine-tuning overall remains to be found.


\appendix
\crefalias{section}{appendix}

\section{Group theory of \texorpdfstring{$(S_3\times S_3)\rtimes \mathbb{Z}_2$}{S3xS3:Z2}}
\label{app:groupth}

\subsection{Properties and irreducible representations}
\label{app:propirreps}

The restricted finite Siegel modular group $N_2(H)$ associated with the invariant region $\mathcal{T}_1$ is isomorphic to $(S_3\times S_3)\rtimes \mathbb{Z}_2$ and
can be presented in terms of five generators $G_{1,2}$, $G'_{1,2}$ and $G_3$ satisfying the relations given if~\cref{eq:S3xS3:Z2}. Each of the two pairs $G_{1,2}^{(\prime)}$ describes one of the $S_3$ factors, while $G_3$ is associated with the action of the $\mathbb{Z}_2$ factor.
It is a group of 72 elements with group ID \texttt{[72,40]} within the computer algebra system GAP~\cite{GAP4,SmallGroups}.
It admits 9 irreducible representations -- 4 one-dimensional, 1 two-dimensional, and 4 four-dimensional -- which we denote by: $\mathbf{1},\, \mathbf{1'},\, \mathbf{1''},\, \mathbf{1'''},\, \mathbf{2},\, \mathbf{4},\, \mathbf{4'},\, \mathbf{4''},\, \mathbf{4'''}$.

\subsection{Representation basis}
\label{app:sym_basis}

In this work, we have adopted the symmetric and real basis for the representations of the generators summarized in~\cref{tab:basis}. Such a basis is convenient for the study of modular symmetry extended by a gCP symmetry, as described in~\cref{sec:gCP}.

\subsection{Tensor products and Clebsch-Gordan coefficients}
\label{app:CGCs}

We present here the decompositions of tensor products of $N_2(H)$ irreps, as well as the corresponding Clebsch-Gordan coefficients, in the basis of~\cref{tab:basis}.
Entries of each multiplet entering the tensor product are denoted by $\alpha_i$ and $\beta_i$. Apart from the trivial products $\mathbf{1} \otimes \mathbf{r} = \mathbf{r}$, these results are collected in~\cref{tab:CGCs}.

\afterpage{
\begin{landscape}
\thispagestyle{empty}
\begin{table}[ht]
\thisfloatpagestyle{empty}
\centering
\begin{tabular}{lccccc}
    \toprule
    $\mathbf{r}$ & $\rho_{\mathbf{r}}(G_1)$ & $\rho_{\mathbf{r}}(G_2)$ & $\rho_{\mathbf{r}}(G_1')$ & $\rho_{\mathbf{r}}(G_2')$ & $\rho_{\mathbf{r}}(G_3)$ \\
    \midrule
    $\mathbf{1}$ & $1$ & $1$ & $1$ & $1$ & $1$ \\
    \addlinespace
    $\mathbf{1'}$ & $-1$ & $-1$ & $-1$ & $-1$ & $-1$ \\
    \addlinespace
    $\mathbf{1''}$ & $-1$ & $-1$ & $-1$ & $-1$ & $1$ \\
    \addlinespace
    $\mathbf{1'''}$ & $1$ & $1$ & $1$ & $1$ & $-1$ \\
    \addlinespace
    $\mathbf{2}$
        & $\begin{pmatrix} 1 & 0 \\ 0 & -1 \end{pmatrix}$
        & $\begin{pmatrix} 1 & 0 \\ 0 & -1 \end{pmatrix}$
        & $\begin{pmatrix} -1 & 0 \\ 0 & 1 \end{pmatrix}$
        & $\begin{pmatrix} -1 & 0 \\ 0 & 1 \end{pmatrix}$
        & $\begin{pmatrix} 0 & 1 \\ 1 & 0 \end{pmatrix}$
    \\
    \addlinespace
    $\mathbf{4}$
         & $\begin{pmatrix} -1&0&0&0\\ 0&-1&0&0\\ 0&0&\frac{1}{2}&\frac{\sqrt{3}}{2}\\ 0&0&\frac{\sqrt{3}}{2}&-\frac{1}{2} \end{pmatrix}$
         & $\begin{pmatrix} -1&0&0&0\\ 0&-1&0&0\\ 0&0&-1&0\\ 0&0&0&1 \end{pmatrix}$
        & $\begin{pmatrix} \frac{1}{2}&\frac{\sqrt{3}}{2}&0&0\\ \frac{\sqrt{3}}{2}&-\frac{1}{2}&0&0\\ 0&0&-1&0\\ 0&0&0&-1 \end{pmatrix}$
        & $\begin{pmatrix} -1&0&0&0\\ 0&1&0&0\\ 0&0&-1&0\\ 0&0&0&-1 \end{pmatrix}$
        & $\begin{pmatrix} 0&0&1&0\\ 0&0&0&1\\ 1&0&0&0\\ 0&1&0&0 \end{pmatrix}$
    \\
    \addlinespace
    $\mathbf{4'}$
        & $\begin{pmatrix} -\frac{1}{2}&-\frac{\sqrt{3}}{2}&0&0\\ -\frac{\sqrt{3}}{2}&\frac{1}{2}&0&0\\ 0&0&-\frac{1}{2}&-\frac{\sqrt{3}}{2}\\ 0&0&-\frac{\sqrt{3}}{2}&\frac{1}{2} \end{pmatrix}$
        & $\begin{pmatrix} 1&0&0&0\\ 0&-1&0&0\\ 0&0&1&0\\ 0&0&0&-1 \end{pmatrix}$
        & $\begin{pmatrix} -1&0&0&0\\ 0&-1&0&0\\ 0&0&1&0\\ 0&0&0&1 \end{pmatrix}$
        & $\begin{pmatrix} \frac{1}{2}&0&-\frac{\sqrt{3}}{2}&0\\ 0&\frac{1}{2}&0&-\frac{\sqrt{3}}{2}\\ -\frac{\sqrt{3}}{2}&0&-\frac{1}{2}&0\\ 0&-\frac{\sqrt{3}}{2}&0&-\frac{1}{2} \end{pmatrix}$
        & $\begin{pmatrix} -\frac{3}{4}&-\frac{\sqrt{3}}{4}&\frac{\sqrt{3}}{4}&\frac{1}{4}\\ -\frac{\sqrt{3}}{4}&-\frac{1}{4}&-\frac{3}{4}&-\frac{\sqrt{3}}{4}\\ \frac{\sqrt{3}}{4}&-\frac{3}{4}&-\frac{1}{4}&\frac{\sqrt{3}}{4}\\ \frac{1}{4}&-\frac{\sqrt{3}}{4}&\frac{\sqrt{3}}{4}&-\frac{3}{4} \end{pmatrix}$
    \\
    \addlinespace
    $\mathbf{4''}$
        & $\begin{pmatrix} -\frac{1}{2}&-\frac{\sqrt{3}}{2}&0&0\\ -\frac{\sqrt{3}}{2}&\frac{1}{2}&0&0\\ 0&0&-\frac{1}{2}&-\frac{\sqrt{3}}{2}\\ 0&0&-\frac{\sqrt{3}}{2}&\frac{1}{2} \end{pmatrix}$
        & $\begin{pmatrix} 1&0&0&0\\ 0&-1&0&0\\ 0&0&1&0\\ 0&0&0&-1 \end{pmatrix}$
        & $\begin{pmatrix} -\frac{1}{2}&0&-\frac{\sqrt{3}}{2}&0\\ 0&-\frac{1}{2}&0&-\frac{\sqrt{3}}{2}\\ -\frac{\sqrt{3}}{2}&0&\frac{1}{2}&0\\ 0&-\frac{\sqrt{3}}{2}&0&\frac{1}{2} \end{pmatrix}$
        & $\begin{pmatrix} 1&0&0&0\\ 0&1&0&0\\ 0&0&-1&0\\ 0&0&0&-1 \end{pmatrix}$
        & $\begin{pmatrix} 1&0&0&0\\ 0&0&1&0\\ 0&1&0&0\\ 0&0&0&1 \end{pmatrix}$
    \\
    \addlinespace
    $\mathbf{4'''}$
       & $\begin{pmatrix} 1&0&0&0\\ 0&1&0&0\\ 0&0&-\frac{1}{2}&-\frac{\sqrt{3}}{2}\\ 0&0&-\frac{\sqrt{3}}{2}&\frac{1}{2} \end{pmatrix}$
       & $\begin{pmatrix} 1&0&0&0\\ 0&1&0&0\\ 0&0&1&0\\ 0&0&0&-1 \end{pmatrix}$
       & $\begin{pmatrix} -\frac{1}{2}&-\frac{\sqrt{3}}{2}&0&0\\ -\frac{\sqrt{3}}{2}&\frac{1}{2}&0&0\\ 0&0&1&0\\ 0&0&0&1 \end{pmatrix}$
       & $\begin{pmatrix} 1&0&0&0\\ 0&-1&0&0\\ 0&0&1&0\\ 0&0&0&1 \end{pmatrix}$
       & $\begin{pmatrix} 0&0&1&0\\ 0&0&0&1\\ 1&0&0&0\\ 0&1&0&0 \end{pmatrix}$
    \\
    \bottomrule
\end{tabular}
\caption{Representation matrices of $(S_3\times S_3)\rtimes \mathbb{Z}_2$ generators, for different irreps $\mathbf{r}$.}
\label{tab:basis}
\end{table}
\end{landscape}

{
\renewcommand{\arraystretch}{1.1}
\begin{longtable}{cc}
\caption{Decomposition of all non-trivial tensor products involving $N_2(H)$ irreps, and corresponding Clebsch-Gordan coefficients. Note that the order is important in matching the results in the left- and right-hand columns.}
\label{tab:CGCs} \\
\toprule
\addlinespace
   ${ }$ \, Tensor product decomposition \, ${ }$
&  ${ }$ \, Clebsch-Gordan coefficients  \, ${ }$\\
\addlinespace
\midrule
\endfirsthead
\caption{(cont.)}\\
\toprule
\addlinespace
   ${ }$ \, Tensor product decomposition \, ${ }$
&  ${ }$ \, Clebsch-Gordan coefficients  \, ${ }$\\
\addlinespace
\midrule \endhead
\endfoot
\addlinespace
$\begin{array}{@{}l@{{}\,\otimes\,{}}l@{{}\,\,=\,\,{}}l@{}}
\mathbf{1'} & \mathbf{1'}   & \mathbf{1}    \\
\mathbf{1'} & \mathbf{1''}  & \mathbf{1'''} \\
\mathbf{1'} & \mathbf{1'''} & \mathbf{1''}  \\
\mathbf{1''}  & \mathbf{1''}  & \mathbf{1} \\
\mathbf{1''}  & \mathbf{1'''} & \mathbf{1'}  \\
\mathbf{1'''} & \mathbf{1'''} & \mathbf{1}
\end{array}$
& $\alpha_1 \beta_1$ \\
\addlinespace
\midrule
\addlinespace
$\begin{array}{@{}l@{{}\,\otimes\,{}}l@{{}\,\,=\,\,{}}l@{}}
\mathbf{1'}       & \mathbf{2}       & \mathbf{2}
\end{array}$
& $\alpha_1 \begin{pmatrix} \beta_2\\ -\beta_1 \end{pmatrix}$ \\
\addlinespace
\midrule
\addlinespace
$\begin{array}{@{}l@{{}\,\otimes\,{}}l@{{}\,\,=\,\,{}}l@{}}
\mathbf{1''}  & \mathbf{2}       & \mathbf{2}
\end{array}$
&
$\alpha_1 \begin{pmatrix} \beta_2\\ \beta_1 \end{pmatrix}$ \\
\addlinespace
\midrule
\addlinespace
$\begin{array}{@{}l@{{}\,\otimes\,{}}l@{{}\,\,=\,\,{}}l@{}}
\mathbf{1'''} & \mathbf{2}       & \mathbf{2}
\end{array}$
& $\alpha_1 \begin{pmatrix} \beta_1\\ -\beta_2 \end{pmatrix}$ \\
\addlinespace
\midrule
\addlinespace
$\begin{array}{@{}l@{{}\,\otimes\,{}}l@{{}\,\,=\,\,{}}l@{}}
\mathbf{1''}       & \mathbf{4}        & \mathbf{4'''}       \\
\mathbf{1''}       & \mathbf{4'''} & \mathbf{4} 
\end{array}$
&
$\alpha_1 \begin{pmatrix} \beta_1\\ \beta_2 \\ \beta_3 \\ \beta_4 \end{pmatrix}$ \\
\addlinespace
\midrule
\addlinespace
$\begin{array}{@{}l@{{}\,\otimes\,{}}l@{{}\,\,=\,\,{}}l@{}}
\mathbf{1'}  & \mathbf{4}        & \mathbf{4'''}  \\
\mathbf{1'}  & \mathbf{4'''} & \mathbf{4}        \\
\mathbf{1'''} & \mathbf{4}        & \mathbf{4} \\
\mathbf{1'''} & \mathbf{4'''} & \mathbf{4'''}
\end{array}$
&
$\alpha_1 \begin{pmatrix} \beta_1\\ \beta_2 \\ -\beta_3 \\ -\beta_4 \end{pmatrix}$ \\
\addlinespace
\midrule
\addlinespace
$\begin{array}{@{}l@{{}\,\otimes\,{}}l@{{}\,\,=\,\,{}}l@{}}
\mathbf{1''}  & \mathbf{4'}       & \mathbf{4'} \\
\mathbf{1''}  & \mathbf{4''}  & \mathbf{4''}    
\end{array}$
&
$\alpha_1 \begin{pmatrix} \beta_4\\ -\beta_3 \\ -\beta_2 \\ \beta_1 \end{pmatrix}$ \\
\addlinespace
\midrule
\addlinespace
$\begin{array}{@{}l@{{}\,\otimes\,{}}l@{{}\,\,=\,\,{}}l@{}}
\mathbf{1'}       & \mathbf{4'}       & \mathbf{4''}       
\end{array}$
&
$\alpha_1 \begin{pmatrix} 
\beta_2 + \sqrt{3}\, \beta_4 \\
-\beta_1 - \sqrt{3}\, \beta_3 \\
-\sqrt{3}\, \beta_2 +  \beta_4 \\
\sqrt{3}\, \beta_1 -  \beta_3 
\end{pmatrix}$ \\
\addlinespace
\midrule
\addlinespace
$\begin{array}{@{}l@{{}\,\otimes\,{}}l@{{}\,\,=\,\,{}}l@{}}
\mathbf{1'}       & \mathbf{4''}  & \mathbf{4'}
\end{array}$
&
$\alpha_1  \begin{pmatrix} 
\beta_2 - \sqrt{3}\, \beta_4 \\
-\beta_1 + \sqrt{3}\, \beta_3 \\
\sqrt{3}\, \beta_2 +  \beta_4 \\
-\sqrt{3}\, \beta_1 -  \beta_3 
\end{pmatrix}$ \\
\addlinespace
\midrule
\addlinespace
$\begin{array}{@{}l@{{}\,\otimes\,{}}l@{{}\,\,=\,\,{}}l@{}}
\mathbf{1'''} & \mathbf{4'}       & \mathbf{4''}
\end{array}$
&
$\alpha_1 \begin{pmatrix} 
\sqrt{3}\, \beta_1 -  \beta_3 \\
\sqrt{3}\, \beta_2 -  \beta_4 \\
\beta_1 + \sqrt{3}\, \beta_3 \\
\beta_2 + \sqrt{3}\, \beta_4 
\end{pmatrix}$ \\
\addlinespace
\midrule
\addlinespace
$\begin{array}{@{}l@{{}\,\otimes\,{}}l@{{}\,\,=\,\,{}}l@{}}
\mathbf{1'''} & \mathbf{4''}  & \mathbf{4'}        
\end{array}$
&
$\alpha_1 \begin{pmatrix} 
\sqrt{3}\, \beta_1 +  \beta_3 \\
\sqrt{3}\, \beta_2 +  \beta_4 \\
-\beta_1 + \sqrt{3}\, \beta_3 \\
-\beta_2 + \sqrt{3}\, \beta_4 
\end{pmatrix}$ \\
\addlinespace
\midrule
\addlinespace
$\begin{array}{@{}l@{{}\,\otimes\,{}}l@{{}\,\,=\,\,{}}l@{}}
\mathbf{2} & \mathbf{2}  & \mathbf{1} \,\oplus\, \mathbf{1'} \,\oplus\, \mathbf{1''} \,\oplus\, \mathbf{1'''}        
\end{array}$
&
$\begin{array}{rl}
& \alpha_1\beta_1 + \alpha_2\beta_2 \\
\oplus & \alpha_1\beta_2 -\alpha_2\beta_1 \\
\oplus & \alpha_1\beta_2 + \alpha_2\beta_1 \\
\oplus & \alpha_1\beta_1 - \alpha_2\beta_2
\end{array}$ \\
\addlinespace
\midrule
\addlinespace
$\begin{array}{@{}l@{{}\,\otimes\,{}}l@{{}\,\,=\,\,{}}l@{{}\,\oplus\,{}}l@{}}
\mathbf{2} & \mathbf{4}  & \mathbf{4} & \mathbf{4'''}\\
\mathbf{2} & \mathbf{4'''}  & \mathbf{4'''} & \mathbf{4}
\end{array}$
&
$\begin{pmatrix}
    \alpha_1\beta_2 \\ -\alpha_1\beta_1 \\ \alpha_2\beta_4 \\ -\alpha_2\beta_3
\end{pmatrix} \,\oplus\, \begin{pmatrix}
    \alpha_2\beta_2 \\ -\alpha_2\beta_1 \\ \alpha_1\beta_4 \\ -\alpha_1\beta_3
\end{pmatrix} $ \\
\addlinespace
\midrule
\addlinespace
$\begin{array}{@{}l@{{}\,\otimes\,{}}l@{{}\,\,=\,\,{}}l@{{}\,\oplus\,{}}l@{}}
\mathbf{2} & \mathbf{4'}  & \mathbf{4'} & \mathbf{4''}
\end{array}$
&
$\begin{array}{rl}
&  \begin{pmatrix}
     \alpha_1\beta_3 + \alpha_2\beta_2 \\ \alpha_1\beta_4-\alpha_2\beta_1 \\ -\alpha_1\beta_1 + \alpha_2\beta_4 \\ -\alpha_1\beta_2 - \alpha_2\beta_3
\end{pmatrix} \\[9mm]
\oplus & \begin{pmatrix}
    \alpha_1\beta_1 - \sqrt{3}\,\alpha_2\beta_2 + \sqrt{3}\,\alpha_1\beta_3 + \alpha_2\beta_4 \\ \sqrt{3}\,\alpha_2\beta_1 + \alpha_1\beta_2 - \alpha_2\beta_3 + \sqrt{3}\,\alpha_1\beta_4 \\ -\sqrt{3}\,\alpha_1\beta_1 - \alpha_2\beta_2 + \alpha_1\beta_3 - \sqrt{3}\,\alpha_2\beta_4 \\ \alpha_2\beta_1 - \sqrt{3}\,\alpha_1\beta_2 + \sqrt{3}\,\alpha_2\beta_3 + \alpha_1\beta_4
\end{pmatrix}
\end{array}$ \\
\addlinespace
\midrule
\addlinespace
$\begin{array}{@{}l@{{}\,\otimes\,{}}l@{{}\,\,=\,\,{}}l@{{}\,\oplus\,{}}l@{}}
\mathbf{2} & \mathbf{4''}  & \mathbf{4'} & \mathbf{4''}
\end{array}$
&
$\begin{array}{rl}
&  \begin{pmatrix}
    \alpha_1\beta_1 + \sqrt{3}\,\alpha_2\beta_2 - \sqrt{3}\,\alpha_1\beta_3 + \alpha_2\beta_4 \\ -\sqrt{3}\,\alpha_2\beta_1 + \alpha_1\beta_2 - \alpha_2\beta_3 - \sqrt{3}\,\alpha_1\beta_4 \\ \sqrt{3}\,\alpha_1\beta_1 - \alpha_2\beta_2 + \alpha_1\beta_3 + \sqrt{3}\,\alpha_2\beta_4 \\ \alpha_2\beta_1 + \sqrt{3}\,\alpha_1\beta_2 - \sqrt{3}\,\alpha_2\beta_3 + \alpha_1\beta_4
\end{pmatrix} \\[9mm]
\oplus & \begin{pmatrix}
     \alpha_1\beta_3 + \alpha_2\beta_2 \\ \alpha_1\beta_4-\alpha_2\beta_1 \\ -\alpha_1\beta_1 + \alpha_2\beta_4 \\ -\alpha_1\beta_2 - \alpha_2\beta_3
\end{pmatrix}
\end{array}$ \\
\addlinespace
\midrule
\addlinespace
%
$\begin{array}{rl}
\mathbf{4} \,\otimes\, \mathbf{4} &=\,\, \mathbf{1} \,\oplus\, \mathbf{1'''} \,\oplus\, \mathbf{2}\\ &\quad\oplus\, \,\mathbf{4'} \,\oplus\, \mathbf{4''} \,\oplus\, \mathbf{4'''} \\[4mm]
\mathbf{4'''} \,\otimes\, \mathbf{4'''}\!\!\!  &=\,\, \mathbf{1} \,\oplus\, \mathbf{1'''} \,\oplus\, \mathbf{2}\\ &\quad\oplus\, \,\mathbf{4'} \,\oplus\, \mathbf{4''} \,\oplus\, \mathbf{4'''} 
\end{array}$
&
$\begin{array}{rl}
&  \alpha_1\beta_1 + \alpha_2\beta_2 + \alpha_3\beta_3 + \alpha_4\beta_4 \\
\oplus & \alpha_1\beta_1 + \alpha_2\beta_2 - \alpha_3\beta_3 - \alpha_4\beta_4 \\[1mm]
\oplus & \begin{pmatrix}
     \alpha_1\beta_2 - \alpha_2\beta_1 \\ \alpha_3\beta_4 - \alpha_4\beta_3
\end{pmatrix} \\[4mm]
\oplus & \begin{pmatrix}
     \sqrt{3}\,\alpha_3\beta_1 + \alpha_3\beta_2 - \sqrt{3}\,\alpha_1\beta_3 - \alpha_2\beta_3 \\
\sqrt{3}\,\alpha_4\beta_1 + \alpha_4\beta_2 - \sqrt{3}\,\alpha_1\beta_4 - \alpha_2\beta_4 \\
-\alpha_3\beta_1 + \sqrt{3}\,\alpha_3\beta_2 + \alpha_1\beta_3 - \sqrt{3}\,\alpha_2\beta_3 \\
-\alpha_4\beta_1 + \sqrt{3}\,\alpha_4\beta_2 + \alpha_1\beta_4 - \sqrt{3}\,\alpha_2\beta_4
\end{pmatrix} \\[9mm]
\oplus & \begin{pmatrix}
    \alpha_1\beta_3 + \alpha_3\beta_1 \\
    \alpha_1\beta_4 + \alpha_4\beta_1 \\
    \alpha_2\beta_3 + \alpha_3\beta_2 \\
    \alpha_2\beta_4 + \alpha_4\beta_2 
\end{pmatrix} \, \oplus\, \begin{pmatrix}
    \alpha_1\beta_1 - \alpha_2\beta_2 \\
    - \alpha_1\beta_2-\alpha_2\beta_1 \\
    \alpha_3\beta_3 - \alpha_4\beta_4 \\
    - \alpha_3\beta_4-\alpha_4\beta_3
\end{pmatrix}
\end{array}$ \\
\addlinespace
\midrule
\addlinespace
$\begin{array}{rl}
\mathbf{4''} \,\otimes\, \mathbf{4''}\!\!  &=\,\, \mathbf{1} \,\oplus\, \mathbf{1''} \,\oplus\, \mathbf{2}\\ &\quad\oplus\, \,\mathbf{4} \,\oplus\, \mathbf{4''} \,\oplus\, \mathbf{4'''} 
\end{array}$
&
$\begin{array}{rl}
&  \alpha_1\beta_1 + \alpha_2\beta_2 + \alpha_3\beta_3 + \alpha_4\beta_4 \\
\oplus & \alpha_1\beta_4 - \alpha_2\beta_3 - \alpha_3\beta_2 + \alpha_4\beta_1 \\[1mm]
\oplus & \begin{pmatrix}
     \alpha_1\beta_3 + \alpha_2\beta_4 -\alpha_3\beta_1 - \alpha_4\beta_2  \\
     \alpha_1\beta_2 -\alpha_2\beta_1 + \alpha_3\beta_4  - \alpha_4\beta_3
\end{pmatrix} \\[4mm]
\oplus & \begin{pmatrix}
\alpha_1\beta_4 -\alpha_2\beta_3 + \alpha_3\beta_2  - \alpha_4\beta_1 \\
\alpha_1\beta_2 -\alpha_2\beta_1 - \alpha_3\beta_4  + \alpha_4\beta_3 \\
\alpha_1\beta_4 +\alpha_2\beta_3 - \alpha_3\beta_2  - \alpha_4\beta_1 \\
\alpha_1\beta_3 - \alpha_2\beta_4 -\alpha_3\beta_1 + \alpha_4\beta_2
\end{pmatrix} \\[9mm]
\oplus & \begin{pmatrix}
\alpha_1\beta_1 -\alpha_2\beta_2 - \alpha_3\beta_3  + \alpha_4\beta_4 \\
-\alpha_1\beta_2 -\alpha_2\beta_1 + \alpha_3\beta_4  + \alpha_4\beta_3 \\
-\alpha_1\beta_3 +\alpha_2\beta_4 - \alpha_3\beta_1  + \alpha_4\beta_2 \\
\alpha_1\beta_4 +\alpha_2\beta_3 + \alpha_3\beta_2  + \alpha_4\beta_1 
\end{pmatrix} \\[9mm]
\oplus & \begin{pmatrix}
\alpha_1\beta_1 +\alpha_2\beta_2 - \alpha_3\beta_3  - \alpha_4\beta_4 \\
-\alpha_1\beta_3 -\alpha_2\beta_4 - \alpha_3\beta_1  - \alpha_4\beta_2 \\
\alpha_1\beta_1 - \alpha_2\beta_2 + \alpha_3\beta_3  - \alpha_4\beta_4 \\
-\alpha_1\beta_2 -\alpha_2\beta_1 - \alpha_3\beta_4  - \alpha_4\beta_3
\end{pmatrix}
\end{array}$ \\
\addlinespace
\midrule
\addlinespace
$\begin{array}{rl}
\mathbf{4'} \,\otimes\, \mathbf{4'}\! &=\,\, \mathbf{1} \,\oplus\, \mathbf{1''} \,\oplus\, \mathbf{2}\\ &\quad\oplus\, \,\mathbf{4} \,\oplus\, \mathbf{4''} \,\oplus\, \mathbf{4'''} 
\end{array}$
&
$\begin{array}{rl}
&  \alpha_1\beta_1 + \alpha_2\beta_2 + \alpha_3\beta_3 + \alpha_4\beta_4 \\
\oplus & \alpha_1\beta_4 - \alpha_2\beta_3 - \alpha_3\beta_2 + \alpha_4\beta_1 \\[1mm]
\oplus & \begin{pmatrix}
     \alpha_1\beta_3 + \alpha_2\beta_4 -\alpha_3\beta_1 - \alpha_4\beta_2  \\
     \alpha_1\beta_2 -\alpha_2\beta_1 + \alpha_3\beta_4  - \alpha_4\beta_3
\end{pmatrix} \\[4mm]
\oplus & \begin{pmatrix*}[l]
-\sqrt{3}\,\alpha_2\beta_1
-{\alpha_4\beta_1}
+\sqrt{3}\,\alpha_1\beta_2
+{\alpha_3\beta_2}\\ \qquad
-{\alpha_2\beta_3}
+\sqrt{3}\,\alpha_4\beta_3
+{\alpha_1\beta_4}
-\sqrt{3}\,\alpha_3\beta_4
\\[1mm]
-{\alpha_2\beta_1}
+\sqrt{3}\,\alpha_4\beta_1
+{\alpha_1\beta_2}
-\sqrt{3}\,\alpha_3\beta_2 \\ \qquad
+\sqrt{3}\,\alpha_2\beta_3
+{\alpha_4\beta_3}
-\sqrt{3}\,\alpha_1\beta_4
-{\alpha_3\beta_4}
\\[1mm]
{2\alpha_1\beta_4}
+{2\alpha_2\beta_3}
-{2\alpha_3\beta_2}
-{2\alpha_4\beta_1}
\\[1mm]
{2\alpha_1\beta_3}
-{2\alpha_2\beta_4}
-{2\alpha_3\beta_1}
+{2\alpha_4\beta_2}
\end{pmatrix*} \\[1.6cm]
\oplus & \begin{pmatrix*}[l]
-\alpha_1\beta_1
+\sqrt{3}\,\alpha_3\beta_1
+\alpha_2\beta_2
-\sqrt{3}\,\alpha_4\beta_2\\ \qquad
+\sqrt{3}\,\alpha_1\beta_3
+\alpha_3\beta_3
-\sqrt{3}\,\alpha_2\beta_4
-\alpha_4\beta_4
\\[1mm]
\alpha_2\beta_1
-\sqrt{3}\,\alpha_4\beta_1
+\alpha_1\beta_2
-\sqrt{3}\,\alpha_3\beta_2\\ \qquad
-\sqrt{3}\,\alpha_2\beta_3
-\alpha_4\beta_3
-\sqrt{3}\,\alpha_1\beta_4
-\alpha_3\beta_4
\\[1mm]
\sqrt{3}\,\alpha_1\beta_1
+\alpha_3\beta_1
-\sqrt{3}\,\alpha_2\beta_2
-\alpha_4\beta_2\\ \qquad
+\alpha_1\beta_3
-\sqrt{3}\,\alpha_3\beta_3
-\alpha_2\beta_4
+\sqrt{3}\,\alpha_4\beta_4
\\[1mm]
-\sqrt{3}\,\alpha_2\beta_1
-\alpha_4\beta_1
-\sqrt{3}\,\alpha_1\beta_2
-\alpha_3\beta_2\\ \qquad
-\alpha_2\beta_3
+\sqrt{3}\,\alpha_4\beta_3
-\alpha_1\beta_4
+\sqrt{3}\,\alpha_3\beta_4
\end{pmatrix*} \\[2.1cm]
\oplus & \begin{pmatrix*}[l]
\alpha_1\beta_1
-\sqrt{3}\,\alpha_3\beta_1
+\alpha_2\beta_2
-\sqrt{3}\,\alpha_4\beta_2\\ \qquad
-\sqrt{3}\,\alpha_1\beta_3
-\alpha_3\beta_3
-\sqrt{3}\,\alpha_2\beta_4
-\alpha_4\beta_4
\\[1mm]
-\sqrt{3}\,\alpha_1\beta_1
-\alpha_3\beta_1
-\sqrt{3}\,\alpha_2\beta_2
-\alpha_4\beta_2\\ \qquad
-\alpha_1\beta_3
+\sqrt{3}\,\alpha_3\beta_3
-\alpha_2\beta_4
+\sqrt{3}\,\alpha_4\beta_4
\\[1mm]
2\alpha_1\beta_1
-2\alpha_2\beta_2
+2\alpha_3\beta_3
-2\alpha_4\beta_4
\\[1mm]
-2\alpha_2\beta_1
-2\alpha_1\beta_2
-2\alpha_4\beta_3
-2\alpha_3\beta_4
\end{pmatrix*}
\end{array}$ \\
\addlinespace
\midrule
\addlinespace
$\begin{array}{rl}
\mathbf{4} \,\otimes\, \mathbf{4'''} \!\!\!&=\,\, \mathbf{1'} \,\oplus\, \mathbf{1''} \,\oplus\, \mathbf{2}\\ &\quad\oplus\, \,\mathbf{4'} \,\oplus\, \mathbf{4} \,\oplus\, \mathbf{4''}  
\end{array}$
&
$\begin{array}{rl}
&  \alpha_1\beta_1 + \alpha_2\beta_2 - \alpha_3\beta_3 - \alpha_4\beta_4 \\
\oplus & \alpha_1\beta_1 + \alpha_2\beta_2 + \alpha_3\beta_3 + \alpha_4\beta_4 \\[1mm]
\oplus & \begin{pmatrix}
    \alpha_3\beta_4 - \alpha_4\beta_3 \\ \alpha_1\beta_2 - \alpha_2\beta_1
\end{pmatrix} \\[4mm]
\oplus & \begin{pmatrix}
    -\alpha_4\beta_1 + \sqrt{3}\,\alpha_4\beta_2 + \alpha_1\beta_4 - \sqrt{3}\,\alpha_2\beta_4 \\
    \alpha_3\beta_1 - \sqrt{3}\,\alpha_3\beta_2 - \alpha_1\beta_3 + \sqrt{3}\,\alpha_2\beta_3 \\
    -\sqrt{3}\,\alpha_4\beta_1 - \alpha_4\beta_2 + \sqrt{3}\,\alpha_1\beta_4 + \alpha_2\beta_4 \\
     \sqrt{3}\,\alpha_3\beta_1 + \alpha_3\beta_2 - \sqrt{3}\,\alpha_1\beta_3 - \alpha_2\beta_3
\end{pmatrix} \\[9mm]
\oplus & \begin{pmatrix}
    \alpha_1\beta_1 - \alpha_2\beta_2 \\
    - \alpha_1\beta_2-\alpha_2\beta_1 \\
    \alpha_3\beta_3 - \alpha_4\beta_4 \\
    - \alpha_3\beta_4-\alpha_4\beta_3
\end{pmatrix} \, \oplus\, \begin{pmatrix}
    \alpha_2\beta_4 + \alpha_4\beta_2 \\
    -\alpha_2\beta_3 - \alpha_3\beta_2 \\
    -\alpha_1\beta_4 - \alpha_4\beta_1 \\
    \alpha_1\beta_3 + \alpha_3\beta_1
\end{pmatrix}
\end{array}$ \\
\addlinespace
\midrule
\addlinespace
$\begin{array}{rl}
\mathbf{4'} \,\otimes\, \mathbf{4''} \!\!&=\,\, \mathbf{1'} \,\oplus\, \mathbf{1'''} \,\oplus\, \mathbf{2}\\ &\quad\oplus\, \,\mathbf{4} \,\oplus\, \mathbf{4'} \,\oplus\, \mathbf{4'''}  
\end{array}$
&
$\begin{array}{rl}
&  -\alpha_2\beta_1
-\sqrt{3}\,\alpha_4\beta_1
+\alpha_1\beta_2
+\sqrt{3}\,\alpha_3\beta_2\\ & \quad
+\sqrt{3}\,\alpha_2\beta_3
-\alpha_4\beta_3
-\sqrt{3}\,\alpha_1\beta_4
+\alpha_3\beta_4 \\[2mm]
\oplus & 
\sqrt{3}\,\alpha_1\beta_1
-{\alpha_3\beta_1}
+\sqrt{3}\,\alpha_2\beta_2
-{\alpha_4\beta_2}\\ & \quad
+{\alpha_1\beta_3}
+\sqrt{3}\,\alpha_3\beta_3
+{\alpha_2\beta_4}
+\sqrt{3}\,\alpha_4\beta_4
\\[2mm]
\oplus & \begin{pmatrix*}[l]
    \alpha_1\beta_1
    +\sqrt{3}\,\alpha_3\beta_1
    +\alpha_2\beta_2
    +\sqrt{3}\,\alpha_4\beta_2\\ \qquad
    -\sqrt{3}\,\alpha_1\beta_3
    +\alpha_3\beta_3
    -\sqrt{3}\,\alpha_2\beta_4
    +\alpha_4\beta_4
    \\[1mm]
    -\sqrt{3}\,\alpha_2\beta_1
    +\alpha_4\beta_1
    +\sqrt{3}\,\alpha_1\beta_2
    -\alpha_3\beta_2\\ \qquad
    -\alpha_2\beta_3
    -\sqrt{3}\,\alpha_4\beta_3
    +\alpha_1\beta_4
    +\sqrt{3}\,\alpha_3\beta_4
\end{pmatrix*} \\[9mm]
\oplus & \begin{pmatrix*}[l]
    -{\alpha_2\beta_1}
    -\sqrt{3}\,\alpha_4\beta_1
    +{\alpha_1\beta_2}
    +\sqrt{3}\,\alpha_3\beta_2\\ \qquad
    -\sqrt{3}\,\alpha_2\beta_3
    +{\alpha_4\beta_3}
    +\sqrt{3}\,\alpha_1\beta_4
    -{\alpha_3\beta_4}
    \\[1mm]
    -\sqrt{3}\,\alpha_2\beta_1
    +{\alpha_4\beta_1}
    +\sqrt{3}\,\alpha_1\beta_2
    -{\alpha_3\beta_2}\\ \qquad
    +{\alpha_2\beta_3}
    +\sqrt{3}\,\alpha_4\beta_3
    -{\alpha_1\beta_4}
    -\sqrt{3}\,\alpha_3\beta_4
    \\[1mm]
    {\alpha_2\beta_1}
    +\sqrt{3}\,\alpha_4\beta_1
    +{\alpha_1\beta_2}
    +\sqrt{3}\,\alpha_3\beta_2\\ \qquad
    -\sqrt{3}\,\alpha_2\beta_3
    +{\alpha_4\beta_3}
    -\sqrt{3}\,\alpha_1\beta_4
    +{\alpha_3\beta_4}
    \\[1mm]
    {\alpha_1\beta_1}
    +\sqrt{3}\,\alpha_3\beta_1
    -{\alpha_2\beta_2}
    -\sqrt{3}\,\alpha_4\beta_2\\ \qquad
    -\sqrt{3}\,\alpha_1\beta_3
    +{\alpha_3\beta_3}
    +\sqrt{3}\,\alpha_2\beta_4
    -{\alpha_4\beta_4}
\end{pmatrix*} \\[2.1cm]
\oplus & \begin{pmatrix*}[l]
    -{\alpha_1\beta_1}
    +\sqrt{3}\,\alpha_3\beta_1
    +{\alpha_2\beta_2}
    -\sqrt{3}\,\alpha_4\beta_2\\ \qquad
    +\sqrt{3}\,\alpha_1\beta_3
    +{\alpha_3\beta_3}
    -\sqrt{3}\,\alpha_2\beta_4
    -{\alpha_4\beta_4}
    \\[1mm]
    {\alpha_2\beta_1}
    -\sqrt{3}\,\alpha_4\beta_1
    +{\alpha_1\beta_2}
    -\sqrt{3}\,\alpha_3\beta_2\\ \qquad
    -\sqrt{3}\,\alpha_2\beta_3
    -{\alpha_4\beta_3}
    -\sqrt{3}\,\alpha_1\beta_4
    -{\alpha_3\beta_4}
    \\[1mm]
    \sqrt{3}\,\alpha_1\beta_1
    +{\alpha_3\beta_1}
    -\sqrt{3}\,\alpha_2\beta_2
    -{\alpha_4\beta_2}\\ \qquad
    +{\alpha_1\beta_3}
    -\sqrt{3}\,\alpha_3\beta_3
    -{\alpha_2\beta_4}
    +\sqrt{3}\,\alpha_4\beta_4
    \\[1mm]
    -\sqrt{3}\,\alpha_2\beta_1
    -{\alpha_4\beta_1}
    -\sqrt{3}\,\alpha_1\beta_2
    -{\alpha_3\beta_2}\\ \qquad
    -{\alpha_2\beta_3}
    +\sqrt{3}\,\alpha_4\beta_3
    -{\alpha_1\beta_4}
    +\sqrt{3}\,\alpha_3\beta_4
\end{pmatrix*}\\[2.1cm]
\oplus & \begin{pmatrix*}[l]
    \sqrt{3}\,\alpha_1\beta_1
    -{\alpha_3\beta_1}
    +\sqrt{3}\,\alpha_2\beta_2
    -{\alpha_4\beta_2}\\ \qquad
    -{\alpha_1\beta_3}
    -\sqrt{3}\,\alpha_3\beta_3
    -{\alpha_2\beta_4}
    -\sqrt{3}\,\alpha_4\beta_4
    \\[1mm]
    -{\alpha_1\beta_1}
    -\sqrt{3}\,\alpha_3\beta_1
    -{\alpha_2\beta_2}
    -\sqrt{3}\,\alpha_4\beta_2\\ \qquad
    -\sqrt{3}\,\alpha_1\beta_3
    +{\alpha_3\beta_3}
    -\sqrt{3}\,\alpha_2\beta_4
    +{\alpha_4\beta_4}
    \\[1mm]
    -\sqrt{3}\,\alpha_1\beta_1
    +{\alpha_3\beta_1}
    +\sqrt{3}\,\alpha_2\beta_2
    -{\alpha_4\beta_2}\\ \qquad
    -{\alpha_1\beta_3}
    -\sqrt{3}\,\alpha_3\beta_3
    +{\alpha_2\beta_4}
    +\sqrt{3}\,\alpha_4\beta_4
    \\[1mm]
    \sqrt{3}\,\alpha_2\beta_1
    -{\alpha_4\beta_1}
    +\sqrt{3}\,\alpha_1\beta_2
    -{\alpha_3\beta_2}\\ \qquad
    +{\alpha_2\beta_3}
    +\sqrt{3}\,\alpha_4\beta_3
    +{\alpha_1\beta_4}
    +\sqrt{3}\,\alpha_3\beta_4
\end{pmatrix*}
\end{array}$ \\
\addlinespace
\midrule
\addlinespace
$\begin{array}{@{}l@{{}\,\otimes\,{}}l@{{}\,\,=\,\,{}}l@{{}\,\oplus\,{}}l@{{}\,\oplus\,{}}l@{{}\,\oplus\,{}}l@{}}
\mathbf{4} & \mathbf{4'}  & \mathbf{4} & \mathbf{4'} & \mathbf{4''} & \mathbf{4'''}
\end{array}$
&
$\begin{array}{rl}
&  \begin{pmatrix}
    \sqrt{3}\,\alpha_3\beta_1
    + \sqrt{3}\,\alpha_4\beta_2
    - {\alpha_3\beta_3}
    - {\alpha_4\beta_4} \\
    {\alpha_3\beta_1}
    + {\alpha_4\beta_2}
    + \sqrt{3}\,\alpha_3\beta_3
    + \sqrt{3}\,\alpha_4\beta_4 \\
    - \sqrt{3}\,\alpha_1\beta_1
    - {\alpha_2\beta_1}
    + {\alpha_1\beta_3}
    - \sqrt{3}\,\alpha_2\beta_3 \\
    - \sqrt{3}\,\alpha_1\beta_2
    - {\alpha_2\beta_2}
    + {\alpha_1\beta_4}
    - \sqrt{3}\,\alpha_2\beta_4
\end{pmatrix} \\[9mm]
\oplus & \begin{pmatrix*}[l]
    \sqrt{3}\,\alpha_1\beta_2
    + {\alpha_2\beta_2}
    + {2\alpha_4\beta_3}\\ \qquad
    + {\alpha_1\beta_4}
    - \sqrt{3}\,\alpha_2\beta_4
    + {2\alpha_3\beta_4} 
    \\[1mm]
    - \sqrt{3}\,\alpha_1\beta_1
    - {\alpha_2\beta_1}
    - {\alpha_1\beta_3}\\ \qquad
    + \sqrt{3}\,\alpha_2\beta_3
    + {2\alpha_3\beta_3}
    - {2\alpha_4\beta_4}
    \\[1mm]
    - {2\alpha_4\beta_1}
    + {\alpha_1\beta_2}
    - \sqrt{3}\,\alpha_2\beta_2\\ \qquad
    - {2\alpha_3\beta_2}
    - \sqrt{3}\,\alpha_1\beta_4
    - {\alpha_2\beta_4}
    \\[1mm]
    - {\alpha_1\beta_1}
    + \sqrt{3}\,\alpha_2\beta_1
    - {2\alpha_3\beta_1}\\ \qquad
    + {2\alpha_4\beta_2}
    + \sqrt{3}\,\alpha_1\beta_3
    + {\alpha_2\beta_3}
\end{pmatrix*} \\[2.1cm]
\oplus & \begin{pmatrix*}[l]
    \alpha_4\beta_1 - \alpha_1\beta_2 - \sqrt{3}\,\alpha_2\beta_2 + \alpha_3\beta_2\\ \qquad
    + \sqrt{3}\,\alpha_4\beta_3 - \sqrt{3}\,\alpha_1\beta_4 + \alpha_2\beta_4 + \sqrt{3}\,\alpha_3\beta_4 \\[1mm]
    \alpha_1\beta_1 + \sqrt{3}\,\alpha_2\beta_1 + \alpha_3\beta_1 - \alpha_4\beta_2\\ \qquad
    + \sqrt{3}\,\alpha_1\beta_3 - \alpha_2\beta_3 + \sqrt{3}\,\alpha_3\beta_3 - \sqrt{3}\,\alpha_4\beta_4 \\[1mm]
    -\sqrt{3}\,\alpha_4\beta_1 - \sqrt{3}\,\alpha_1\beta_2 + \alpha_2\beta_2 - \sqrt{3}\,\alpha_3\beta_2\\ \qquad
    + \alpha_4\beta_3 + \alpha_1\beta_4 + \sqrt{3}\,\alpha_2\beta_4 + \alpha_3\beta_4 \\[1mm]
    \sqrt{3}\,\alpha_1\beta_1 - \alpha_2\beta_1 - \sqrt{3}\,\alpha_3\beta_1 + \sqrt{3}\,\alpha_4\beta_2\\ \qquad
    - \alpha_1\beta_3 - \sqrt{3}\,\alpha_2\beta_3 + \alpha_3\beta_3 - \alpha_4\beta_4
\end{pmatrix*} \\[2.1cm]
\oplus & \begin{pmatrix}
    -\alpha_4\beta_1 + \alpha_3\beta_2 - \sqrt{3}\,\alpha_4\beta_3 + \sqrt{3}\,\alpha_3\beta_4 \\
    \sqrt{3}\,\alpha_4\beta_1 - \sqrt{3}\,\alpha_3\beta_2 - \alpha_4\beta_3 + \alpha_3\beta_4 \\
    -\alpha_1\beta_2 + \sqrt{3}\,\alpha_2\beta_2 - \sqrt{3}\,\alpha_1\beta_4 - \alpha_2\beta_4 \\
    \alpha_1\beta_1 - \sqrt{3}\,\alpha_2\beta_1 + \sqrt{3}\,\alpha_1\beta_3 + \alpha_2\beta_3
\end{pmatrix}
\end{array}$ \\
\addlinespace
\midrule
\addlinespace
$\begin{array}{@{}l@{{}\,\otimes\,{}}l@{{}\,\,=\,\,{}}l@{{}\,\oplus\,{}}l@{{}\,\oplus\,{}}l@{{}\,\oplus\,{}}l@{}}
\mathbf{4'''} & \mathbf{4'}  & \mathbf{4} & \mathbf{4'} & \mathbf{4''} & \mathbf{4'''}
\end{array}$
&
$\begin{array}{rl}
&  \begin{pmatrix}
    -\alpha_4\beta_1 + \alpha_3\beta_2 - \sqrt{3}\,\alpha_4\beta_3 + \sqrt{3}\,\alpha_3\beta_4 \\
    \sqrt{3}\,\alpha_4\beta_1 - \sqrt{3}\,\alpha_3\beta_2 - \alpha_4\beta_3 + \alpha_3\beta_4 \\
    -\alpha_1\beta_2 + \sqrt{3}\,\alpha_2\beta_2 - \sqrt{3}\,\alpha_1\beta_4 - \alpha_2\beta_4 \\
    \alpha_1\beta_1 - \sqrt{3}\,\alpha_2\beta_1 + \sqrt{3}\,\alpha_1\beta_3 + \alpha_2\beta_3
\end{pmatrix} \\[9mm]
\oplus & \begin{pmatrix*}[l]
    - {\alpha_1\beta_1}
    + \sqrt{3}\,\alpha_2\beta_1
    - {2\alpha_3\beta_1}\\ \qquad
    + {2\alpha_4\beta_2}
    + \sqrt{3}\,\alpha_1\beta_3
    + {\alpha_2\beta_3}
    \\[1mm]
    {2\alpha_4\beta_1}
    - {\alpha_1\beta_2}
    + \sqrt{3}\,\alpha_2\beta_2\\ \qquad
    + {2\alpha_3\beta_2}
    + \sqrt{3}\,\alpha_1\beta_4
    + {\alpha_2\beta_4}
    \\[1mm]
    \sqrt{3}\,\alpha_1\beta_1
    + {\alpha_2\beta_1}
    + {\alpha_1\beta_3}\\ \qquad
    - \sqrt{3}\,\alpha_2\beta_3
    - {2\alpha_3\beta_3}
    + {2\alpha_4\beta_4}
    \\[1mm]
    \sqrt{3}\,\alpha_1\beta_2
    + {\alpha_2\beta_2}
    + {2\alpha_4\beta_3}\\ \qquad
    + {\alpha_1\beta_4}
    - \sqrt{3}\,\alpha_2\beta_4
    + {2\alpha_3\beta_4} 
\end{pmatrix*} \\[2.1cm]
\oplus & \begin{pmatrix*}[l]   
    \sqrt{3}\,\alpha_1\beta_1 - \alpha_2\beta_1 - \sqrt{3}\,\alpha_3\beta_1 + \sqrt{3}\,\alpha_4\beta_2\\ \qquad
    - \alpha_1\beta_3 - \sqrt{3}\,\alpha_2\beta_3 + \alpha_3\beta_3 - \alpha_4\beta_4 \\[1mm]
    \sqrt{3}\,\alpha_4\beta_1 + \sqrt{3}\,\alpha_1\beta_2 - \alpha_2\beta_2 + \sqrt{3}\,\alpha_3\beta_2\\ \qquad
    - \alpha_4\beta_3 - \alpha_1\beta_4 - \sqrt{3}\,\alpha_2\beta_4 - \alpha_3\beta_4 \\[1mm]
    -\alpha_1\beta_1 - \sqrt{3}\,\alpha_2\beta_1 - \alpha_3\beta_1 + \alpha_4\beta_2\\ \qquad
    - \sqrt{3}\,\alpha_1\beta_3 + \alpha_2\beta_3 - \sqrt{3}\,\alpha_3\beta_3 + \sqrt{3}\,\alpha_4\beta_4 \\[1mm]
    \alpha_4\beta_1 - \alpha_1\beta_2 - \sqrt{3}\,\alpha_2\beta_2 + \alpha_3\beta_2\\ \qquad
    + \sqrt{3}\,\alpha_4\beta_3 - \sqrt{3}\,\alpha_1\beta_4 + \alpha_2\beta_4 + \sqrt{3}\,\alpha_3\beta_4 
\end{pmatrix*} \\[2.1cm]
\oplus & \begin{pmatrix}
    \sqrt{3}\,\alpha_3\beta_1
    + \sqrt{3}\,\alpha_4\beta_2
    - {\alpha_3\beta_3}
    - {\alpha_4\beta_4} \\
    {\alpha_3\beta_1}
    + {\alpha_4\beta_2}
    + \sqrt{3}\,\alpha_3\beta_3
    + \sqrt{3}\,\alpha_4\beta_4 \\
    - \sqrt{3}\,\alpha_1\beta_1
    - {\alpha_2\beta_1}
    + {\alpha_1\beta_3}
    - \sqrt{3}\,\alpha_2\beta_3 \\
    - \sqrt{3}\,\alpha_1\beta_2
    - {\alpha_2\beta_2}
    + {\alpha_1\beta_4}
    - \sqrt{3}\,\alpha_2\beta_4
\end{pmatrix}
\end{array}$ \\
\addlinespace
\midrule
\addlinespace
$\begin{array}{@{}l@{{}\,\otimes\,{}}l@{{}\,\,=\,\,{}}l@{{}\,\oplus\,{}}l@{{}\,\oplus\,{}}l@{{}\,\oplus\,{}}l@{}}
\mathbf{4} & \mathbf{4''}  & \mathbf{4} & \mathbf{4'''} & \mathbf{4'} & \mathbf{4''} 
\end{array}$
&
$\begin{array}{rl}
&  \begin{pmatrix}
\alpha_3\beta_1 + \alpha_4\beta_2 \\
\alpha_3\beta_3 + \alpha_4\beta_4 \\
\alpha_1\beta_1 + \alpha_2\beta_3 \\
\alpha_1\beta_2 + \alpha_2\beta_4
\end{pmatrix} 
\,\oplus\,
\begin{pmatrix}
\alpha_3\beta_4 -\alpha_4\beta_3 \\
- \alpha_3\beta_2 + \alpha_4\beta_1  \\
 \alpha_1\beta_4 -\alpha_2\beta_2 \\
- \alpha_1\beta_3 + \alpha_2\beta_1 
\end{pmatrix} 
\\[9mm]
\oplus & \begin{pmatrix*}[l]
    \alpha_4\beta_1 + \alpha_1\beta_2 + \sqrt{3}\,\alpha_2\beta_2 + \alpha_3\beta_2\\ \qquad
    - \sqrt{3}\,\alpha_4\beta_3 + \sqrt{3}\,\alpha_1\beta_4 - \alpha_2\beta_4 - \sqrt{3}\,\alpha_3\beta_4 
    \\[1mm]
    -\alpha_1\beta_1 - \sqrt{3}\,\alpha_2\beta_1 + \alpha_3\beta_1 - \alpha_4\beta_2\\ \qquad
    - \sqrt{3}\,\alpha_1\beta_3 + \alpha_2\beta_3 - \sqrt{3}\,\alpha_3\beta_3 + \sqrt{3}\,\alpha_4\beta_4
    \\[1mm]
    \sqrt{3}\,\alpha_4\beta_1 + \sqrt{3}\,\alpha_1\beta_2 - \alpha_2\beta_2 + \sqrt{3}\,\alpha_3\beta_2\\ \qquad
    + \alpha_4\beta_3 - \alpha_1\beta_4 - \sqrt{3}\,\alpha_2\beta_4 + \alpha_3\beta_4
    \\[1mm]
    -\sqrt{3}\,\alpha_1\beta_1 + \alpha_2\beta_1 + \sqrt{3}\,\alpha_3\beta_1 - \sqrt{3}\,\alpha_4\beta_2\\ \qquad
    + \alpha_1\beta_3 + \sqrt{3}\,\alpha_2\beta_3 + \alpha_3\beta_3 - \alpha_4\beta_4
\end{pmatrix*} \\[2.1cm]
\oplus & \begin{pmatrix}
\alpha_2\beta_2 + \alpha_4\beta_3 + \alpha_1\beta_4 + \alpha_3\beta_4 \\
-\alpha_2\beta_1 - \alpha_1\beta_3 + \alpha_3\beta_3 - \alpha_4\beta_4 \\
-\alpha_4\beta_1 + \alpha_1\beta_2 - \alpha_3\beta_2 - \alpha_2\beta_4 \\
-\alpha_1\beta_1 - \alpha_3\beta_1 + \alpha_4\beta_2 + \alpha_2\beta_3
\end{pmatrix}
\end{array}$ \\
\addlinespace
\midrule
\addlinespace
$\begin{array}{@{}l@{{}\,\otimes\,{}}l@{{}\,\,=\,\,{}}l@{{}\,\oplus\,{}}l@{{}\,\oplus\,{}}l@{{}\,\oplus\,{}}l@{}}
\mathbf{4'''} & \mathbf{4''}  & \mathbf{4} & \mathbf{4'''} & \mathbf{4'} & \mathbf{4''}
\end{array}$
&
$\begin{array}{rl}
& 
\begin{pmatrix}
\alpha_3\beta_4 -\alpha_4\beta_3 \\
- \alpha_3\beta_2 + \alpha_4\beta_1  \\
 \alpha_1\beta_4 -\alpha_2\beta_2 \\
- \alpha_1\beta_3 + \alpha_2\beta_1 
\end{pmatrix} 
\,\oplus\, \begin{pmatrix}
\alpha_3\beta_1 + \alpha_4\beta_2 \\
\alpha_3\beta_3 + \alpha_4\beta_4 \\
\alpha_1\beta_1 + \alpha_2\beta_3 \\
\alpha_1\beta_2 + \alpha_2\beta_4
\end{pmatrix} 
\\[9mm]
\oplus & \begin{pmatrix*}[l]    
    -\sqrt{3}\,\alpha_1\beta_1 + \alpha_2\beta_1 + \sqrt{3}\,\alpha_3\beta_1 - \sqrt{3}\,\alpha_4\beta_2\\ \qquad
    + \alpha_1\beta_3 + \sqrt{3}\,\alpha_2\beta_3 + \alpha_3\beta_3 - \alpha_4\beta_4
    \\[1mm]
    -\sqrt{3}\,\alpha_4\beta_1 - \sqrt{3}\,\alpha_1\beta_2 + \alpha_2\beta_2 - \sqrt{3}\,\alpha_3\beta_2\\ \qquad
    - \alpha_4\beta_3 + \alpha_1\beta_4 + \sqrt{3}\,\alpha_2\beta_4 - \alpha_3\beta_4
    \\[1mm]
    \alpha_1\beta_1 + \sqrt{3}\,\alpha_2\beta_1 - \alpha_3\beta_1 + \alpha_4\beta_2\\ \qquad
    + \sqrt{3}\,\alpha_1\beta_3 - \alpha_2\beta_3 + \sqrt{3}\,\alpha_3\beta_3 - \sqrt{3}\,\alpha_4\beta_4
    \\[1mm]
    \alpha_4\beta_1 + \alpha_1\beta_2 + \sqrt{3}\,\alpha_2\beta_2 + \alpha_3\beta_2\\ \qquad
    - \sqrt{3}\,\alpha_4\beta_3 + \sqrt{3}\,\alpha_1\beta_4 - \alpha_2\beta_4 - \sqrt{3}\,\alpha_3\beta_4 
\end{pmatrix*} \\[2.1cm]
\oplus & \begin{pmatrix}
-\alpha_1\beta_1 - \alpha_3\beta_1 + \alpha_4\beta_2 + \alpha_2\beta_3 \\
\alpha_4\beta_1 - \alpha_1\beta_2 + \alpha_3\beta_2 + \alpha_2\beta_4 \\
\alpha_2\beta_1 + \alpha_1\beta_3 - \alpha_3\beta_3 + \alpha_4\beta_4 \\
\alpha_2\beta_2 + \alpha_4\beta_3 + \alpha_1\beta_4 + \alpha_3\beta_4 
\end{pmatrix}
\end{array}$ \\
\addlinespace
\bottomrule
\end{longtable}
}
}

\vfill
\clearpage
\clearpage

\section{Modular forms under \texorpdfstring{$(S_3\times S_3)\rtimes \mathbb{Z}_2$}{S3xS3:Z2}}
\label{app:modforms}

Within the invariant region $\mathcal{T}_1$, one has $\tau_3 = 0$ and the forms of~\cref{eq:w2formsgen} reduce to
\begin{equation} \label{eq:Y4}
\begin{aligned}
p_0(\tau) &= e_1(\tau_1)\,e_1(\tau_2)\,,\qquad
p_1(\tau) = e_1(\tau_1)\,\tilde e_2(\tau_2)\,,\\
p_2(\tau) &= \tilde e_2(\tau_1)\,e_1(\tau_2)\,,\qquad
p_3(\tau)  = \tilde e_2(\tau_1)\,\tilde e_2(\tau_2) = p_4(\tau)\,,
\end{aligned}
\end{equation}
with $\tilde e_2 \equiv e_2/ \sqrt{3}$ and the $e_i(\tau_j)$ being the weight-2 modular forms of $\Gamma_1(2)$ making up a doublet of the genus-one modular $S_3$ finite group~\cite{Kobayashi:2018vbk,Meloni:2023aru}:\footnote{Here $\eta(\tau)$ is the Dedekind eta function and $\eta'(\tau)$ its first derivative.} 
\begin{equation}
\label{multiplets1}
\begin{cases}
e_1(\tau_i)=\displaystyle\frac{c}{2}i\left[\frac{\eta'(\tau_i/2)}{\eta(\tau_i/2)}+\frac{\eta'\left(\frac{\tau_i+1}{2}\right)}{\eta\left(\frac{\tau_i+1}{2}\right)}-\frac{8\eta'(2\tau_i)}{\eta(2\tau_i)}\right]\\ \\
e_2(\tau_i)=\displaystyle\frac{c}{2}i\sqrt{3}\left[\frac{\eta'(\tau_i/2)}{\eta(\tau_i/2)}-\frac{\eta'\left(\frac{\tau_i+1}{2}\right)}{\eta\left(\frac{\tau_i+1}{2}\right)}\right]
\end{cases}\,,
\end{equation}
where $c$ is an arbitrary normalization constant. The corresponding $q$-expansion is:
\begin{equation}\frac{1}{c_k}
 \begin{pmatrix}
     e_1 (\tau_i) \\[1mm]
     e_2 (\tau_i)
 \end{pmatrix}  =
 \begin{pmatrix}
     1+ 24\,q_i+24\,q_i^2+96\,q_i^3+24\,q_i^4+144\,q_i^5 +96 \,q_i^6  + \ldots \\[1mm]
     8 \sqrt{3\,q_i} \left(1 + 4 \,q_i + 6 \,q_i^2 + 8 \,q_i^3 + 13 \,q_i^4 + 12 \,q_i^5 + 
  14 q_i^6 + \ldots\right)
 \end{pmatrix}
 \,,
\end{equation}
where $q_i = e^{2\pi i \tau_i}$ and $c_k \in \mathbb{R}$ is the ratio between our normalization ($c \simeq 0.9$, reported in~\cref{sec:model}) and the one used in~\cite{Ding:2020zxw}. One finds that the tensor product of two weight-2 $\Gamma_{1,2}\cong S_3$ doublets produces a $\mathbf{4''}$  as the lowest-weight ($k=2$) modular multiplet of $N_2(H)$. It explicitly reads
\begin{equation}
    Y^{(2)}_\mathbf{4''}(\tau_1,\tau_2)=
\begin{pmatrix}
e_1(\tau_1)\\
e_2(\tau_1)
\end{pmatrix}\otimes \begin{pmatrix}
e_1(\tau_2)\\
e_2(\tau_2)
\end{pmatrix}=
\begin{pmatrix}
e_1(\tau_1)\,e_1(\tau_2)\\
e_1(\tau_1)\,e_2(\tau_2)\\
e_2(\tau_1)\,e_1(\tau_2)\\
e_2(\tau_1)\,e_2(\tau_2)
\end{pmatrix}
=
\begin{pmatrix}
p_0(\tau)\\
\sqrt{3}\,p_1(\tau)\\
\sqrt{3}\,p_2(\tau)\\
3\,p_3(\tau)
\end{pmatrix}\,,
\end{equation}
in the basis of~\cref{app:groupth}.
%
Via repeated tensor products of $Y^{(2)}_\mathbf{4''}$ with itself, one may obtain modular form multiplets of higher weight. At weight $k=4$ one finds:
\begin{equation}
\begin{aligned}
    Y^{(4)}_\mathbf{1}(\tau_1,\tau_2)
    &=
    \left(e_1(\tau_1)^2+e_2(\tau_1)^2\right)
    \left(e_1(\tau_2)^2+e_2(\tau_2)^2\right)\,,
    \\[2mm]
    Y^{(4)}_\mathbf{4''}(\tau_1,\tau_2)
    &=
    \begin{pmatrix}
    \left(e_1(\tau_1)^2-e_2(\tau_1)^2\right)
    \left(e_1(\tau_2)^2-e_2(\tau_2)^2\right)\\[2mm]
    -2\left(e_1(\tau_1)^2-e_2(\tau_1)^2\right)e_1(\tau_2)\,e_2(\tau_2)\\[2mm]
    -2\,e_1(\tau_1)\, e_2(\tau_1)\left(e_1(\tau_2)^2-e_2(\tau_2)^2\right)\\[2mm]
    4\,e_1(\tau_1)\, e_2(\tau_1)\, e_1(\tau_2)\, e_2(\tau_2)
    \end{pmatrix}\,,
    \\[2mm]
    Y^{(4)}_\mathbf{4'''}(\tau_1,\tau_2)
    &=
    \begin{pmatrix}
    \left(e_1(\tau_1)^2-e_2(\tau_1)^2\right)
    \left(e_1(\tau_2)^2+e_2(\tau_2)^2\right)\\[2mm]
    -2\,e_1(\tau_1)\, e_2(\tau_1)
    \left(e_1(\tau_2)^2+e_2(\tau_2)^2\right)\\[2mm]
    \left(e_1(\tau_1)^2+e_2(\tau_1)^2\right)
    \left(e_1(\tau_2)^2-e_2(\tau_2)^2\right)\\[2mm]
    -2
    \left(e_1(\tau_1)^2+e_2(\tau_1)^2\right)
    e_1(\tau_2)\,e_2(\tau_2)
    \end{pmatrix}\,,
\end{aligned}
\end{equation}
%
For weight $k=6$, one finds:
\begin{equation}
\begin{aligned}
    Y^{(6)}_\mathbf{1}(\tau_1,\tau_2)&=
    e_1(\tau_1)
    \left(e_1(\tau_1)^2-3\,e_2(\tau_1)^2\right)
    e_1(\tau_2)
    \left(e_1(\tau_2)^2-3\,e_2(\tau_2)^2\right)\,,
    \\[2mm]
    Y^{(6)}_\mathbf{1''}(\tau_1,\tau_2)&=
    e_2(\tau_1)
    \left(3\,e_1(\tau_1)^2-e_2(\tau_1)^2\right)
    e_2(\tau_2)
    \left(3\,e_1(\tau_2)^2-e_2(\tau_2)^2\right)\,,
    \\[2mm]
    Y^{(6)}_\mathbf{2}(\tau_1,\tau_2)&=
    \begin{pmatrix}
    e_2(\tau_1)
    \left(3\,e_1(\tau_1)^2-e_2(\tau_1)^2\right)
    e_1(\tau_2)
    \left(e_1(\tau_2)^2-3\,e_2(\tau_2)^2\right)
    \\[2mm]
    e_1(\tau_1)
    \left(e_1(\tau_1)^2-3\,e_2(\tau_1)^2\right)
    e_2(\tau_2)
    \left(3\,e_1(\tau_2)^2-e_2(\tau_2)^2\right)
    \end{pmatrix}\,,
    \\[2mm]
    Y^{(6)}_\mathbf{4}(\tau_1,\tau_2)&=
    \begin{pmatrix}
    \phantom{+}e_2(\tau_1)
    \left(e_1(\tau_1)^2+e_2(\tau_1)^2\right)
    e_2(\tau_2)
    \left(3\,e_1(\tau_2)^2-e_2(\tau_2)^2\right)
    \\[2mm]
    -e_1(\tau_1)
    \left(e_1(\tau_1)^2+e_2(\tau_1)^2\right)
    e_2(\tau_2)
    \left(3\,e_1(\tau_2)^2-e_2(\tau_2)^2\right)
    \\[2mm]
    \phantom{+}e_2(\tau_1)
    \left(3\,e_1(\tau_1)^2-e_2(\tau_1)^2\right)
    e_2(\tau_2)
    \left(e_1(\tau_2)^2+e_2(\tau_2)^2\right)
    \\[2mm]
    - e_2(\tau_1)
    \left(3\,e_1(\tau_1)^2-e_2(\tau_1)^2\right)
    e_1(\tau_2)
    \left(e_1(\tau_2)^2+e_2(\tau_2)^2\right)
    \end{pmatrix}\,,
    \\[2mm]
    Y^{(6)}_\mathbf{4''}(\tau_1,\tau_2)&=
    \begin{pmatrix}
    e_1(\tau_1)
    \left(e_1(\tau_1)^2+e_2(\tau_1)^2\right)
    e_1(\tau_2)
    \left(e_1(\tau_2)^2+e_2(\tau_2)^2\right)
    \\[2mm]
    e_1(\tau_1)
    \left(e_1(\tau_1)^2+e_2(\tau_1)^2\right)
    e_2(\tau_2)
    \left(e_1(\tau_2)^2+e_2(\tau_2)^2\right)
    \\[2mm]
    e_2(\tau_1)
    \left(e_1(\tau_1)^2+e_2(\tau_1)^2\right)
    e_1(\tau_2)
    \left(e_1(\tau_2)^2+e_2(\tau_2)^2\right)
    \\[2mm]
    e_2(\tau_1)
    \left(e_1(\tau_1)^2+e_2(\tau_1)^2\right)
    e_2(\tau_2)
    \left(e_1(\tau_2)^2+e_2(\tau_2)^2\right)
    \end{pmatrix}\,,
    \\[2mm]
    Y^{(6)}_\mathbf{4'''}(\tau_1,\tau_2)&=
    \begin{pmatrix}
    e_1(\tau_1)
    \left(e_1(\tau_1)^2+e_2(\tau_1)^2\right)
    e_1(\tau_2)
    \left(e_1(\tau_2)^2-3\,e_2(\tau_2)^2\right)
    \\[2mm]
    e_2(\tau_1)
    \left(e_1(\tau_1)^2+e_2(\tau_1)^2\right)
    e_1(\tau_2)
    \left(e_1(\tau_2)^2-3\,e_2(\tau_2)^2\right)
    \\[2mm]
    e_1(\tau_1)
    \left(e_1(\tau_1)^2-3\,e_2(\tau_1)^2\right)
    e_1(\tau_2)
    \left(e_1(\tau_2)^2+e_2(\tau_2)^2\right)
    \\[2mm]
    e_1(\tau_1)
    \left(e_1(\tau_1)^2-3\,e_2(\tau_1)^2\right)
    e_2(\tau_2)
    \left(e_1(\tau_2)^2+e_2(\tau_2)^2\right)
    \end{pmatrix}\,.
\end{aligned}
\end{equation}
%
For weight $k=8$, one finds:
\begin{equation}
\begin{aligned}
    Y^{(8)}_\mathbf{1}(\tau_1,\tau_2) &= \left(e_1(\tau_1)^2 + e_2(\tau_1)^2\right)^2 \left(e_1(\tau_2)^2 + e_2(\tau_2)^2\right)^2\,,
    \\[2mm]
    Y^{(8)}_\mathbf{4'}(\tau_1,\tau_2)
    &=
    \begin{pmatrix*}[l]
    3
    \left(e_1(\tau_1)^2+e_2(\tau_1)^2\right)
    \left(\sqrt{3}\,e_1(\tau_1)^2
    -2 e_1(\tau_1)\, e_2(\tau_1)
    -\sqrt{3}\,e_2(\tau_1)^2\right)
    e_1(\tau_2)^2\,e_2(\tau_2)^2
    \\ \qquad
    + e_2(\tau_1)
    \left(e_1(\tau_1)-\sqrt{3}\,e_2(\tau_1)\right)
    \left(3\,e_1(\tau_1)^2-e_2(\tau_1)^2\right)
    e_1(\tau_2)^4
    \\ \qquad
    - e_1(\tau_1)
    \left(\sqrt{3}\,e_1(\tau_1)+e_2(\tau_1)\right)
    \left(e_1(\tau_1)^2-3\,e_2(\tau_1)^2\right)
    e_2(\tau_2)^4
    \\[2mm]
    \Big[
    \sqrt{3}\, e_2(\tau_1)^4
    \left(e_1(\tau_2)^2-3\,e_2(\tau_2)^2\right)
    - \sqrt{3}\, e_1(\tau_1)^4
    \left(3\,e_1(\tau_2)^2-e_2(\tau_2)^2\right)
    \\ \qquad
    +8\, e_1(\tau_1) \, e_2(\tau_1)^3 e_1(\tau_2)^2
    -8\, e_1(\tau_1)^3  e_2(\tau_1) \,e_2(\tau_2)^2
    \\ \qquad
    + 6\sqrt{3}\,
    e_1(\tau_1)^2 e_2(\tau_1)^2
    \left(e_1(\tau_2)^2+e_2(\tau_2)^2\right)
    \Big]
    e_1(\tau_2)\,e_2(\tau_2)
    \\[2mm]
    -3
    \left(e_1(\tau_1)^2+2\sqrt{3}\,e_1(\tau_1)\, e_2(\tau_1)-e_2(\tau_1)^2\right)
    \left(e_1(\tau_1)^2+e_2(\tau_1)^2\right)
    e_1(\tau_2)^2 e_2(\tau_2)^2
    \\ \qquad
    + e_2(\tau_1)
    \left(\sqrt{3}\,e_1(\tau_1)+e_2(\tau_1)\right)
    \left(3\,e_1(\tau_1)^2-e_2(\tau_1)^2\right)
    e_1(\tau_2)^4 
    \\ \qquad
    + e_1(\tau_1)
    \left(e_1(\tau_1)-\sqrt{3}\,e_2(\tau_1)\right)
    \left(e_1(\tau_1)^2-3\,e_2(\tau_1)^2\right)
    e_2(\tau_2)^4
    \\[2mm]
    \left(3\,e_1(\tau_1)^4
    -6\, e_1(\tau_1)^2 e_2(\tau_1)^2
    +8\sqrt{3}\, e_1(\tau_1) e_2(\tau_1)^3
    -e_2(\tau_1)^4\right)
    e_1(\tau_2)^3 e_2(\tau_2)
    \\ \qquad
    - 
    \left(e_1(\tau_1)^4
    +8\sqrt{3}\, e_1(\tau_1)^3 e_2(\tau_1)
    +6\, e_1(\tau_1)^2 e_2(\tau_1)^2
    -3\,e_2(\tau_1)^4\right)
    e_1(\tau_2)\,e_2(\tau_2)^3
    \end{pmatrix*}\,,
    \\[2mm]
    Y^{(8)}_{\mathbf{4''},1}(\tau_1,\tau_2)
    &=
    \begin{pmatrix*}[l]
    \phantom{+++}\!\!\! e_1(\tau_1)^2 \left(e_1(\tau_1)^2-3\,e_2(\tau_1)^2\right) e_1(\tau_2)^2 \left(e_1(\tau_2)^2-3\,e_2(\tau_2)^2\right)
    \\[2mm]
    \phantom{+++}\!\!\! e_1(\tau_1)^2 \left(e_1(\tau_1)^2-3\,e_2(\tau_1)^2\right) e_1(\tau_2) \, e_2(\tau_2) \left(e_1(\tau_2)^2-3\,e_2(\tau_2)^2\right)
    \\[2mm]
    e_1(\tau_1) \,e_2(\tau_1) \left(e_1(\tau_1)^2-3\,e_2(\tau_1)^2\right) e_1(\tau_2)^2 \left(e_1(\tau_2)^2-3\,e_2(\tau_2)^2\right)
    \\[2mm]
    e_1(\tau_1) \,e_2(\tau_1) \left(e_1(\tau_1)^2-3\,e_2(\tau_1)^2\right) e_1(\tau_2) \, e_2(\tau_2) \left(e_1(\tau_2)^2-3\,e_2(\tau_2)^2\right)
    \end{pmatrix*}\,,
    \\[2mm]
    Y^{(8)}_{\mathbf{4''},2}(\tau_1,\tau_2)
    &=
    \begin{pmatrix*}[l]
    \phantom{++} - e_2(\tau_1)^2 \left(3\,e_1(\tau_1)^2-e_2(\tau_1)^2 \right) e_2(\tau_2)^2 \left(3\,e_1(\tau_2)^2-e_2(\tau_2)^2\right)
    \\[2mm]
    \phantom{+++} e_2(\tau_1)^2 \left(3\,e_1(\tau_1)^2-e_2(\tau_1)^2\right) e_1(\tau_2) \, e_2(\tau_2) \left(3\,e_1(\tau_2)^2-e_2(\tau_2)^2\right)
    \\[2mm]
    \phantom{+} e_1(\tau_1) e_2(\tau_1) \left(3\,e_1(\tau_1)^2-e_2(\tau_1)^2\right) e_2(\tau_2)^2 \left(3\,e_1(\tau_2)^2-e_2(\tau_2)^2\right)
    \\[2mm]
    - e_1(\tau_1) e_2(\tau_1) \left(3\,e_1(\tau_1)^2-e_2(\tau_1)^2\right) e_1(\tau_2) e_2(\tau_2) \left(3\,e_1(\tau_2)^2-e_2(\tau_2)^2\right)
    \end{pmatrix*}\,,
    \\[2mm]
    Y^{(8)}_{\mathbf{4''},3}(\tau_1,\tau_2)
    &=
    \begin{pmatrix*}[l]
    e_2(\tau_1)^2 \left(3\,e_1(\tau_1)^2 - e_2(\tau_1)^2\right) e_1(\tau_2)^2 \left(e_1(\tau_2)^2 - 3\,e_2(\tau_2)^2\right)
    \\ \qquad
    + e_1(\tau_1)^2 \left(e_1(\tau_1)^2 - 3\,e_2(\tau_1)^2\right) e_2(\tau_2)^2 \left(3\,e_1(\tau_2)^2 - e_2(\tau_2)^2\right)
    \\[2mm]
    \left(e_1(\tau_1)^4 - 12\, e_1(\tau_1)^2 e_2(\tau_1)^2 + 3\,e_2(\tau_1)^4\right) e_1(\tau_2) \, e_2(\tau_2)^3
    \\ \qquad
    - \left(3\,e_1(\tau_1)^4 - 12\, e_1(\tau_1)^2 e_2(\tau_1)^2 + e_2(\tau_1)^4\right) e_1(\tau_2)^3 e_2(\tau_2)
    \\[2mm]
    e_1(\tau_1)\, e_2(\tau_1)^3 \left(e_1(\tau_2)^4 - 12\, e_1(\tau_2)^2 e_2(\tau_2)^2 + 3\,e_2(\tau_2)^4\right)
    \\ \qquad
    - e_1(\tau_1)^3 e_2(\tau_1) \left(3\,e_1(\tau_2)^4 - 12\, e_1(\tau_2)^2 e_2(\tau_2)^2 + e_2(\tau_2)^4\right)
    \\[2mm]
    2\, e_1(\tau_1)\, e_2(\tau_1) \Big[
    e_2(\tau_1)^2 \left(5\, e_1(\tau_2)^2 - 3\,e_2(\tau_2)^2\right)
    \\ \qquad\qquad\qquad\!\!
    - e_1(\tau_1)^2 \left(3\,e_1(\tau_2)^2 - 5\, e_2(\tau_2)^2\right)
    \Big] e_1(\tau_2)\, e_2(\tau_2)
    \end{pmatrix*}\,,
    \\[2mm]
    Y^{(8)}_{\mathbf{4'''},1}(\tau_1,\tau_2)
    &=
    \begin{pmatrix*}[l]
    \phantom{++}\!\!\! -e_2(\tau_1)^2 \left(3\,e_1(\tau_1)^2-e_2(\tau_1)^2\right) \left(e_1(\tau_2)^2+e_2(\tau_2)^2\right)^2
    \\[2mm]
    e_1(\tau_1) \,e_2(\tau_1) \left(3\,e_1(\tau_1)^2-e_2(\tau_1)^2\right) \left(e_1(\tau_2)^2+e_2(\tau_2)^2\right)^2
    \\[2mm]
    -\left(e_1(\tau_1)^2+e_2(\tau_1)^2\right)^2 e_2(\tau_2)^2 \left(3\,e_1(\tau_2)^2-e_2(\tau_2)^2\right)
    \\[2mm]
    \phantom{+}\left(e_1(\tau_1)^2+e_2(\tau_1)^2\right)^2 e_1(\tau_2)\, e_2(\tau_2) \left(3\,e_1(\tau_2)^2-e_2(\tau_2)^2\right)
    \end{pmatrix*}\,,
    \\[2mm]
    Y^{(8)}_{\mathbf{4'''},2}(\tau_1,\tau_2)
    &=
    \begin{pmatrix}
    \left(e_1(\tau_1)^4-e_2(\tau_1)^4\right) \left(e_1(\tau_2)^2+e_2(\tau_2)^2\right)^2
    \\[6pt]
    -2\, e_1(\tau_1) \,e_2(\tau_1) \left(e_1(\tau_1)^2+e_2(\tau_1)^2\right) \left(e_1(\tau_2)^2+e_2(\tau_2)^2\right)^2
    \\[6pt]
    \left(e_1(\tau_1)^2+e_2(\tau_1)^2\right)^2 \left(e_1(\tau_2)^4-e_2(\tau_2)^4\right)
    \\[6pt]
    -2  \left(e_1(\tau_1)^2+e_2(\tau_1)^2\right)^2 e_1(\tau_2)\,e_2(\tau_2) \left(e_1(\tau_2)^2+e_2(\tau_2)^2\right)
    \end{pmatrix}\,.
\end{aligned}
\end{equation}
Finally, and for brevity, we only present the modular forms of weights $k=10$ and $k=12$ explicitly employed in our construction, since the complete basis would be unnecessarily long for our purposes. These read, for $k=10$:
\begin{equation}
\begin{aligned}
    Y^{(10)}_\mathbf{1}(\tau_1,\tau_2) &= 
    e_1(\tau_1) 
    \left(e_1(\tau_1)^2 - 3\, e_2(\tau_1)^2\right)
    \left(e_1(\tau_1)^2 + e_2(\tau_1)^2\right) \\
    &\,\times
    e_1(\tau_2)
    \left(e_1(\tau_2)^2 - 3\, e_2(\tau_2)^2\right)
    \left(e_1(\tau_2)^2 + e_2(\tau_2)^2\right)\,,
    \\[2mm]
    Y^{(10)}_\mathbf{1''}(\tau_1,\tau_2) &= 
    e_2(\tau_1) \left(3\,e_1(\tau_1)^4+2\,e_1(\tau_1)^2e_2(\tau_1)^2-e_2(\tau_1)^4 \right)\\
    &\,\times
    e_2(\tau_2) \left(3\,e_1(\tau_2)^4+2\,e_1(\tau_2)^2e_2(\tau_2)^2-e_2(\tau_2)^4 \right)\,,
    \\[2mm]
    Y^{(10)}_\mathbf{2}(\tau_1,\tau_2) &=
    \begin{pmatrix*}[l]
    e_2(\tau_1)
    \left(3\,e_1(\tau_1)^4 + 2 e_1(\tau_1)^2 e_2(\tau_1)^2 - e_2(\tau_1)^4\right)
    \\ \qquad \times \,
    e_1(\tau_2)
    \left(e_1(\tau_2)^2 - 3\, e_2(\tau_2)^2\right)
    \left(e_1(\tau_2)^2 + e_2(\tau_2)^2\right) \\[2mm]
    e_2(\tau_2)
    \left(3\,e_1(\tau_2)^4 + 2 e_1(\tau_2)^2 e_2(\tau_2)^2 - e_2(\tau_2)^4\right)
    \\ \qquad \times \,
    e_1(\tau_1)
    \left(e_1(\tau_1)^2 - 3\, e_2(\tau_1)^2\right)
    \left(e_1(\tau_1)^2 + e_2(\tau_1)^2\right)
    \end{pmatrix*}\,,
\end{aligned}
\end{equation}
and, for $k=12$:
\begin{equation}
\begin{aligned}
    Y^{(12)}_{\mathbf{2},1}(\tau_1,\tau_2) &=
    \begin{pmatrix*}[l]
    e_1(\tau_1)\, e_2(\tau_1)
    \left(3\,e_1(\tau_1)^4 - 10\, e_1(\tau_1)^2 e_2(\tau_1)^2 + 3\,e_2(\tau_1)^4\right)
    \\ \qquad \times \,
    \left(e_1(\tau_2)^2 + e_2(\tau_2)^2\right)^3 
    \\[2mm]
    e_1(\tau_2) \, e_2(\tau_2)
    \left(3\,e_1(\tau_2)^4 - 10\, e_1(\tau_2)^2 e_2(\tau_2)^2 + 3\,e_2(\tau_2)^4\right)
    \\ \qquad \times \,
    \left(e_1(\tau_1)^2 + e_2(\tau_1)^2\right)^3
    \end{pmatrix*}\,,
    \\[2mm]
    Y^{(12)}_{\mathbf{2},2}(\tau_1,\tau_2) &=
    \begin{pmatrix*}[l]
    e_1(\tau_1)\, e_2(\tau_1)
    \left(3\,e_1(\tau_1)^4 - 10\, e_1(\tau_1)^2 e_2(\tau_1)^2 + 3\,e_2(\tau_1)^4\right)
    \\ \qquad \times \,
    e_1(\tau_2)^2
    \left(e_1(\tau_2)^2 - 3\, e_2(\tau_2)^2\right)^2 
    \\[2mm]
    e_1(\tau_2) \, e_2(\tau_2)
    \left(3\,e_1(\tau_2)^4 - 10\, e_1(\tau_2)^2 e_2(\tau_2)^2 + 3\,e_2(\tau_2)^4\right)
    \\ \qquad \times \,
    e_1(\tau_1)^2
    \left(e_1(\tau_1)^2 - 3\,e_2(\tau_1)^2\right)^2
    \end{pmatrix*}\,.
\end{aligned}
\end{equation}

\footnotesize
\bibliographystyle{JHEPwithnote}
\bibliography{bibliography}

\providecommand{\noopsort}[1]{}\providecommand{\singleletter}[1]{#1}%

\providecommand{\href}[2]{#2}\begingroup\raggedright\begin{thebibliography}{100}

\bibitem{ParticleDataGroup:2024cfk}
{\scshape Particle Data Group} collaboration, S.~Navas et~al., \emph{{Review of
  particle physics}},
  \href{https://doi.org/10.1103/PhysRevD.110.030001}{\emph{Phys. Rev. D}
  {\bfseries 110} (2024) 030001}.

\bibitem{Xing:2020ijf}
Z.-z. Xing, \emph{{Flavor structures of charged fermions and massive
  neutrinos}}, \href{https://doi.org/10.1016/j.physrep.2020.02.001}{\emph{Phys.
  Rept.} {\bfseries 854} (2020) 1}
  [\href{https://arxiv.org/abs/1909.09610}{{\ttfamily 1909.09610}}].

\bibitem{Feruglio:2025ztj}
F.~Feruglio and S.~Ramos-Sanchez, \emph{{Quark and lepton masses}},
  \href{https://arxiv.org/abs/2506.20755}{{\ttfamily 2506.20755}}.

\bibitem{Altarelli:2010gt}
G.~Altarelli and F.~Feruglio, \emph{{Discrete Flavor Symmetries and Models of
  Neutrino Mixing}},
  \href{https://doi.org/10.1103/RevModPhys.82.2701}{\emph{Rev. Mod. Phys.}
  {\bfseries 82} (2010) 2701}
  [\href{https://arxiv.org/abs/1002.0211}{{\ttfamily 1002.0211}}].

\bibitem{Ishimori:2010au}
H.~Ishimori, T.~Kobayashi, H.~Ohki, Y.~Shimizu, H.~Okada and M.~Tanimoto,
  \emph{{Non-Abelian Discrete Symmetries in Particle Physics}},
  \href{https://doi.org/10.1143/PTPS.183.1}{\emph{Prog. Theor. Phys. Suppl.}
  {\bfseries 183} (2010) 1} [\href{https://arxiv.org/abs/1003.3552}{{\ttfamily
  1003.3552}}].

\bibitem{King:2014nza}
S.~F. King, A.~Merle, S.~Morisi, Y.~Shimizu and M.~Tanimoto, \emph{{Neutrino
  Mass and Mixing: from Theory to Experiment}},
  \href{https://doi.org/10.1088/1367-2630/16/4/045018}{\emph{New J. Phys.}
  {\bfseries 16} (2014) 045018}
  [\href{https://arxiv.org/abs/1402.4271}{{\ttfamily 1402.4271}}].

\bibitem{Petcov:2017ggy}
S.~T. Petcov, \emph{{Discrete Flavour Symmetries, Neutrino Mixing and Leptonic
  CP Violation}},
  \href{https://doi.org/10.1140/epjc/s10052-018-6158-5}{\emph{Eur. Phys. J. C}
  {\bfseries 78} (2018) 709}
  [\href{https://arxiv.org/abs/1711.10806}{{\ttfamily 1711.10806}}].

\bibitem{Feruglio:2019ybq}
F.~Feruglio and A.~Romanino, \emph{{Lepton flavor symmetries}},
  \href{https://doi.org/10.1103/RevModPhys.93.015007}{\emph{Rev. Mod. Phys.}
  {\bfseries 93} (2021) 015007}
  [\href{https://arxiv.org/abs/1912.06028}{{\ttfamily 1912.06028}}].

\bibitem{Ding:2024ozt}
G.-J. Ding and J.~W.~F. Valle, \emph{{The symmetry approach to quark and lepton
  masses and mixing}},
  \href{https://doi.org/10.1016/j.physrep.2024.12.005}{\emph{Phys. Rept.}
  {\bfseries 1109} (2025) 1}
  [\href{https://arxiv.org/abs/2402.16963}{{\ttfamily 2402.16963}}].

\bibitem{Feruglio:2022cgv}
F.~Feruglio, \emph{{Automorphic Forms and~Fermion Masses}},
  \href{https://doi.org/10.1007/978-981-19-4751-3_41}{\emph{Springer Proc.
  Math. Stat.} {\bfseries 396} (2022) 449}.

\bibitem{Rovelli:2014ssa}
C.~Rovelli and F.~Vidotto, \emph{{Covariant Loop Quantum Gravity}: {An
  Elementary Introduction to Quantum Gravity and Spinfoam Theory}}, Cambridge
  Monographs on Mathematical Physics. Cambridge University Press, 11, 2014.

\bibitem{Feruglio:2017spp}
F.~Feruglio, \emph{{Are neutrino masses modular forms?}},  in \emph{From My
  Vast Repertoire...: Guido Altarelli's Legacy} (A.~Levy, S.~Forte and
  G.~Ridolfi, eds.), pp.~227--266.
\newblock World Scientific Publishing, 2019.
\newblock [\href{https://arxiv.org/abs/1706.08749}{{\ttfamily 1706.08749}}].

\bibitem{Kobayashi:2023zzc}
T.~Kobayashi and M.~Tanimoto, \emph{{Modular flavor symmetric models}},
  \href{https://doi.org/10.1142/S0217751X24410124}{\emph{Int. J. Mod. Phys. A}
  {\bfseries 39} (2024) 2441012}
  [\href{https://arxiv.org/abs/2307.03384}{{\ttfamily 2307.03384}}].

\bibitem{Ding:2023htn}
G.-J. Ding and S.~F. King, \emph{{Neutrino mass and mixing with modular
  symmetry}}, \href{https://doi.org/10.1088/1361-6633/ad52a3}{\emph{Rept. Prog.
  Phys.} {\bfseries 87} (2024) 084201}
  [\href{https://arxiv.org/abs/2311.09282}{{\ttfamily 2311.09282}}].

\bibitem{Blumenhagen:2013fgp}
R.~Blumenhagen, D.~L{\"u}st and S.~Theisen, \emph{{Basic concepts of string
  theory}}, Theoretical and Mathematical Physics. Springer, Heidelberg,
  Germany, 2013,
  \href{https://doi.org/10.1007/978-3-642-29497-6}{10.1007/978-3-642-29497-6}.

\bibitem{Greene:1986qva}
B.~R. Greene, \emph{{Superstrings : Topology, geometry and phenomenology and
  astrophysical implications of supersymmetric models}}, Ph.D. thesis, Oxford
  U., Theor. Phys., 1986.

\bibitem{Bailin:1999nk}
D.~Bailin and A.~Love, \emph{{Orbifold compactifications of string theory}},
  \href{https://doi.org/10.1016/S0370-1573(98)00126-4}{\emph{Phys. Rept.}
  {\bfseries 315} (1999) 285}.

\bibitem{Dine:1992ya}
M.~Dine, R.~G. Leigh and D.~A. MacIntire, \emph{{Of CP and other gauge
  symmetries in string theory}},
  \href{https://doi.org/10.1103/PhysRevLett.69.2030}{\emph{Phys. Rev. Lett.}
  {\bfseries 69} (1992) 2030}
  [\href{https://arxiv.org/abs/hep-th/9205011}{{\ttfamily hep-th/9205011}}].

\bibitem{Choi:1992xp}
K.-w. Choi, D.~B. Kaplan and A.~E. Nelson, \emph{{Is CP a gauge symmetry?}},
  \href{https://doi.org/10.1016/0550-3213(93)90082-Z}{\emph{Nucl. Phys. B}
  {\bfseries 391} (1993) 515}
  [\href{https://arxiv.org/abs/hep-ph/9205202}{{\ttfamily hep-ph/9205202}}].

\bibitem{Leigh:1993ae}
R.~G. Leigh, \emph{{The Strong CP problem, string theory and the Nelson-Barr
  mechanism}},  in \emph{{International Workshop on Recent Advances in the
  Superworld}}, 6, 1993, \href{https://arxiv.org/abs/hep-ph/9307214}{{\ttfamily
  hep-ph/9307214}}.

\bibitem{Acharya:1995ag}
B.~S. Acharya, D.~Bailin, A.~Love, W.~A. Sabra and S.~Thomas,
  \emph{{Spontaneous breaking of CP symmetry by orbifold moduli}},
  \href{https://doi.org/10.1016/0370-2693(95)00945-H}{\emph{Phys. Lett. B}
  {\bfseries 357} (1995) 387}
  [\href{https://arxiv.org/abs/hep-th/9506143}{{\ttfamily hep-th/9506143}}],
  [Erratum: Phys.Lett.B 407, 451--451 (1997)].

\bibitem{Dent:2001cc}
T.~Dent, \emph{{CP violation and modular symmetries}},
  \href{https://doi.org/10.1103/PhysRevD.64.056005}{\emph{Phys. Rev. D}
  {\bfseries 64} (2001) 056005}
  [\href{https://arxiv.org/abs/hep-ph/0105285}{{\ttfamily hep-ph/0105285}}].

\bibitem{Giedt:2002ns}
J.~Giedt, \emph{{CP violation and moduli stabilization in heterotic models}},
  \href{https://doi.org/10.1142/S0217732302007879}{\emph{Mod. Phys. Lett. A}
  {\bfseries 17} (2002) 1465}
  [\href{https://arxiv.org/abs/hep-ph/0204017}{{\ttfamily hep-ph/0204017}}].

\bibitem{Baur:2019kwi}
A.~Baur, H.~P. Nilles, A.~Trautner and P.~K.~S. Vaudrevange, \emph{{Unification
  of Flavor, CP, and Modular Symmetries}},
  \href{https://doi.org/10.1016/j.physletb.2019.03.066}{\emph{Phys. Lett. B}
  {\bfseries 795} (2019) 7} [\href{https://arxiv.org/abs/1901.03251}{{\ttfamily
  1901.03251}}].

\bibitem{Novichkov:2019sqv}
P.~P. Novichkov, J.~T. Penedo, S.~T. Petcov and A.~V. Titov, \emph{{Generalised
  CP Symmetry in Modular-Invariant Models of Flavour}},
  \href{https://doi.org/10.1007/JHEP07(2019)165}{\emph{JHEP} {\bfseries 07}
  (2019) 165} [\href{https://arxiv.org/abs/1905.11970}{{\ttfamily
  1905.11970}}].

\bibitem{Feruglio:2021dte}
F.~Feruglio, V.~Gherardi, A.~Romanino and A.~Titov, \emph{{Modular invariant
  dynamics and fermion mass hierarchies around $\tau = i$}},
  \href{https://doi.org/10.1007/JHEP05(2021)242}{\emph{JHEP} {\bfseries 05}
  (2021) 242} [\href{https://arxiv.org/abs/2101.08718}{{\ttfamily
  2101.08718}}].

\bibitem{Novichkov:2021evw}
P.~P. Novichkov, J.~T. Penedo and S.~T. Petcov, \emph{{Fermion mass
  hierarchies, large lepton mixing and residual modular symmetries}},
  \href{https://doi.org/10.1007/JHEP04(2021)206}{\emph{JHEP} {\bfseries 04}
  (2021) 206} [\href{https://arxiv.org/abs/2102.07488}{{\ttfamily
  2102.07488}}].

\bibitem{Froggatt:1978nt}
C.~D. Froggatt and H.~B. Nielsen, \emph{{Hierarchy of Quark Masses, Cabibbo
  Angles and CP Violation}},
  \href{https://doi.org/10.1016/0550-3213(79)90316-X}{\emph{Nucl. Phys. B}
  {\bfseries 147} (1979) 277}.

\bibitem{Feruglio:2023uof}
F.~Feruglio, A.~Strumia and A.~Titov, \emph{{Modular invariance and the QCD
  angle}}, \href{https://doi.org/10.1007/JHEP07(2023)027}{\emph{JHEP}
  {\bfseries 07} (2023) 027}
  [\href{https://arxiv.org/abs/2305.08908}{{\ttfamily 2305.08908}}].

\bibitem{Penedo:2024gtb}
J.~T. Penedo and S.~T. Petcov, \emph{{Finite modular symmetries and the strong
  CP problem}}, \href{https://doi.org/10.1007/JHEP10(2024)172}{\emph{JHEP}
  {\bfseries 10} (2024) 172}
  [\href{https://arxiv.org/abs/2404.08032}{{\ttfamily 2404.08032}}].

\bibitem{Feruglio:2024ytl}
F.~Feruglio, M.~Parriciatu, A.~Strumia and A.~Titov, \emph{{Solving the strong
  CP problem without axions}},
  \href{https://doi.org/10.1007/JHEP08(2024)214}{\emph{JHEP} {\bfseries 08}
  (2024) 214} [\href{https://arxiv.org/abs/2406.01689}{{\ttfamily
  2406.01689}}].

\bibitem{Feruglio:2025ajb}
F.~Feruglio, A.~Marrone, A.~Strumia and A.~Titov, \emph{{Solving the strong CP
  problem in string-inspired theories with modular invariance}},
  \href{https://doi.org/10.1007/JHEP08(2025)076}{\emph{JHEP} {\bfseries 08}
  (2025) 076} [\href{https://arxiv.org/abs/2505.20395}{{\ttfamily
  2505.20395}}].

\bibitem{Duch:2025abl}
M.~Duch, A.~Strumia and A.~Titov, \emph{{Baryogenesis from cosmological CP
  breaking}}, \href{https://doi.org/10.1007/JHEP11(2025)109}{\emph{JHEP}
  {\bfseries 11} (2025) 109}
  [\href{https://arxiv.org/abs/2504.03506}{{\ttfamily 2504.03506}}].

\bibitem{Abe:2023ylh}
Y.~Abe, T.~Higaki, F.~Kaneko, T.~Kobayashi and H.~Otsuka, \emph{{Moduli
  inflation from modular flavor symmetries}},
  \href{https://doi.org/10.1007/JHEP06(2023)187}{\emph{JHEP} {\bfseries 06}
  (2023) 187} [\href{https://arxiv.org/abs/2303.02947}{{\ttfamily
  2303.02947}}].

\bibitem{Ding:2024neh}
G.-J. Ding, S.-Y. Jiang and W.~Zhao, \emph{{Modular invariant slow roll
  inflation}}, \href{https://doi.org/10.1088/1475-7516/2024/10/016}{\emph{JCAP}
  {\bfseries 10} (2024) 016}
  [\href{https://arxiv.org/abs/2405.06497}{{\ttfamily 2405.06497}}].

\bibitem{Ding:2024euc}
G.-J. Ding, S.-Y. Jiang, Y.~Xu and W.~Zhao, \emph{{Modular invariant inflation
  and reheating}}, \href{https://doi.org/10.1007/JHEP11(2025)141}{\emph{JHEP}
  {\bfseries 11} (2025) 141}
  [\href{https://arxiv.org/abs/2411.18603}{{\ttfamily 2411.18603}}].

\bibitem{Aoki:2025wld}
S.~Aoki, H.~Otsuka and R.~Yanagita, \emph{{Higgs-modular inflation}},
  \href{https://doi.org/10.1103/v4z9-676d}{\emph{Phys. Rev. D} {\bfseries 112}
  (2025) 043505} [\href{https://arxiv.org/abs/2504.01622}{{\ttfamily
  2504.01622}}].

\bibitem{Granelli:2025lds}
A.~Granelli, D.~Meloni, M.~Parriciatu, J.~T. Penedo and S.~T. Petcov,
  \emph{{Modular-symmetry-protected seesaw}},
  \href{https://doi.org/10.1007/JHEP12(2025)035}{\emph{JHEP} {\bfseries 12}
  (2025) 035} [\href{https://arxiv.org/abs/2505.21405}{{\ttfamily
  2505.21405}}].

\bibitem{Petcov:2022fjf}
S.~T. Petcov and M.~Tanimoto, \emph{{$A_4$ modular flavour model of quark mass
  hierarchies close to the fixed point $\tau = \omega $}},
  \href{https://doi.org/10.1140/epjc/s10052-023-11727-0}{\emph{Eur. Phys. J. C}
  {\bfseries 83} (2023) 579}
  [\href{https://arxiv.org/abs/2212.13336}{{\ttfamily 2212.13336}}].

\bibitem{Petcov:2023vws}
S.~T. Petcov and M.~Tanimoto, \emph{{A$_{4}$ modular flavour model of quark
  mass hierarchies close to the fixed point {\ensuremath{\tau}} =
  i{\ensuremath{\infty}}}},
  \href{https://doi.org/10.1007/JHEP08(2023)086}{\emph{JHEP} {\bfseries 08}
  (2023) 086} [\href{https://arxiv.org/abs/2306.05730}{{\ttfamily
  2306.05730}}].

\bibitem{deMedeirosVarzielas:2023crv}
I.~de~Medeiros~Varzielas, M.~Levy, J.~T. Penedo and S.~T. Petcov, \emph{{Quarks
  at the modular S$_{4}$ cusp}},
  \href{https://doi.org/10.1007/JHEP09(2023)196}{\emph{JHEP} {\bfseries 09}
  (2023) 196} [\href{https://arxiv.org/abs/2307.14410}{{\ttfamily
  2307.14410}}].

\bibitem{Petcov:2026mdx}
S.~T. Petcov and M.~Tanimoto, \emph{{{\(S'_4\)} Quark Flavour Model in the
  Vicinity of the Fixed Point {\(\tau= i\infty\)}}},
  \href{https://arxiv.org/abs/2601.04529}{{\ttfamily 2601.04529}}.

\bibitem{Kikuchi:2023jap}
S.~Kikuchi, T.~Kobayashi, K.~Nasu, S.~Takada and H.~Uchida, \emph{{Quark mass
  hierarchies and CP violation in A$_{4}$ {\texttimes} A$_{4}$ {\texttimes}
  A$_{4}$ modular symmetric flavor models}},
  \href{https://doi.org/10.1007/JHEP07(2023)134}{\emph{JHEP} {\bfseries 07}
  (2023) 134} [\href{https://arxiv.org/abs/2302.03326}{{\ttfamily
  2302.03326}}].

\bibitem{deMedeirosVarzielas:2026lfw}
I.~de~Medeiros~Varzielas and M.~Paiva, \emph{{Quark masses and mixing from
  Modular $S'_4$ with Canonical K{\"a}hler Effects}},
  \href{https://arxiv.org/abs/2604.01422}{{\ttfamily 2604.01422}}.

\bibitem{Ding:2020zxw}
G.-J. Ding, F.~Feruglio and X.-G. Liu, \emph{{Automorphic Forms and Fermion
  Masses}}, \href{https://doi.org/10.1007/JHEP01(2021)037}{\emph{JHEP}
  {\bfseries 01} (2021) 037}
  [\href{https://arxiv.org/abs/2010.07952}{{\ttfamily 2010.07952}}].

\bibitem{Ding:2021iqp}
G.-J. Ding, F.~Feruglio and X.-G. Liu, \emph{{CP symmetry and symplectic
  modular invariance}},
  \href{https://doi.org/10.21468/SciPostPhys.10.6.133}{\emph{SciPost Phys.}
  {\bfseries 10} (2021) 133}
  [\href{https://arxiv.org/abs/2102.06716}{{\ttfamily 2102.06716}}].

\bibitem{Ding:2024xhz}
G.-J. Ding, F.~Feruglio and X.-G. Liu, \emph{{Universal predictions of Siegel
  modular invariant theories near the fixed points}},
  \href{https://doi.org/10.1007/JHEP05(2024)052}{\emph{JHEP} {\bfseries 05}
  (2024) 052} [\href{https://arxiv.org/abs/2402.14915}{{\ttfamily
  2402.14915}}].

\bibitem{Jiang:2025qbi}
S.-Y. Jiang, W.~Zhao and G.-J. Ding, \emph{{$Sp(4,\mathbb{Z})$ modular
  inflation}},  \href{https://arxiv.org/abs/2512.21597}{{\ttfamily
  2512.21597}}.

\bibitem{Kikuchi:2023dow}
S.~Kikuchi, T.~Kobayashi, K.~Nasu, S.~Takada and H.~Uchida, \emph{{Sp(6, Z)
  modular symmetry in flavor structures: quark flavor models and Siegel modular
  forms for $\widetilde{\Delta }\left(96\right)$}},
  \href{https://doi.org/10.1007/JHEP04(2024)045}{\emph{JHEP} {\bfseries 04}
  (2024) 045} [\href{https://arxiv.org/abs/2310.17978}{{\ttfamily
  2310.17978}}].

\bibitem{Fleig:2015vky}
P.~Fleig, H.~P.~A. Gustafsson, A.~Kleinschmidt and D.~Persson,
  \emph{{Eisenstein series and automorphic representations: With Applications
  in String Theory}}, vol.~176. Cambridge University Press, 6, 2018,
  \href{https://doi.org/10.1017/9781316995860}{10.1017/9781316995860},
  [\href{https://arxiv.org/abs/1511.04265}{{\ttfamily 1511.04265}}].

\bibitem{Hamidi:1986vh}
S.~Hamidi and C.~Vafa, \emph{{Interactions on Orbifolds}},
  \href{https://doi.org/10.1016/0550-3213(87)90006-X}{\emph{Nucl. Phys. B}
  {\bfseries 279} (1987) 465}.

\bibitem{Alvarez-Gaume:1986bwm}
L.~Alvarez-Gaume and P.~C. Nelson, \emph{{Riemann surfaces and string
  theories}},  in \emph{{4th Trieste Spring School on Supersymmetry,
  Supergravity, Superstrings}: {(followed by 3 day Workshop)}}, 12, 1986.

\bibitem{Siegel:1943}
C.~L. Siegel, \emph{Symplectic geometry}, {\emph{American Journal of
  Mathematics} {\bfseries 65} (1943) 1}.

\bibitem{Gaillard:1981rj}
M.~K. Gaillard and B.~Zumino, \emph{{Duality Rotations for Interacting
  Fields}}, \href{https://doi.org/10.1016/0550-3213(81)90527-7}{\emph{Nucl.
  Phys. B} {\bfseries 193} (1981) 221}.

\bibitem{Castellani:1991et}
L.~Castellani, R.~D'Auria and P.~Fre, \emph{{Supergravity and superstrings: A
  Geometric perspective. Vol. 1: Mathematical foundations}}. World Scientific,
  1991.

\bibitem{Castellani:1991eu}
L.~Castellani, R.~D'Auria and P.~Fre, \emph{{Supergravity and superstrings: A
  Geometric perspective. Vol. 2: Supergravity}}. World Scientific, 1991.

\bibitem{Castellani:1991ev}
L.~Castellani, R.~D'Auria and P.~Fre, \emph{{Supergravity and superstrings: A
  Geometric perspective. Vol. 3: Superstrings}}. World Scientific, 1991.

\bibitem{Yau:1977ms}
S.-T. Yau, \emph{{Calabi's Conjecture and some new results in algebraic
  geometry}}, \href{https://doi.org/10.1073/pnas.74.5.1798}{\emph{Proc. Nat.
  Acad. Sci.} {\bfseries 74} (1977) 1798}.

\bibitem{Marcus:1982fr}
N.~Marcus and A.~Sagnotti, \emph{{Tree Level Constraints on Gauge Groups for
  Type I Superstrings}},
  \href{https://doi.org/10.1016/0370-2693(82)90253-2}{\emph{Phys. Lett. B}
  {\bfseries 119} (1982) 97}.

\bibitem{Candelas:1985en}
P.~Candelas, G.~T. Horowitz, A.~Strominger and E.~Witten, \emph{{Vacuum
  configurations for superstrings}},
  \href{https://doi.org/10.1016/0550-3213(85)90602-9}{\emph{Nucl. Phys. B}
  {\bfseries 258} (1985) 46}.

\bibitem{Narain:1986am}
K.~S. Narain, M.~H. Sarmadi and E.~Witten, \emph{{A Note on Toroidal
  Compactification of Heterotic String Theory}},
  \href{https://doi.org/10.1016/0550-3213(87)90001-0}{\emph{Nucl. Phys. B}
  {\bfseries 279} (1987) 369}.

\bibitem{Bianchi:1991eu}
M.~Bianchi, G.~Pradisi and A.~Sagnotti, \emph{{Toroidal compactification and
  symmetry breaking in open string theories}},
  \href{https://doi.org/10.1016/0550-3213(92)90129-Y}{\emph{Nucl. Phys. B}
  {\bfseries 376} (1992) 365}.

\bibitem{Giveon:1994fu}
A.~Giveon, M.~Porrati and E.~Rabinovici, \emph{{Target space duality in string
  theory}}, \href{https://doi.org/10.1016/0370-1573(94)90070-1}{\emph{Phys.
  Rept.} {\bfseries 244} (1994) 77}
  [\href{https://arxiv.org/abs/hep-th/9401139}{{\ttfamily hep-th/9401139}}].

\bibitem{Sen:1995ff}
A.~Sen and C.~Vafa, \emph{{Dual pairs of type II string compactification}},
  \href{https://doi.org/10.1016/0550-3213(95)00498-H}{\emph{Nucl. Phys. B}
  {\bfseries 455} (1995) 165}
  [\href{https://arxiv.org/abs/hep-th/9508064}{{\ttfamily hep-th/9508064}}].

\bibitem{Witten:1997bs}
E.~Witten, \emph{{Toroidal compactification without vector structure}},
  \href{https://doi.org/10.1088/1126-6708/1998/02/006}{\emph{JHEP} {\bfseries
  02} (1998) 006} [\href{https://arxiv.org/abs/hep-th/9712028}{{\ttfamily
  hep-th/9712028}}].

\bibitem{Hollowood:2003cv}
T.~J. Hollowood, A.~Iqbal and C.~Vafa, \emph{{Matrix models, geometric
  engineering and elliptic genera}},
  \href{https://doi.org/10.1088/1126-6708/2008/03/069}{\emph{JHEP} {\bfseries
  03} (2008) 069} [\href{https://arxiv.org/abs/hep-th/0310272}{{\ttfamily
  hep-th/0310272}}].

\bibitem{Blumenhagen:2006ci}
R.~Blumenhagen, B.~Kors, D.~Lust and S.~Stieberger, \emph{{Four-dimensional
  String Compactifications with D-Branes, Orientifolds and Fluxes}},
  \href{https://doi.org/10.1016/j.physrep.2007.04.003}{\emph{Phys. Rept.}
  {\bfseries 445} (2007) 1}
  [\href{https://arxiv.org/abs/hep-th/0610327}{{\ttfamily hep-th/0610327}}].

\bibitem{Dijkgraaf:2007sw}
R.~Dijkgraaf, L.~Hollands, P.~Sulkowski and C.~Vafa, \emph{{Supersymmetric
  gauge theories, intersecting branes and free fermions}},
  \href{https://doi.org/10.1088/1126-6708/2008/02/106}{\emph{JHEP} {\bfseries
  02} (2008) 106} [\href{https://arxiv.org/abs/0709.4446}{{\ttfamily
  0709.4446}}].

\bibitem{Antoniadis:2009bg}
I.~Antoniadis, A.~Kumar and B.~Panda, \emph{{Fermion Wavefunctions in
  Magnetized branes: Theta identities and Yukawa couplings}},
  \href{https://doi.org/10.1016/j.nuclphysb.2009.08.002}{\emph{Nucl. Phys. B}
  {\bfseries 823} (2009) 116}
  [\href{https://arxiv.org/abs/0904.0910}{{\ttfamily 0904.0910}}].

\bibitem{Cordova:2012xk}
C.~Cordova, S.~Espahbodi, B.~Haghighat, A.~Rastogi and C.~Vafa, \emph{{Tangles,
  Generalized Reidemeister Moves, and Three-Dimensional Mirror Symmetry}},
  \href{https://doi.org/10.1007/JHEP05(2014)014}{\emph{JHEP} {\bfseries 05}
  (2014) 014} [\href{https://arxiv.org/abs/1211.3730}{{\ttfamily 1211.3730}}].

\bibitem{Hebecker:2017lxm}
A.~Hebecker, P.~Henkenjohann and L.~T. Witkowski, \emph{{Flat Monodromies and a
  Moduli Space Size Conjecture}},
  \href{https://doi.org/10.1007/JHEP12(2017)033}{\emph{JHEP} {\bfseries 12}
  (2017) 033} [\href{https://arxiv.org/abs/1708.06761}{{\ttfamily
  1708.06761}}].

\bibitem{Reffert:2006du}
S.~Reffert, \emph{{Toroidal Orbifolds: Resolutions, Orientifolds and
  Applications in String Phenomenology}}, Ph.D. thesis, Munich U., 2006.
\newblock \href{https://arxiv.org/abs/hep-th/0609040}{{\ttfamily
  hep-th/0609040}}.

\bibitem{Dixon:1985jw}
L.~J. Dixon, J.~A. Harvey, C.~Vafa and E.~Witten, \emph{{Strings on
  Orbifolds}}, \href{https://doi.org/10.1016/0550-3213(85)90593-0}{\emph{Nucl.
  Phys. B} {\bfseries 261} (1985) 678}.

\bibitem{Dixon:1986jc}
L.~J. Dixon, J.~A. Harvey, C.~Vafa and E.~Witten, \emph{{Strings on Orbifolds.
  2.}}, \href{https://doi.org/10.1016/0550-3213(86)90287-7}{\emph{Nucl. Phys.
  B} {\bfseries 274} (1986) 285}.

\bibitem{Cremades:2004wa}
D.~Cremades, L.~E. Ibanez and F.~Marchesano, \emph{{Computing Yukawa couplings
  from magnetized extra dimensions}},
  \href{https://doi.org/10.1088/1126-6708/2004/05/079}{\emph{JHEP} {\bfseries
  05} (2004) 079} [\href{https://arxiv.org/abs/hep-th/0404229}{{\ttfamily
  hep-th/0404229}}].

\bibitem{Kikuchi:2020frp}
S.~Kikuchi, T.~Kobayashi, S.~Takada, T.~H. Tatsuishi and H.~Uchida,
  \emph{{Revisiting modular symmetry in magnetized torus and orbifold
  compactifications}},
  \href{https://doi.org/10.1103/PhysRevD.102.105010}{\emph{Phys. Rev. D}
  {\bfseries 102} (2020) 105010}
  [\href{https://arxiv.org/abs/2005.12642}{{\ttfamily 2005.12642}}].

\bibitem{Kikuchi:2020nxn}
S.~Kikuchi, T.~Kobayashi, H.~Otsuka, S.~Takada and H.~Uchida, \emph{{Modular
  symmetry by orbifolding magnetized $T^2\times T^2$: realization of double
  cover of $\Gamma_N$}},
  \href{https://doi.org/10.1007/JHEP11(2020)101}{\emph{JHEP} {\bfseries 11}
  (2020) 101} [\href{https://arxiv.org/abs/2007.06188}{{\ttfamily
  2007.06188}}].

\bibitem{Obers:1999es}
N.~A. Obers and B.~Pioline, \emph{{Eisenstein series in string theory}},
  \href{https://doi.org/10.1088/0264-9381/17/5/330}{\emph{Class. Quant. Grav.}
  {\bfseries 17} (2000) 1215}
  [\href{https://arxiv.org/abs/hep-th/9910115}{{\ttfamily hep-th/9910115}}].

\bibitem{DHoker:1988pdl}
E.~D'Hoker and D.~H. Phong, \emph{{The Geometry of String Perturbation
  Theory}}, \href{https://doi.org/10.1103/RevModPhys.60.917}{\emph{Rev. Mod.
  Phys.} {\bfseries 60} (1988) 917}.

\bibitem{DHoker:2001kkt}
E.~D'Hoker and D.~H. Phong, \emph{{Two loop superstrings. 1. Main formulas}},
  \href{https://doi.org/10.1016/S0370-2693(02)01255-8}{\emph{Phys. Lett. B}
  {\bfseries 529} (2002) 241}
  [\href{https://arxiv.org/abs/hep-th/0110247}{{\ttfamily hep-th/0110247}}].

\bibitem{DHoker:2001qqx}
E.~D'Hoker and D.~H. Phong, \emph{{Two loop superstrings. 2. The Chiral measure
  on moduli space}},
  \href{https://doi.org/10.1016/S0550-3213(02)00431-5}{\emph{Nucl. Phys. B}
  {\bfseries 636} (2002) 3}
  [\href{https://arxiv.org/abs/hep-th/0110283}{{\ttfamily hep-th/0110283}}].

\bibitem{DHoker:2001foj}
E.~D'Hoker and D.~H. Phong, \emph{{Two loop superstrings. 3. Slice independence
  and absence of ambiguities}},
  \href{https://doi.org/10.1016/S0550-3213(02)00432-7}{\emph{Nucl. Phys. B}
  {\bfseries 636} (2002) 61}
  [\href{https://arxiv.org/abs/hep-th/0111016}{{\ttfamily hep-th/0111016}}].

\bibitem{DHoker:2001jaf}
E.~D'Hoker and D.~H. Phong, \emph{{Two loop superstrings 4: The Cosmological
  constant and modular forms}},
  \href{https://doi.org/10.1016/S0550-3213(02)00516-3}{\emph{Nucl. Phys. B}
  {\bfseries 639} (2002) 129}
  [\href{https://arxiv.org/abs/hep-th/0111040}{{\ttfamily hep-th/0111040}}].

\bibitem{Igusaa}
I.~Jun-Ichi, \emph{On siegel modular forms of genus two},
  \href{https://doi.org/10.2307/2372812}{\emph{American Journal of Mathematics}
  {\bfseries 84} (1962) 175}.

\bibitem{DallaPiazza:2008qoi}
F.~Dalla~Piazza and B.~van Geemen, \emph{{Siegel modular forms and finite
  symplectic groups}},
  \href{https://doi.org/10.4310/ATMP.2009.v13.n6.a4}{\emph{Adv. Theor. Math.
  Phys.} {\bfseries 13} (2009) 1771}
  [\href{https://arxiv.org/abs/0804.3769}{{\ttfamily 0804.3769}}].

\bibitem{Cacciatori:2007vk}
S.~L. Cacciatori and F.~Dalla~Piazza, \emph{{Two loop superstring amplitudes
  and S(6) representations}},
  \href{https://doi.org/10.1007/s11005-007-0213-8}{\emph{Lett. Math. Phys.}
  {\bfseries 83} (2008) 127} [\href{https://arxiv.org/abs/0707.0646}{{\ttfamily
  0707.0646}}].

\bibitem{Nilles:2021glx}
H.~P. Nilles, S.~Ramos-Sanchez, A.~Trautner and P.~K.~S. Vaudrevange,
  \emph{{Orbifolds from Sp(4,Z) and their modular symmetries}},
  \href{https://doi.org/10.1016/j.nuclphysb.2021.115534}{\emph{Nucl. Phys. B}
  {\bfseries 971} (2021) 115534}
  [\href{https://arxiv.org/abs/2105.08078}{{\ttfamily 2105.08078}}].

\bibitem{gottschling1961fixpunkte}
E.~Gottschling, \emph{{\"U}ber die fixpunkte der siegelschen modulgruppe},
  {\emph{Mathematische Annalen} {\bfseries 143} (1961) 111}.

\bibitem{RickyDevi:2024ijc}
M.~Ricky~Devi, \emph{{Neutrino Masses and Higher Degree Siegel Modular Forms}},
   \href{https://arxiv.org/abs/2401.16257}{{\ttfamily 2401.16257}}.

\bibitem{Novichkov:2022wvg}
P.~P. Novichkov, J.~T. Penedo and S.~T. Petcov, \emph{{Modular flavour
  symmetries and modulus stabilisation}},
  \href{https://doi.org/10.1007/JHEP03(2022)149}{\emph{JHEP} {\bfseries 03}
  (2022) 149} [\href{https://arxiv.org/abs/2201.02020}{{\ttfamily
  2201.02020}}].

\bibitem{Feruglio:2022koo}
F.~Feruglio, \emph{{Universal Predictions of Modular Invariant Flavor Models
  near the Self-Dual Point}},
  \href{https://doi.org/10.1103/PhysRevLett.130.101801}{\emph{Phys. Rev. Lett.}
  {\bfseries 130} (2023) 101801}
  [\href{https://arxiv.org/abs/2211.00659}{{\ttfamily 2211.00659}}].

\bibitem{Feruglio:2023mii}
F.~Feruglio, \emph{{Fermion masses, critical behavior and universality}},
  \href{https://doi.org/10.1007/JHEP03(2023)236}{\emph{JHEP} {\bfseries 03}
  (2023) 236} [\href{https://arxiv.org/abs/2302.11580}{{\ttfamily
  2302.11580}}].

\bibitem{Okada:2020rjb}
H.~Okada and M.~Tanimoto, \emph{{Quark and lepton flavors with common modulus
  {\ensuremath{\tau}} in A4 modular symmetry}},
  \href{https://doi.org/10.1016/j.dark.2023.101204}{\emph{Phys. Dark Univ.}
  {\bfseries 40} (2023) 101204}
  [\href{https://arxiv.org/abs/2005.00775}{{\ttfamily 2005.00775}}].

\bibitem{Novichkov:2018ovf}
P.~P. Novichkov, J.~T. Penedo, S.~T. Petcov and A.~V. Titov, \emph{{Modular
  S$_{4}$ models of lepton masses and mixing}},
  \href{https://doi.org/10.1007/JHEP04(2019)005}{\emph{JHEP} {\bfseries 04}
  (2019) 005} [\href{https://arxiv.org/abs/1811.04933}{{\ttfamily
  1811.04933}}].

\bibitem{Novichkov:2021cgl}
P.~Novichkov, \emph{{Aspects of the Modular Symmetry Approach to Lepton
  Flavour}}, Ph.D. thesis, SISSA, Trieste, 2021.

\bibitem{Gatto:1968ss}
R.~Gatto, G.~Sartori and M.~Tonin, \emph{{Weak Selfmasses, Cabibbo Angle, and
  Broken SU(2) x SU(2)}},
  \href{https://doi.org/10.1016/0370-2693(68)90150-0}{\emph{Phys. Lett. B}
  {\bfseries 28} (1968) 128}.

\bibitem{kato:1949ts}
T.~Kato, \emph{On the convergence of the perturbation method.},
  \href{https://doi.org/10.1143/ptp/4.4.514}{\emph{Progress of Theoretical
  Physics} {\bfseries 4} (1949) 514}.

\bibitem{bloch:1958cl}
C.~Bloch, \emph{Sur la théorie des perturbations des états liés},
  \href{https://doi.org/https://doi.org/10.1016/0029-5582(58)90116-0}{\emph{Nuclear
  Physics} {\bfseries 6} (1958) 329}.

\bibitem{GAP4}
{The GAP Group}, \emph{{GAP -- Groups, Algorithms, and Programming}},  Version
  4.10.2, 2019, \url{https://www.gap-system.org}.

\bibitem{SmallGroups}
H.~U. Besche, B.~Eick and E.~O'Brien, \emph{{SmallGrp -- a GAP package}},
  Version 1.3, 2018, \url{https://gap-packages.github.io/smallgrp}.

\bibitem{Kobayashi:2018vbk}
T.~Kobayashi, K.~Tanaka and T.~H. Tatsuishi, \emph{{Neutrino mixing from finite
  modular groups}},
  \href{https://doi.org/10.1103/PhysRevD.98.016004}{\emph{Phys. Rev. D}
  {\bfseries 98} (2018) 016004}
  [\href{https://arxiv.org/abs/1803.10391}{{\ttfamily 1803.10391}}].

\bibitem{Meloni:2023aru}
D.~Meloni and M.~Parriciatu, \emph{{A simplest modular S$_{3}$ model for
  leptons}}, \href{https://doi.org/10.1007/JHEP09(2023)043}{\emph{JHEP}
  {\bfseries 09} (2023) 043}
  [\href{https://arxiv.org/abs/2306.09028}{{\ttfamily 2306.09028}}].

\end{thebibliography}\endgroup

\end{document}